\documentclass[aps,prc,
twocolumn, reprint,tightenlines,
floatfix,showpacs,preprintnumbers,amsmath,amssymb,nofootinbib,superscriptaddress]{revtex4-2}
\usepackage{graphicx}
\usepackage{placeins}
\usepackage{dcolumn}
\usepackage{bm}
\usepackage{subcaption}
\usepackage{color}
\usepackage{bm}

\usepackage{amsmath,amssymb}
\usepackage{hyperref}
\usepackage{cleveref}
\usepackage{comment}
\usepackage{url}
\usepackage[normalem]{ulem}
\usepackage{physics}
\usepackage{diagbox}
\usepackage{parskip}
\newcommand{\sat}{\textrm{sat}}
\newcommand{\sym}{\textrm{sym}}

\renewcommand{\L}{\mathcal{L}}
\newcommand{\psib}{\bar{\psi}}

\begin{document}
\newcommand{\MsLatNEP}{$866^{+92}_{-116}$}
\newcommand{\gsigmaLatNEP}{$11.81^{+2.00}_{-2.13}$}
\newcommand{\gomegaLatNEP}{$6.59^{+0.94}_{-1.00}$}
\newcommand{\CnsLatNEP}{$1.459^{+0.153}_{-0.180}$}
\newcommand{\grhoLatNEP}{$5.23^{+3.92}_{-1.21}$}
\newcommand{\gdeltaLatNEP}{$4.18^{+6.77}_{-3.74}$}
\newcommand{\MeffecLatNEP}{$0.85^{+0.025}_{-0.024}$}
\newcommand{\nsatLatNEP}{$0.16^{+0.010}_{-0.010}$}
\newcommand{\EsatLatNEP}{$-16.1^{+0.67}_{-0.58}$}
\newcommand{\KsatLatNEP}{$273^{+18}_{-21}$}
\newcommand{\QsatLatNEP}{$-560^{+87}_{-90}$}
\newcommand{\ZsatLatNEP}{$628^{+762}_{-578}$}
\newcommand{\JsymLatNEP}{$31.9^{+3.30}_{-3.45}$}
\newcommand{\LsymLatNEP}{$89^{+21}_{-12}$}
\newcommand{\KsymLatNEP}{$-13^{+135}_{-15}$}
\newcommand{\QsymLatNEP}{$51^{+48}_{-18}$}
\newcommand{\ZsymLatNEP}{$-471^{+62}_{-386}$}
\newcommand{\atwoLatNEP}{$1.475^{+0.156}_{-0.073}$}
\newcommand{\afourLatNEP}{$-0.514^{+0.053}_{-0.045}$}
\newcommand{\MsLatNEPtwoMsun}{$963^{+29}_{-33}$}
\newcommand{\gsigmaLatNEPtwoMsun}{$13.97^{+0.76}_{-0.62}$}
\newcommand{\gomegaLatNEPtwoMsun}{$7.62^{+0.38}_{-0.30}$}
\newcommand{\CnsLatNEPtwoMsun}{$1.626^{+0.063}_{-0.058}$}
\newcommand{\grhoLatNEPtwoMsun}{$6.76^{+3.44}_{-2.42}$}
\newcommand{\gdeltaLatNEPtwoMsun}{$6.94^{+5.54}_{-6.03}$}
\newcommand{\MeffecLatNEPtwoMsun}{$0.82^{+0.009}_{-0.011}$}
\newcommand{\nsatLatNEPtwoMsun}{$0.15^{+0.005}_{-0.006}$}
\newcommand{\EsatLatNEPtwoMsun}{$-16.2^{+0.64}_{-0.59}$}
\newcommand{\KsatLatNEPtwoMsun}{$289^{+10}_{-10}$}
\newcommand{\QsatLatNEPtwoMsun}{$-472^{+40}_{-38}$}
\newcommand{\ZsatLatNEPtwoMsun}{$42^{+194}_{-180}$}
\newcommand{\JsymLatNEPtwoMsun}{$31.8^{+3.32}_{-3.21}$}
\newcommand{\LsymLatNEPtwoMsun}{$95^{+26}_{-14}$}
\newcommand{\KsymLatNEPtwoMsun}{$24^{+188}_{-45}$}
\newcommand{\QsymLatNEPtwoMsun}{$39^{+176}_{-10}$}
\newcommand{\ZsymLatNEPtwoMsun}{$-585^{+128}_{-1093}$}
\newcommand{\atwoLatNEPtwoMsun}{$1.415^{+0.049}_{-0.017}$}
\newcommand{\afourLatNEPtwoMsun}{$-0.522^{+0.056}_{-0.038}$}
\newcommand{\MsLatNEPAstro}{$967^{+35}_{-36}$}
\newcommand{\gsigmaLatNEPAstro}{$14.08^{+0.93}_{-0.73}$}
\newcommand{\gomegaLatNEPAstro}{$7.67^{+0.45}_{-0.32}$}
\newcommand{\CnsLatNEPAstro}{$1.622^{+0.077}_{-0.059}$}
\newcommand{\grhoLatNEPAstro}{$5.11^{+2.37}_{-1.01}$}
\newcommand{\gdeltaLatNEPAstro}{$3.71^{+4.57}_{-3.38}$}
\newcommand{\MeffecLatNEPAstro}{$0.82^{+0.010}_{-0.012}$}
\newcommand{\nsatLatNEPAstro}{$0.15^{+0.005}_{-0.006}$}
\newcommand{\EsatLatNEPAstro}{$-16.2^{+0.61}_{-0.60}$}
\newcommand{\KsatLatNEPAstro}{$291^{+10}_{-10}$}
\newcommand{\QsatLatNEPAstro}{$-468^{+43}_{-38}$}
\newcommand{\ZsatLatNEPAstro}{$12^{+202}_{-201}$}
\newcommand{\JsymLatNEPAstro}{$31.5^{+3.24}_{-3.25}$}
\newcommand{\LsymLatNEPAstro}{$88^{+12}_{-11}$}
\newcommand{\KsymLatNEPAstro}{$-10^{+60}_{-12}$}
\newcommand{\QsymLatNEPAstro}{$34^{+11}_{-7}$}
\newcommand{\ZsymLatNEPAstro}{$-487^{+34}_{-188}$}
\newcommand{\atwoLatNEPAstro}{$1.413^{+0.047}_{-0.016}$}
\newcommand{\afourLatNEPAstro}{$-0.533^{+0.060}_{-0.028}$}
\newcommand{\MsNEPAstro}{$1088^{+188}_{-212}$}
\newcommand{\gsigmaNEPAstro}{$16.41^{+2.97}_{-3.50}$}
\newcommand{\gomegaNEPAstro}{$8.46^{+1.01}_{-0.73}$}
\newcommand{\CnsNEPAstro}{$1.608^{+0.526}_{-0.363}$}
\newcommand{\grhoNEPAstro}{$4.60^{+2.19}_{-0.88}$}
\newcommand{\gdeltaNEPAstro}{$3.04^{+4.38}_{-2.74}$}
\newcommand{\MeffecNEPAstro}{$0.78^{+0.030}_{-0.045}$}
\newcommand{\nsatNEPAstro}{$0.16^{+0.009}_{-0.009}$}
\newcommand{\EsatNEPAstro}{$-16^{+0.65}_{-0.63}$}
\newcommand{\KsatNEPAstro}{$309^{+21}_{-17}$}
\newcommand{\QsatNEPAstro}{$-353^{+131}_{-90}$}
\newcommand{\ZsatNEPAstro}{$-521^{+402}_{-431}$}
\newcommand{\JsymNEPAstro}{$31.6^{+3.26}_{-3.28}$}
\newcommand{\LsymNEPAstro}{$89^{+13}_{-11}$}
\newcommand{\KsymNEPAstro}{$-1^{+66}_{-16}$}
\newcommand{\QsymNEPAstro}{$22^{+25}_{-17}$}
\newcommand{\ZsymNEPAstro}{$-536^{+91}_{-350}$}
\newcommand{\atwoNEPAstro}{$1.309^{+0.281}_{-0.209}$}
\newcommand{\afourNEPAstro}{$-0.573^{+0.328}_{-0.534}$}
\newcommand{\MsNEP}{$930^{+290}_{-267}$}
\newcommand{\gsigmaNEP}{$12.57^{+5.88}_{-4.10}$}
\newcommand{\gomegaNEP}{$6.39^{+3.28}_{-2.33}$}
\newcommand{\CnsNEP}{$1.683^{+0.875}_{-0.661}$}
\newcommand{\grhoNEP}{$5.20^{+3.99}_{-1.23}$}
\newcommand{\gdeltaNEP}{$4.19^{+6.81}_{-3.78}$}
\newcommand{\MeffecNEP}{$0.85^{+0.055}_{-0.118}$}
\newcommand{\nsatNEP}{$0.16^{+0.010}_{-0.010}$}
\newcommand{\EsatNEP}{$-16.0^{+0.63}_{-0.64}$}
\newcommand{\KsatNEP}{$269^{+57}_{-61}$}
\newcommand{\QsatNEP}{$-574^{+339}_{-198}$}
\newcommand{\ZsatNEP}{$783^{+2389}_{-1762}$}
\newcommand{\JsymNEP}{$31.9^{+3.26}_{-3.23}$}
\newcommand{\LsymNEP}{$89^{+24}_{-12}$}
\newcommand{\KsymNEP}{$-8^{+173}_{-21}$}
\newcommand{\QsymNEP}{$60^{+111}_{-42}$}
\newcommand{\ZsymNEP}{$-508^{+92}_{-930}$}
\newcommand{\atwoNEP}{$1.381^{+0.496}_{-0.289}$}
\newcommand{\afourNEP}{$-0.239^{+1.270}_{-1.464}$}

\newcommand{\MsLatNEPGWtwoMsun}{$963^{+19}_{-27}$}
\newcommand{\gsigmaLatNEPGWtwoMsun}{$13.93^{+0.50}_{-0.50}$}
\newcommand{\gomegaLatNEPGWtwoMsun}{$7.58^{+0.25}_{-0.22}$}
\newcommand{\CnsLatNEPGWtwoMsun}{$1.616^{+0.052}_{-0.049}$}
\newcommand{\grhoLatNEPGWtwoMsun}{$5.10^{+2.08}_{-1.00}$}
\newcommand{\gdeltaLatNEPGWtwoMsun}{$3.69^{+4.08}_{-3.32}$}
\newcommand{\MeffecLatNEPGWtwoMsun}{$0.82^{+0.007}_{-0.009}$}
\newcommand{\nsatLatNEPGWtwoMsun}{$0.15^{+0.004}_{-0.005}$}
\newcommand{\EsatLatNEPGWtwoMsun}{$-16.2^{+0.55}_{-0.58}$}
\newcommand{\KsatLatNEPGWtwoMsun}{$289^{+8}_{-9}$}
\newcommand{\QsatLatNEPGWtwoMsun}{$-474^{+31}_{-35}$}
\newcommand{\ZsatLatNEPGWtwoMsun}{$44^{+147}_{-155}$}
\newcommand{\JsymLatNEPGWtwoMsun}{$31.7^{+2.87}_{-3.35}$}
\newcommand{\LsymLatNEPGWtwoMsun}{$87^{+11}_{-11}$}
\newcommand{\KsymLatNEPGWtwoMsun}{$-10^{+50}_{-12}$}
\newcommand{\QsymLatNEPGWtwoMsun}{$36^{+8}_{-8}$}
\newcommand{\ZsymLatNEPGWtwoMsun}{$-483^{+28}_{-190}$}
\newcommand{\atwoLatNEPGWtwoMsun}{$1.413^{+0.035}_{-0.015}$}
\newcommand{\afourLatNEPGWtwoMsun}{$-0.526^{+0.051}_{-0.035}$}


\newcommand{\MsRMFCCNEPGWtwoMsun}{$1054^{+202}_{-219}$}
\newcommand{\gsigmaRMFCCNEPGWtwoMsun}{$15.56^{+3.14}_{-3.46}$}
\newcommand{\gomegaRMFCCNEPGWtwoMsun}{$8.05^{+0.97}_{-0.50}$}
\newcommand{\CnsRMFCCNEPGWtwoMsun}{$1.593^{+0.542}_{-0.423}$}
\newcommand{\grhoRMFCCNEPGWtwoMsun}{$4.62^{+1.88}_{-0.87}$}
\newcommand{\gdeltaRMFCCNEPGWtwoMsun}{$3.04^{+3.87}_{-2.74}$}
\newcommand{\MeffecRMFCCNEPGWtwoMsun}{$0.79^{+0.022}_{-0.047}$}
\newcommand{\nsatRMFCCNEPGWtwoMsun}{$0.16^{+0.008}_{-0.009}$}
\newcommand{\EsatRMFCCNEPGWtwoMsun}{$-16.0^{+0.57}_{-0.65}$}
\newcommand{\KsatRMFCCNEPGWtwoMsun}{$302^{+19}_{-14}$}
\newcommand{\QsatRMFCCNEPGWtwoMsun}{$-394^{+114}_{-74}$}
\newcommand{\ZsatRMFCCNEPGWtwoMsun}{$-323^{+300}_{-508}$}
\newcommand{\JsymRMFCCNEPGWtwoMsun}{$31.7^{+2.88}_{-3.18}$}
\newcommand{\LsymRMFCCNEPGWtwoMsun}{$89^{+11}_{-11}$}
\newcommand{\KsymRMFCCNEPGWtwoMsun}{$-5^{+49}_{-14}$}
\newcommand{\QsymRMFCCNEPGWtwoMsun}{$26^{+12}_{-17}$}
\newcommand{\ZsymRMFCCNEPGWtwoMsun}{$-505^{+68}_{-251}$}
\newcommand{\atwoRMFCCNEPGWtwoMsun}{$1.321^{+0.286}_{-0.235}$}
\newcommand{\afourRMFCCNEPGWtwoMsun}{$-0.560^{+0.331}_{-0.704}$}
\newcommand{\MsRMFCCLwLatNEP}{$860^{+85}_{-111}$}
\newcommand{\gsigmaRMFCCLwLatNEP}{$11.54^{+1.93}_{-1.82}$}
\newcommand{\gomegaRMFCCLwLatNEP}{$6.46^{+0.91}_{-0.87}$}
\newcommand{\CnsRMFCCLwLatNEP}{$1.436^{+0.150}_{-0.155}$}
\newcommand{\grhoRMFCCLwLatNEP}{$5.27^{+1.94}_{-1.08}$}
\newcommand{\gdeltaRMFCCLwLatNEP}{$2.50^{+3.36}_{-2.28}$}
\newcommand{\LambdaomegaRMFCCLwLatNEP}{$0.502^{+0.440}_{-0.461}$}
\newcommand{\MeffecRMFCCLwLatNEP}{$0.85^{+0.022}_{-0.023}$}
\newcommand{\nsatRMFCCLwLatNEP}{$0.16^{+0.009}_{-0.009}$}
\newcommand{\EsatRMFCCLwLatNEP}{$-16.0^{+0.67}_{-0.62}$}
\newcommand{\KsatRMFCCLwLatNEP}{$272^{+16}_{-19}$}
\newcommand{\QsatRMFCCLwLatNEP}{$-563^{+81}_{-86}$}
\newcommand{\ZsatRMFCCLwLatNEP}{$675^{+705}_{-557}$}
\newcommand{\JsymRMFCCLwLatNEP}{$31.8^{+3.33}_{-3.16}$}
\newcommand{\LsymRMFCCLwLatNEP}{$59^{+24}_{-31}$}
\newcommand{\KsymRMFCCLwLatNEP}{$-171^{+125}_{-94}$}
\newcommand{\QsymRMFCCLwLatNEP}{$261^{+778}_{-290}$}
\newcommand{\ZsymRMFCCLwLatNEP}{$102^{+833}_{-2860}$}
\newcommand{\atwoRMFCCLwLatNEP}{$1.474^{+0.155}_{-0.072}$}
\newcommand{\afourRMFCCLwLatNEP}{$-0.512^{+0.052}_{-0.047}$}
\newcommand{\MsRMFCCLwLatNEPAstro}{$1015^{+37}_{-38}$}
\newcommand{\gsigmaRMFCCLwLatNEPAstro}{$15.47^{+1.04}_{-0.79}$}
\newcommand{\gomegaRMFCCLwLatNEPAstro}{$8.35^{+0.47}_{-0.32}$}
\newcommand{\CnsRMFCCLwLatNEPAstro}{$1.728^{+0.066}_{-0.063}$}
\newcommand{\grhoRMFCCLwLatNEPAstro}{$6.52^{+1.34}_{-1.46}$}
\newcommand{\gdeltaRMFCCLwLatNEPAstro}{$1.18^{+2.23}_{-1.07}$}
\newcommand{\LambdaomegaRMFCCLwLatNEPAstro}{$0.584^{+0.350}_{-0.384}$}
\newcommand{\MeffecRMFCCLwLatNEPAstro}{$0.81^{+0.011}_{-0.015}$}
\newcommand{\nsatRMFCCLwLatNEPAstro}{$0.14^{+0.005}_{-0.002}$}
\newcommand{\EsatRMFCCLwLatNEPAstro}{$-16.2^{+0.63}_{-0.60}$}
\newcommand{\KsatRMFCCLwLatNEPAstro}{$296^{+11}_{-10}$}
\newcommand{\QsatRMFCCLwLatNEPAstro}{$-417^{+43}_{-41}$}
\newcommand{\ZsatRMFCCLwLatNEPAstro}{$-159^{+191}_{-193}$}
\newcommand{\JsymRMFCCLwLatNEPAstro}{$31.4^{+3.31}_{-3.23}$}
\newcommand{\LsymRMFCCLwLatNEPAstro}{$28^{+33}_{-23}$}
\newcommand{\KsymRMFCCLwLatNEPAstro}{$-201^{+77}_{-75}$}
\newcommand{\QsymRMFCCLwLatNEPAstro}{$1378^{+1502}_{-1121}$}
\newcommand{\ZsymRMFCCLwLatNEPAstro}{$-1841^{+10247}_{-7647}$}
\newcommand{\atwoRMFCCLwLatNEPAstro}{$1.411^{+0.040}_{-0.014}$}
\newcommand{\afourRMFCCLwLatNEPAstro}{$-0.538^{+0.062}_{-0.024}$}
\newcommand{\MsRMFCCLwNEPAstro}{$1112^{+135}_{-156}$}
\newcommand{\gsigmaRMFCCLwNEPAstro}{$17.69^{+1.95}_{-2.95}$}
\newcommand{\gomegaRMFCCLwNEPAstro}{$9.21^{+1.10}_{-0.76}$}
\newcommand{\CnsRMFCCLwNEPAstro}{$1.587^{+0.390}_{-0.238}$}
\newcommand{\grhoRMFCCLwNEPAstro}{$6.49^{+2.29}_{-1.89}$}
\newcommand{\gdeltaRMFCCLwNEPAstro}{$1.52^{+3.68}_{-1.37}$}
\newcommand{\LambdaomegaRMFCCLwNEPAstro}{$0.372^{+0.432}_{-0.306}$}
\newcommand{\MeffecRMFCCLwNEPAstro}{$0.75^{+0.035}_{-0.053}$}
\newcommand{\nsatRMFCCLwNEPAstro}{$0.16^{+0.010}_{-0.010}$}
\newcommand{\EsatRMFCCLwNEPAstro}{$-16.0^{+0.64}_{-0.63}$}
\newcommand{\KsatRMFCCLwNEPAstro}{$320^{+21}_{-17}$}
\newcommand{\QsatRMFCCLwNEPAstro}{$-278^{+158}_{-97}$}
\newcommand{\ZsatRMFCCLwNEPAstro}{$-832^{+370}_{-308}$}
\newcommand{\JsymRMFCCLwNEPAstro}{$31.4^{+3.28}_{-3.26}$}
\newcommand{\LsymRMFCCLwNEPAstro}{$29^{+40}_{-24}$}
\newcommand{\KsymRMFCCLwNEPAstro}{$-147^{+278}_{-106}$}
\newcommand{\QsymRMFCCLwNEPAstro}{$1366^{+1376}_{-1440}$}
\newcommand{\ZsymRMFCCLwNEPAstro}{$-5312^{+6534}_{-36652}$}
\newcommand{\atwoRMFCCLwNEPAstro}{$1.340^{+0.208}_{-0.174}$}
\newcommand{\afourRMFCCLwNEPAstro}{$-0.644^{+0.289}_{-0.373}$}

\newcommand{\MsRMFCCLwLatNEPGWtwoMsun}{$1004^{+31}_{-33}$}
\newcommand{\gsigmaRMFCCLwLatNEPGWtwoMsun}{$15.17^{+0.87}_{-0.62}$}
\newcommand{\gomegaRMFCCLwLatNEPGWtwoMsun}{$8.20^{+0.40}_{-0.23}$}
\newcommand{\CnsRMFCCLwLatNEPGWtwoMsun}{$1.713^{+0.066}_{-0.057}$}
\newcommand{\grhoRMFCCLwLatNEPGWtwoMsun}{$6.66^{+1.22}_{-1.67}$}
\newcommand{\gdeltaRMFCCLwLatNEPGWtwoMsun}{$1.52^{+2.35}_{-1.37}$}
\newcommand{\LambdaomegaRMFCCLwLatNEPGWtwoMsun}{$0.599^{+0.321}_{-0.432}$}
\newcommand{\MeffecRMFCCLwLatNEPGWtwoMsun}{$0.81^{+0.008}_{-0.014}$}
\newcommand{\nsatRMFCCLwLatNEPGWtwoMsun}{$0.14^{+0.005}_{-0.003}$}
\newcommand{\EsatRMFCCLwLatNEPGWtwoMsun}{$-16.2^{+0.57}_{-0.59}$}
\newcommand{\KsatRMFCCLwLatNEPGWtwoMsun}{$295^{+9}_{-9}$}
\newcommand{\QsatRMFCCLwLatNEPGWtwoMsun}{$-428^{+38}_{-37}$}
\newcommand{\ZsatRMFCCLwLatNEPGWtwoMsun}{$-103^{+158}_{-188}$}
\newcommand{\JsymRMFCCLwLatNEPGWtwoMsun}{$31.7^{+2.80}_{-3.40}$}
\newcommand{\LsymRMFCCLwLatNEPGWtwoMsun}{$27^{+31}_{-22}$}
\newcommand{\KsymRMFCCLwLatNEPGWtwoMsun}{$-216^{+70}_{-88}$}
\newcommand{\QsymRMFCCLwLatNEPGWtwoMsun}{$1349^{+1077}_{-1168}$}
\newcommand{\ZsymRMFCCLwLatNEPGWtwoMsun}{$-1416^{+9160}_{-4296}$}
\newcommand{\atwoRMFCCLwLatNEPGWtwoMsun}{$1.415^{+0.038}_{-0.017}$}
\newcommand{\afourRMFCCLwLatNEPGWtwoMsun}{$-0.533^{+0.057}_{-0.029}$}

\newcommand{\MsRMFCCLwNEPGWtwoMsun}{$1088^{+161}_{-173}$}
\newcommand{\gsigmaRMFCCLwNEPGWtwoMsun}{$16.89^{+2.46}_{-3.01}$}
\newcommand{\gomegaRMFCCLwNEPGWtwoMsun}{$8.77^{+1.10}_{-0.53}$}
\newcommand{\CnsRMFCCLwNEPGWtwoMsun}{$1.583^{+0.431}_{-0.289}$}
\newcommand{\grhoRMFCCLwNEPGWtwoMsun}{$6.65^{+1.58}_{-1.96}$}
\newcommand{\gdeltaRMFCCLwNEPGWtwoMsun}{$1.75^{+4.00}_{-1.59}$}
\newcommand{\LambdaomegaRMFCCLwNEPGWtwoMsun}{$0.413^{+0.409}_{-0.343}$}
\newcommand{\MeffecRMFCCLwNEPGWtwoMsun}{$0.77^{+0.025}_{-0.055}$}
\newcommand{\nsatRMFCCLwNEPGWtwoMsun}{$0.16^{+0.009}_{-0.010}$}
\newcommand{\EsatRMFCCLwNEPGWtwoMsun}{$-16.0^{+0.56}_{-0.64}$}
\newcommand{\KsatRMFCCLwNEPGWtwoMsun}{$313^{+19}_{-15}$}
\newcommand{\QsatRMFCCLwNEPGWtwoMsun}{$-323^{+134}_{-79}$}
\newcommand{\ZsatRMFCCLwNEPGWtwoMsun}{$-655^{+278}_{-434}$}
\newcommand{\JsymRMFCCLwNEPGWtwoMsun}{$31.6^{+2.97}_{-3.26}$}
\newcommand{\LsymRMFCCLwNEPGWtwoMsun}{$26^{+37}_{-22}$}
\newcommand{\KsymRMFCCLwNEPGWtwoMsun}{$-171^{+183}_{-132}$}
\newcommand{\QsymRMFCCLwNEPGWtwoMsun}{$1442^{+1355}_{-1463}$}
\newcommand{\ZsymRMFCCLwNEPGWtwoMsun}{$-3796^{+4799}_{-35612}$}
\newcommand{\atwoRMFCCLwNEPGWtwoMsun}{$1.338^{+0.215}_{-0.206}$}
\newcommand{\afourRMFCCLwNEPGWtwoMsun}{$-0.629^{+0.308}_{-0.462}$}
\newcommand{\MsRMFCCLwZetaLatNEP}{$878^{+89}_{-116}$}
\newcommand{\gsigmaRMFCCLwZetaLatNEP}{$11.97^{+2.10}_{-2.05}$}
\newcommand{\gomegaRMFCCLwZetaLatNEP}{$6.67^{+1.01}_{-0.94}$}
\newcommand{\CnsRMFCCLwZetaLatNEP}{$1.473^{+0.164}_{-0.172}$}
\newcommand{\grhoRMFCCLwZetaLatNEP}{$5.45^{+1.90}_{-1.19}$}
\newcommand{\gdeltaRMFCCLwZetaLatNEP}{$2.57^{+3.33}_{-2.32}$}
\newcommand{\LambdaomegaRMFCCLwZetaLatNEP}{$0.504^{+0.444}_{-0.460}$}
\newcommand{\zetaRMFCCLwZetaLatNEP}{$0.025^{+0.022}_{-0.023}$}
\newcommand{\MeffecRMFCCLwZetaLatNEP}{$0.84^{+0.024}_{-0.025}$}
\newcommand{\nsatRMFCCLwZetaLatNEP}{$0.16^{+0.009}_{-0.010}$}
\newcommand{\EsatRMFCCLwZetaLatNEP}{$-16.1^{+0.65}_{-0.60}$}
\newcommand{\KsatRMFCCLwZetaLatNEP}{$273^{+15}_{-19}$}
\newcommand{\QsatRMFCCLwZetaLatNEP}{$-561^{+68}_{-81}$}
\newcommand{\ZsatRMFCCLwZetaLatNEP}{$571^{+712}_{-501}$}
\newcommand{\JsymRMFCCLwZetaLatNEP}{$32.0^{+3.13}_{-3.34}$}
\newcommand{\LsymRMFCCLwZetaLatNEP}{$57^{+26}_{-33}$}
\newcommand{\KsymRMFCCLwZetaLatNEP}{$-179^{+131}_{-88}$}
\newcommand{\QsymRMFCCLwZetaLatNEP}{$346^{+840}_{-372}$}
\newcommand{\ZsymRMFCCLwZetaLatNEP}{$-4^{+997}_{-3188}$}
\newcommand{\atwoRMFCCLwZetaLatNEP}{$1.462^{+0.153}_{-0.061}$}
\newcommand{\afourRMFCCLwZetaLatNEP}{$-0.516^{+0.054}_{-0.043}$}
\newcommand{\MsRMFCCLwZetaLatNEPAstro}{$1028^{+40}_{-41}$}
\newcommand{\gsigmaRMFCCLwZetaLatNEPAstro}{$15.88^{+1.16}_{-0.94}$}
\newcommand{\gomegaRMFCCLwZetaLatNEPAstro}{$8.52^{+0.54}_{-0.39}$}
\newcommand{\CnsRMFCCLwZetaLatNEPAstro}{$1.756^{+0.063}_{-0.062}$}
\newcommand{\grhoRMFCCLwZetaLatNEPAstro}{$6.62^{+1.15}_{-1.18}$}
\newcommand{\gdeltaRMFCCLwZetaLatNEPAstro}{$1.18^{+2.17}_{-1.07}$}
\newcommand{\LambdaomegaRMFCCLwZetaLatNEPAstro}{$0.564^{+0.336}_{-0.329}$}
\newcommand{\zetaRMFCCLwZetaLatNEPAstro}{$0.003^{+0.007}_{-0.002}$}
\newcommand{\MeffecRMFCCLwZetaLatNEPAstro}{$0.80^{+0.013}_{-0.018}$}
\newcommand{\nsatRMFCCLwZetaLatNEPAstro}{$0.14^{+0.003}_{-0.002}$}
\newcommand{\EsatRMFCCLwZetaLatNEPAstro}{$-16.4^{+0.55}_{-0.45}$}
\newcommand{\KsatRMFCCLwZetaLatNEPAstro}{$300^{+10}_{-9}$}
\newcommand{\QsatRMFCCLwZetaLatNEPAstro}{$-422^{+40}_{-39}$}
\newcommand{\ZsatRMFCCLwZetaLatNEPAstro}{$-192^{+203}_{-199}$}
\newcommand{\JsymRMFCCLwZetaLatNEPAstro}{$31.3^{+3.19}_{-3.10}$}
\newcommand{\LsymRMFCCLwZetaLatNEPAstro}{$27^{+28}_{-20}$}
\newcommand{\KsymRMFCCLwZetaLatNEPAstro}{$-199^{+80}_{-62}$}
\newcommand{\QsymRMFCCLwZetaLatNEPAstro}{$1494^{+1282}_{-979}$}
\newcommand{\ZsymRMFCCLwZetaLatNEPAstro}{$-2529^{+4944}_{-10605}$}
\newcommand{\atwoRMFCCLwZetaLatNEPAstro}{$1.411^{+0.038}_{-0.014}$}
\newcommand{\afourRMFCCLwZetaLatNEPAstro}{$-0.538^{+0.060}_{-0.024}$}
\newcommand{\MsRMFCCLwZetaNEPAstro}{$1114^{+120}_{-156}$}
\newcommand{\gsigmaRMFCCLwZetaNEPAstro}{$18.40^{+1.44}_{-3.04}$}
\newcommand{\gomegaRMFCCLwZetaNEPAstro}{$9.63^{+2.40}_{-0.98}$}
\newcommand{\CnsRMFCCLwZetaNEPAstro}{$1.567^{+0.355}_{-0.249}$}
\newcommand{\grhoRMFCCLwZetaNEPAstro}{$6.75^{+2.85}_{-2.07}$}
\newcommand{\gdeltaRMFCCLwZetaNEPAstro}{$1.51^{+3.95}_{-1.37}$}
\newcommand{\LambdaomegaRMFCCLwZetaNEPAstro}{$0.329^{+0.428}_{-0.269}$}
\newcommand{\zetaRMFCCLwZetaNEPAstro}{$0.003^{+0.009}_{-0.003}$}
\newcommand{\MeffecRMFCCLwZetaNEPAstro}{$0.74^{+0.044}_{-0.117}$}
\newcommand{\nsatRMFCCLwZetaNEPAstro}{$0.16^{+0.010}_{-0.010}$}
\newcommand{\EsatRMFCCLwZetaNEPAstro}{$-16.0^{+0.62}_{-0.62}$}
\newcommand{\KsatRMFCCLwZetaNEPAstro}{$320^{+20}_{-18}$}
\newcommand{\QsatRMFCCLwZetaNEPAstro}{$-268^{+165}_{-102}$}
\newcommand{\ZsatRMFCCLwZetaNEPAstro}{$-880^{+936}_{-255}$}
\newcommand{\JsymRMFCCLwZetaNEPAstro}{$31.4^{+3.38}_{-3.31}$}
\newcommand{\LsymRMFCCLwZetaNEPAstro}{$29^{+39}_{-24}$}
\newcommand{\KsymRMFCCLwZetaNEPAstro}{$-118^{+296}_{-120}$}
\newcommand{\QsymRMFCCLwZetaNEPAstro}{$1315^{+1318}_{-2371}$}
\newcommand{\ZsymRMFCCLwZetaNEPAstro}{$-8045^{+9405}_{-33496}$}
\newcommand{\atwoRMFCCLwZetaNEPAstro}{$1.376^{+0.339}_{-0.182}$}
\newcommand{\afourRMFCCLwZetaNEPAstro}{$-0.698^{+0.304}_{-0.617}$}

\newcommand{\MsRMFCCLwZetaLatNEPGWtwoMsun}{$1016^{+38}_{-37}$}
\newcommand{\gsigmaRMFCCLwZetaLatNEPGWtwoMsun}{$15.53^{+1.07}_{-0.78}$}
\newcommand{\gomegaRMFCCLwZetaLatNEPGWtwoMsun}{$8.38^{+0.49}_{-0.32}$}
\newcommand{\CnsRMFCCLwZetaLatNEPGWtwoMsun}{$1.733^{+0.067}_{-0.064}$}
\newcommand{\grhoRMFCCLwZetaLatNEPGWtwoMsun}{$6.65^{+1.28}_{-1.64}$}
\newcommand{\gdeltaRMFCCLwZetaLatNEPGWtwoMsun}{$1.27^{+2.38}_{-1.14}$}
\newcommand{\LambdaomegaRMFCCLwZetaLatNEPGWtwoMsun}{$0.578^{+0.334}_{-0.415}$}
\newcommand{\zetaRMFCCLwZetaLatNEPGWtwoMsun}{$0.003^{+0.006}_{-0.003}$}
\newcommand{\MeffecRMFCCLwZetaLatNEPGWtwoMsun}{$0.80^{+0.011}_{-0.016}$}
\newcommand{\nsatRMFCCLwZetaLatNEPGWtwoMsun}{$0.143^{+0.004}_{-0.002}$}
\newcommand{\EsatRMFCCLwZetaLatNEPGWtwoMsun}{$-16.2^{+0.59}_{-0.58}$}
\newcommand{\KsatRMFCCLwZetaLatNEPGWtwoMsun}{$296^{+10}_{-10}$}
\newcommand{\QsatRMFCCLwZetaLatNEPGWtwoMsun}{$-424^{+34}_{-36}$}
\newcommand{\ZsatRMFCCLwZetaLatNEPGWtwoMsun}{$-158^{+173}_{-190}$}
\newcommand{\JsymRMFCCLwZetaLatNEPGWtwoMsun}{$31.5^{+2.95}_{-3.31}$}
\newcommand{\LsymRMFCCLwZetaLatNEPGWtwoMsun}{$26^{+35}_{-22}$}
\newcommand{\KsymRMFCCLwZetaLatNEPGWtwoMsun}{$-204^{+82}_{-89}$}
\newcommand{\QsymRMFCCLwZetaLatNEPGWtwoMsun}{$1423^{+1379}_{-1237}$}
\newcommand{\ZsymRMFCCLwZetaLatNEPGWtwoMsun}{$-1784^{+10364}_{-7631}$}
\newcommand{\atwoRMFCCLwZetaLatNEPGWtwoMsun}{$1.412^{+0.037}_{-0.015}$}
\newcommand{\afourRMFCCLwZetaLatNEPGWtwoMsun}{$-0.537^{+0.054}_{-0.025}$}
\newcommand{\MeffecRMFNEPAstro}{$0.708^{+0.046}_{-0.067}$}
\newcommand{\nsatRMFNEPAstro}{$0.160^{+0.010}_{-0.010}$}
\newcommand{\EsatRMFNEPAstro}{$-16.00^{+0.63}_{-0.63}$}
\newcommand{\ZsatRMFNEPAstro}{$2612.57^{+3056.21}_{-4363.26}$}
\newcommand{\KsatRMFNEPAstro}{$255^{+84}_{-74}$}
\newcommand{\QsatRMFNEPAstro}{$-384^{+304}_{-311}$}
\newcommand{\LsymRMFNEPAstro}{$54^{+34}_{-29}$}
\newcommand{\JsymRMFNEPAstro}{$32.46^{+4.57}_{-4.42}$}
\newcommand{\KsymRMFNEPAstro}{$-116^{+141}_{-88}$}
\newcommand{\QsymRMFNEPAstro}{$811.11^{+1279.03}_{-842.95}$}
\newcommand{\ZsymRMFNEPAstro}{$-816.91^{+2040.24}_{-14449.93}$}

\newcommand{\MmaxRMFNEPAstro}{$2.22^{+0.24}_{-0.15}$}
\newcommand{\RmaxRMFNEPAstro}{$11.12^{+0.75}_{-0.58}$}
\newcommand{\nbMmaxRMFNEPAstro}{$0.99^{+0.12}_{-0.17}$}
\newcommand{\CsMmaxRMFNEPAstro}{$0.75^{+0.04}_{-0.08}$}
\newcommand{\ecMmaxRMFNEPAstro}{$1288^{+175}_{-225}$}
\newcommand{\pcmaxRMFNEPAstro}{$595^{+121}_{-100}$}
\newcommand{\pnsatRMFNEPAstro}{$2.69^{+1.40}_{-1.38}$}
\newcommand{\pTwonsatRMFNEPAstro}{$24.00^{+8.34}_{-6.49}$}
\newcommand{\pThreeZeroRMFNEPAstro}{$19.39^{+4.91}_{-4.51}$}
\newcommand{\csnsatRMFNEPAstro}{$0.05^{+0.02}_{-0.01}$}
\newcommand{\csTwonsatRMFNEPAstro}{$0.23^{+0.10}_{-0.07}$}
\newcommand{\csThreeZeroRMFNEPAstro}{$0.20^{+0.07}_{-0.06}$}
\newcommand{\ensatRMFNEPAstro}{$153.04^{+9.34}_{-10.41}$}
\newcommand{\eTwonsatRMFNEPAstro}{$314.89^{+19.99}_{-20.28}$}
\newcommand{\eThreeZeroRMFNEPAstro}{$292.94^{+2.13}_{-1.81}$}
\newcommand{\nbOnepointFourMsunRMFNEPAstro}{$0.42^{+0.08}_{-0.06}$}
\newcommand{\pcOnepointFourmsunRMFNEPAstro}{$59.34^{+18.60}_{-11.69}$}
\newcommand{\csOnepointFourMsunRMFNEPAstro}{$0.39^{+0.11}_{-0.06}$}
\newcommand{\ecOnepointFourMsunRMFNEPAstro}{$429^{+82}_{-63}$}
\newcommand{\RadiusonepointfourRMFNEPAstro}{$12.50^{+0.69}_{-0.67}$}
\newcommand{\TidalonepointfourRMFNEPAstro}{$484^{+171}_{-151}$}
\newcommand{\ReducedtidalonepointthreeRMFNEPAstro}{$572^{+200}_{-175}$}
\newcommand{\ReducedtidalonepointfourRMFNEPAstro}{$484^{+171}_{-151}$}
\newcommand{\MmaxRMFCCNEPAstro}{$2.10^{+0.10}_{-0.08}$}
\newcommand{\RmaxRMFCCNEPAstro}{$11.57^{+0.58}_{-0.30}$}
\newcommand{\nbMmaxRMFCCNEPAstro}{$0.97^{+0.07}_{-0.18}$}
\newcommand{\CsMmaxRMFCCNEPAstro}{$0.58^{+0.05}_{-0.06}$}
\newcommand{\ecMmaxRMFCCNEPAstro}{$1209^{+100}_{-279}$}
\newcommand{\pcmaxRMFCCNEPAstro}{$434^{+56}_{-138}$}
\newcommand{\pnsatRMFCCNEPAstro}{$4.10^{+0.58}_{-0.44}$}
\newcommand{\pTwonsatRMFCCNEPAstro}{$30.14^{+4.62}_{-3.52}$}
\newcommand{\pThreeZeroRMFCCNEPAstro}{$24.82^{+2.64}_{-2.08}$}
\newcommand{\csnsatRMFCCNEPAstro}{$0.08^{+0.01}_{-0.00}$}
\newcommand{\csTwonsatRMFCCNEPAstro}{$0.23^{+0.03}_{-0.02}$}
\newcommand{\csThreeZeroRMFCCNEPAstro}{$0.21^{+0.02}_{-0.02}$}
\newcommand{\ensatRMFCCNEPAstro}{$153.63^{+8.72}_{-8.70}$}
\newcommand{\eTwonsatRMFCCNEPAstro}{$319.15^{+18.63}_{-18.48}$}
\newcommand{\eThreeZeroRMFCCNEPAstro}{$295.10^{+1.64}_{-1.60}$}
\newcommand{\nbOnepointFourMsunRMFCCNEPAstro}{$0.38^{+0.03}_{-0.03}$}
\newcommand{\pcOnepointFourmsunRMFCCNEPAstro}{$46.39^{+5.71}_{-5.76}$}
\newcommand{\csOnepointFourMsunRMFCCNEPAstro}{$0.29^{+0.02}_{-0.02}$}
\newcommand{\ecOnepointFourMsunRMFCCNEPAstro}{$381.40^{+31.95}_{-33.57}$}
\newcommand{\RadiusonepointfourRMFCCNEPAstro}{$13.46^{+0.41}_{-0.35}$}
\newcommand{\TidalonepointfourRMFCCNEPAstro}{$720^{+145}_{-107}$}
\newcommand{\ReducedtidalonepointthreeRMFCCNEPAstro}{$856^{+168}_{-125}$}
\newcommand{\ReducedtidalonepointfourRMFCCNEPAstro}{$720^{+145}_{-107}$}
\newcommand{\MmaxRMFCCLatNEPAstro}{$2.04^{+0.07}_{-0.03}$}
\newcommand{\RmaxRMFCCLatNEPAstro}{$11.67^{+0.27}_{-0.21}$}
\newcommand{\nbMmaxRMFCCLatNEPAstro}{$0.97^{+0.07}_{-0.05}$}
\newcommand{\CsMmaxRMFCCLatNEPAstro}{$0.52^{+0.03}_{-0.03}$}
\newcommand{\ecMmaxRMFCCLatNEPAstro}{$1207^{+110}_{-64}$}
\newcommand{\pcmaxRMFCCLatNEPAstro}{$398^{+59}_{-32}$}
\newcommand{\pnsatRMFCCLatNEPAstro}{$3.74^{+0.49}_{-0.36}$}
\newcommand{\pTwonsatRMFCCLatNEPAstro}{$25.64^{+2.58}_{-1.67}$}
\newcommand{\pThreeZeroRMFCCLatNEPAstro}{$25.57^{+1.96}_{-1.41}$}
\newcommand{\csnsatRMFCCLatNEPAstro}{$0.08^{+0.01}_{-0.00}$}
\newcommand{\csTwonsatRMFCCLatNEPAstro}{$0.21^{+0.01}_{-0.01}$}
\newcommand{\csThreeZeroRMFCCLatNEPAstro}{$0.21^{+0.01}_{-0.01}$}
\newcommand{\ensatRMFCCLatNEPAstro}{$143.26^{+4.97}_{-5.47}$}
\newcommand{\eTwonsatRMFCCLatNEPAstro}{$296.97^{+10.70}_{-11.51}$}
\newcommand{\eThreeZeroRMFCCLatNEPAstro}{$296.15^{+1.54}_{-1.44}$}
\newcommand{\nbOnepointFourMsunRMFCCLatNEPAstro}{$0.37^{+0.02}_{-0.03}$}
\newcommand{\pcOnepointFourmsunRMFCCLatNEPAstro}{$44.35^{+3.88}_{-4.63}$}
\newcommand{\csOnepointFourMsunRMFCCLatNEPAstro}{$0.26^{+0.01}_{-0.01}$}
\newcommand{\ecOnepointFourMsunRMFCCLatNEPAstro}{$375.59^{+21.08}_{-27.43}$}
\newcommand{\RadiusonepointfourRMFCCLatNEPAstro}{$13.71^{+0.33}_{-0.28}$}
\newcommand{\TidalonepointfourRMFCCLatNEPAstro}{$783^{+128}_{-87}$}
\newcommand{\ReducedtidalonepointthreeRMFCCLatNEPAstro}{$935^{+149}_{-103}$}
\newcommand{\ReducedtidalonepointfourRMFCCLatNEPAstro}{$783^{+128}_{-87}$}

\newcommand{\MmaxRMFCClwLatNEPAstro}{$2.05^{+0.07}_{-0.05}$}
\newcommand{\RmaxRMFCClwLatNEPAstro}{$10.96^{+0.29}_{-0.25}$}
\newcommand{\nbMmaxRMFCClwLatNEPAstro}{$1.04^{+0.06}_{-0.05}$}
\newcommand{\CsMmaxRMFCClwLatNEPAstro}{$0.59^{+0.03}_{-0.02}$}
\newcommand{\ecMmaxRMFCClwLatNEPAstro}{$1298^{+110}_{-73}$}
\newcommand{\pcmaxRMFCClwLatNEPAstro}{$482^{+68}_{-28}$}
\newcommand{\pnsatRMFCClwLatNEPAstro}{$1.20^{+1.30}_{-0.79}$}
\newcommand{\pTwonsatRMFCClwLatNEPAstro}{$16.96^{+1.94}_{-2.14}$}
\newcommand{\pThreeZeroRMFCClwLatNEPAstro}{$19.79^{+2.27}_{-2.19}$}
\newcommand{\csnsatRMFCClwLatNEPAstro}{$0.04^{+0.02}_{-0.01}$}
\newcommand{\csTwonsatRMFCClwLatNEPAstro}{$0.19^{+0.02}_{-0.02}$}
\newcommand{\csThreeZeroRMFCClwLatNEPAstro}{$0.20^{+0.02}_{-0.02}$}
\newcommand{\ensatRMFCClwLatNEPAstro}{$135.70^{+4.31}_{-2.19}$}
\newcommand{\eTwonsatRMFCClwLatNEPAstro}{$276.85^{+9.05}_{-4.81}$}
\newcommand{\eThreeZeroRMFCClwLatNEPAstro}{$292.10^{+2.28}_{-1.42}$}
\newcommand{\nbOnepointFourMsunRMFCClwLatNEPAstro}{$0.44^{+0.03}_{-0.03}$}
\newcommand{\pcOnepointFourmsunRMFCClwLatNEPAstro}{$60.55^{+6.39}_{-6.69}$}
\newcommand{\csOnepointFourMsunRMFCClwLatNEPAstro}{$0.33^{+0.02}_{-0.02}$}
\newcommand{\ecOnepointFourMsunRMFCClwLatNEPAstro}{$442.61^{+28.77}_{-33.74}$}
\newcommand{\RadiusonepointfourRMFCClwLatNEPAstro}{$12.25^{+0.66}_{-0.39}$}
\newcommand{\TidalonepointfourRMFCClwLatNEPAstro}{$468^{+93}_{-66}$}
\newcommand{\ReducedtidalonepointthreeRMFCClwLatNEPAstro}{$555^{+111}_{-77}$}
\newcommand{\ReducedtidalonepointfourRMFCClwLatNEPAstro}{$468^{+93}_{-66}$}
\newcommand{\MmaxRMFCClwNEPAstro}{$2.12^{+0.13}_{-0.09}$}
\newcommand{\RmaxRMFCClwNEPAstro}{$11.04^{+0.91}_{-0.42}$}
\newcommand{\nbMmaxRMFCClwNEPAstro}{$1.00^{+0.10}_{-0.29}$}
\newcommand{\CsMmaxRMFCClwNEPAstro}{$0.64^{+0.06}_{-0.08}$}
\newcommand{\ecMmaxRMFCClwNEPAstro}{$1254^{+138}_{-452}$}
\newcommand{\pcmaxRMFCClwNEPAstro}{$504^{+82}_{-254}$}
\newcommand{\pnsatRMFCClwNEPAstro}{$1.37^{+1.81}_{-0.94}$}
\newcommand{\pTwonsatRMFCClwNEPAstro}{$22.97^{+5.86}_{-4.90}$}
\newcommand{\pThreeZeroRMFCClwNEPAstro}{$19.35^{+3.94}_{-3.67}$}
\newcommand{\csnsatRMFCClwNEPAstro}{$0.05^{+0.02}_{-0.02}$}
\newcommand{\csTwonsatRMFCClwNEPAstro}{$0.23^{+0.05}_{-0.03}$}
\newcommand{\csThreeZeroRMFCClwNEPAstro}{$0.21^{+0.04}_{-0.03}$}
\newcommand{\ensatRMFCClwNEPAstro}{$150.40^{+9.20}_{-9.13}$}
\newcommand{\eTwonsatRMFCClwNEPAstro}{$308.13^{+19.65}_{-19.44}$}
\newcommand{\eThreeZeroRMFCClwNEPAstro}{$291.45^{+2.33}_{-1.74}$}
\newcommand{\nbOnepointFourMsunRMFCClwNEPAstro}{$0.43^{+0.05}_{-0.05}$}
\newcommand{\pcOnepointFourmsunRMFCClwNEPAstro}{$60.06^{+10.90}_{-9.84}$}
\newcommand{\csOnepointFourMsunRMFCClwNEPAstro}{$0.35^{+0.04}_{-0.03}$}
\newcommand{\ecOnepointFourMsunRMFCClwNEPAstro}{$433.76^{+50.23}_{-50.75}$}
\newcommand{\RadiusonepointfourRMFCClwNEPAstro}{$12.16^{+0.73}_{-0.51}$}
\newcommand{\TidalonepointfourRMFCClwNEPAstro}{$464^{+135}_{-97}$}
\newcommand{\ReducedtidalonepointthreeRMFCClwNEPAstro}{$547^{+158}_{-112}$}
\newcommand{\ReducedtidalonepointfourRMFCClwNEPAstro}{$464^{+135}_{-97}$}

\title{Relativistic Mean Field Approach with Chiral Symmetry Breaking and Quark Confinement in the light of Astrophysical Observations}
\author{B.  K. Pradhan}
\email{b-k.pradhan@ip2i.in2p3.fr}
\affiliation{Institut de Physique des 2 infinis de Lyon, CNRS/IN2P3, Universit\'e de Lyon, Universit\'e Claude Bernard Lyon 1, F-69622 Villeurbanne Cedex, France}

\author{M. Chamseddine}
\affiliation{IJCLab, Universit\'e Paris-Saclay, CNRS/IN2P3, 91405 Orsay Cedex, France}

\author{J. Margueron}
\affiliation{International Research Laboratory on Nuclear Physics and Astrophysics, Michigan State University and CNRS, East Lansing, MI 48824, USA}

\author{H. Hansen}
\affiliation{Institut de Physique des 2 infinis de Lyon, CNRS/IN2P3, Universit\'e de Lyon, Universit\'e Claude Bernard Lyon 1, F-69622 Villeurbanne Cedex, France}

\author{G. Chanfray}
\affiliation{Institut de Physique des 2 infinis de Lyon, CNRS/IN2P3, Universit\'e de Lyon, Universit\'e Claude Bernard Lyon 1, F-69622 Villeurbanne Cedex, France}

\author{J.-P. Ebran}
\affiliation{CEA,DAM,DIF, F-91297 Arpajon, France}
\affiliation{Universit\'e Paris-Saclay, CEA, Laboratoire Mati\`ere en Conditions Extr\^emes, 91680, Bruy\`eres-le-Ch\^atel, France}
\author{E. Khan}
\affiliation{IJCLab, Universit\'e Paris-Saclay, CNRS/IN2P3, 91405 Orsay Cedex, France}
\affiliation{Institut Universitaire de France (IUF)}

\begin{abstract}
We perform a Bayesian analysis of a relativistic mean-field approach, which 
is an implementation of the chiral confining model with both chiral symmetry breaking and confinement effects, and which was recently proven to reproduce well the ground state properties of finite nuclei.
We additionally explore the impact of couplings between $\rho$ and $\omega$ mesons as well as a non-linear $\omega$ coupling. Our models are simultaneously constrained by nuclear matter properties near saturation density, multi-messenger neutron star astrophysical observations, and/or lattice QCD predictions of the nucleon mass. It exhibits tension in simultaneously reproducing the $\sim 2M_{\odot}$ massive NS and the tidal deformability inferred from GW170817. We show that an additional $\omega\rho$ coupling, favored by Bayes factor analysis, substantially alleviates this tension, while adding a non-linear $\omega$ self-interaction is not necessary for the RMF-CC model. Owing to the strong constraints on the scalar sector imposed by chiral dynamics and the softening of the equation of state at high densities induced by our treatment of confinement, the RMF-CC approach favors stiff equations of state. Since we do not consider phase transition in the core of neutron stars, this stiffening is obtained with large values of the incompressibility modulus of about $\sim300$ MeV. We finally compare the well-known RMF model with RMF-CC models with the same constraints, and we obtain a preference for the RMF model in the absence of a phase transition in the core of neutron stars.

\end{abstract}
\keywords{Chiral Symmetry Breaking, Confinement,RMF, Neutron Star}                             
                              
\maketitle
\section{Introduction}\label{sec:intro}

Since the discovery of neutron stars (NSs), their interior composition has remained one of the most intriguing open questions in nuclear astrophysics~\cite{Baym2017,Vidana2020,Lattimer2021}. Terrestrial experiments provide valuable constraints on nuclear interactions near saturation density ($n_{\rm sat}$)~\cite{Danielewicz2002,Fevre2016,Roca-Maza2018,nucleardatapy2026} and, for heavy-ion collisions~\cite{Hartnack_KaoS2006, LEFEVRE_FOPI2016,Russotto_ASYEoS2016}, up to a few times $ n_{\rm sat}$. However, the densities reached in the cores of NS exceed those accessible in laboratories by several times. Moreover, laboratory experiments primarily probe nearly isospin symmetric (equal numbers of protons and neutrons ) nuclear matter, whereas NS interiors are characterized by extreme isospin asymmetry in the neutron-rich sector. This necessitates substantial extrapolations of our current understanding of nuclear interactions to regimes of high density and large isospin asymmetry, introducing significant theoretical uncertainties~\cite{Vidana2020,Lattimer2021}.


The equation of state (EoS), which encodes the relation between pressure and energy density, connects the microscopic description of dense matter to macroscopic NS observables, such as mass $M$, radius $R$, and moment of inertia ($I$). Nuclear EoSs can be broadly classified into two main categories~\cite{OertelRMP,CompOSE2022}: The first comprises microscopic or \textit{ab initio} approaches calibrated on $NN$ phase shifts and light nuclei properties~\cite{Hofmann2001,Hagen2010,Tsukiyama2011,Roth2012,Soma2013,Holt2013,Bovermann2019,Jiang2024}, while the second includes phenomenological models calibrated on saturation properties of nuclear matter and/or properties of nuclei, such as the interactions of the Skyrme type~\cite{Skyrme1956,Skyrme1956b,CHABANAT1997,Sagawa2007,Stone2007,Vautherin1972}, Gogny-type forces~\cite{Decharg1980,BergerNPA1989,Sellahewa2014,Chappert2015, Chen2012}, as well as relativistic approaches~\cite{MACHLEIDT19871,Haidenbauer,Shen2010,Cescato1998,Nik2002,Serot1997,Todd2005,Chanfray2007,Chanfray2011,Chen2014,Hornick2018}. In addition to these, several nuclear-physics–motivated EoS frameworks have been employed to describe NS interiors and interpret astrophysical observations, including nuclear meta-models~\cite{Margueron2017a,Margueron2017b,Chatterjee2017c}, Taylor-expansion–based parameterizations~\cite{Chen2009,Li2026,Margueron2019}, and hybrid constructions~\cite{Biswas2021,Suleiman2025}. An alternative strategy is to consider empirical representations of the EoS that are not explicitly tied to microscopic nuclear interactions, such as piecewise polytropes~\cite{Read2009} or the spectral decomposition method~\cite{Spectral}. More recently, model-agnostic approaches have gained prominence, enabling the inference of the NS EoS directly from observations using non-parametric descriptions~\cite{Essick2020,Landry2020,Gorda_2023} or parameterizations based on the speed of sound~\cite{Greif2019}.

 Neutron Stars have been observed for several decades across the electromagnetic spectrum using both ground- and space-based telescopes. In particular, the discovery of massive pulsars (PSRs) with masses $M \geq 2M_{\odot}$~\cite{Demorest,Cromartie,Antoniadis} has ruled out several theoretical EoSs that are unable to support such massive NSs. A new method to probe the NS interior has emerged with the direct detection of gravitational waves (GWs) from binary neutron star (BNS) mergers by the LIGO–VIRGO-KAGRA Collaboration, which provided constraints on the tidal deformability of NSs~\cite{Abbott2017,Abbott2017apjl,Abbott2018,Abbott2019,Abbott2021,GW190425}. These constraints are further complemented by independent mass–radius ($MR$) measurements of millisecond pulsars by NICER~\cite{NICER0740_Riley,NICER0030_Riley,Miller_2021,Miller_2019,NICER}, which significantly constrained the uncertainties in the NS EoS. The combined information from multi-messenger astrophysical observations has therefore revolutionized our understanding of NS interiors. As a result, extensive efforts over the past decade have been devoted to constraining the NS EoS and/or understanding the NS interior composition, either using astrophysical observations alone or through joint analyses that combine multi-messenger astrophysical data with low-density theoretical predictions (such as chiral effective field theory) and/or terrestrial experimental constraints~\cite{Margueron2018,Miyatsu_2022,Lopes2023,Maiti2024,Ghosh2021,Biswas2021,Essick2020,Landry2020,Gorda_2023,Malik2024,Biswas2025,Tewari2025,Hornick2018,Annala2022,Pang2021,Huth2022}.


In this work, we investigate a chiral Lagrangian for nuclear and NS matter with nucleons and mesons as degrees of freedom, where the spontaneous breaking of the chiral symmetry gives rise to a scalar chiral potential, and confinement is reflected in the nucleon polarization. The model is motivated by the features of low-energy quantum chromodynamics (QCD), i.e, the relevant degrees of freedom are hadronic (nucleons and mesons) rather than quark and gluon fields, and the vacuum is nontrivial as a consequence of spontaneous chiral symmetry breaking, with the dynamics being intrinsically relativistic or near-relativistic. As detailed in ~\cite{Chanfray2001,Chanfray2005,Chanfray2006,Chanfray2007,Massot2008,Massot2009,Chanfray2011,Chanfray_Schuck,Chanfray2023,Chanfray2025}, this framework offers a consistent theoretical bridge between QCD and the description of nuclear and NS matter. The chiral confining model (CCM) in its Nambu-Jona-Lasinio (NJL)-based realization,  builds upon the NJL model, whose bosonization procedure and formal structure are presented in detail in~\cite{Chanfray2011,Chanfray2025}. The model elucidates the response of composite nucleons to the nuclear scalar field and the interplay between confinement and chiral symmetry breaking.

The CCM has been systematically investigated in the context of lattice QCD (LQCD) results and nuclear physics constraints within both the relativistic mean field (RMF) and relativistic Hartree–Fock (RHF) frameworks~\cite{Somasundaram:2021hna,Chamseddine2023,Chamseddine_NJL}. A detailed Bayesian analysis of the CCM within the RMF formalism (abbreviated as RMF-CC in Ref.~\cite{Somasundaram:2021hna} ), including comparisons with other RMF-based approaches under LQCD and nuclear physics constraints, was carried out in Ref.~\cite{Somasundaram:2021hna}. In the present work, we extend the analysis of  Ref.~\cite{Somasundaram:2021hna} by examining the CCM within the RMF limit in light of astrophysical observations. Our primary focus is to assess the impact of astrophysical constraints, in conjunction with LQCD and nuclear physics inputs, on the RMF-CC model through a comprehensive Bayesian inference and model comparison.

This work is organized as follows. In \Cref{sec:Formalism}, we introduce the theoretical framework underlying the construction of the NS EoS and the computation of relevant NS properties, and outline the set of physical constraints together with the Bayesian inference strategy adopted in this work. The resulting constraints and model comparisons are presented and discussed in \Cref{sec:results}. We conclude with a summary of our main findings and perspectives in \Cref{sec:conclusion}.

Throughout this work, we employ natural units with  $G=c=\hbar=1$, unless stated otherwise.

\section{Formalism} \label{sec:Formalism}

In this section, we describe the formalism employed in this paper, describing first the model for baryons, and then presenting neutron stars with baryons and leptons.

\subsection{Model for baryons}

Considering only the lightest $u$ and $d$ quarks and the flavor number $N_f=2$, the chiral fields associated to the fluctuations of the quark condensate $\langle \bar q q\rangle$ resulting from chiral symmetry breaking are usually parameterized in term of a $\rm{SU}(2)$ matrix $M$ as:
\begin{equation}
M=\sigma + i\vec{\tau}\cdot\vec{\phi}\equiv S\, U
\label{REPRES}
\end{equation}
with $S = s + f_\pi$ and $U=e^{i\,{\vec{\tau}\cdot\vec{\pi}}/{f_\pi}}$.
The scalar field $\sigma$ ($S$) and pseudo-scalar fields $\vec{\phi}$ ($\vec{\pi}$) written in Cartesian (polar) coordinates appear as the dynamical degrees of freedom. In this study, we identify $\sigma_W$, the sigma meson of the Walecka model,  with the chiral invariant $s$ ($=S-f_\pi$) field, which represents the radial fluctuation of the chiral condensate, for more details see Ref.~\cite{Chanfray_Schuck} and references therein. In this work,  we employ the notation $s$ for the scalar field to avoid confusion with the meson $\sigma$.

The relativistic Lagrangian can generically be written as the sum of a kinetic fermionic term,
\begin{equation}
\label{eq:L_kinetic}
\L_N = \psib \left( i \gamma^{\mu}\partial_{\mu} -M_N(s) \right) \psi \, ,
\end{equation}
where the fields $\psi$ represent the nucleon spinor, and $M_N(s)$ is the polarised nucleon mass defined in the following Eq.~\eqref{eq:nucleon_mass}. In the present approach we consider the following fields: $s$, $\omega$, $\rho$, $\delta$. The meson-nucleon Lagrangian density reads
\begin{align}
\label{eq:L_m}
\L_m \;=\;& 
\frac{1}{2}\partial^\mu s\,\partial_\mu s - V(s) \nonumber \\
&- g_\omega \omega_\mu \bar{\psi}\gamma^\mu\psi
+ \frac{1}{2} m_\omega^2 \omega^\mu\omega_\mu
- \frac{1}{4} F^{\mu\nu}F_{\mu\nu} \nonumber \\
&- g_\rho \rho_{a\mu}\bar{\psi}\gamma^\mu\tau_a\psi
+ g_\rho \frac{\kappa_\rho}{2M_N}
\partial_\nu\rho_{a\mu}\bar{\psi}\sigma^{\mu\nu}\tau_a\psi \nonumber \\
&+ \frac{1}{2} m_\rho^2 \rho_{a\mu}\rho_a^{\mu}
- \frac{1}{4} G_a^{\mu\nu}G_{a\mu\nu} \nonumber \\
&- g_\delta \delta_a \bar{\psi}\tau_a\psi
- \frac{1}{2} m_\delta^2 \delta_a\delta_a
+ \frac{1}{2}\partial^\mu\delta_a\,\partial_\mu\delta_a \nonumber \\
&- V_{\text{MCs}}(\omega,\rho) \nonumber \\
&+ \frac{g_A}{2f_\pi}
\partial_\mu\varphi_{\pi a}\bar{\psi}\gamma^\mu\gamma^5\tau_a\psi
- \frac{1}{2} m_\pi^2 \varphi_{\pi a}\varphi_{\pi a} \nonumber \\
& + \frac{1}{2}\partial^\mu\varphi_{\pi a}\partial_\mu\varphi_{\pi a}\, ,
\end{align}
where the symbols have their usual meaning. We remind that $g_A = 1.25$ is the axial coupling constant and $f_\pi $ is the pion decay constant. We also note the nucleon-$\rho$ tensor coupling $f_\rho = g_\rho \kappa_\rho$. However, the pion and tensor $\rho$ contributions vanish at the mean field level and will thus not be considered later on. The details of the scalar potential $V(s)$ can be found in Refs.~\cite{Chamseddine_NJL,Chamseddine2023,Chanfray2011}. The effective scalar potential $V(s)$, expressed in terms of the scalar field $s$, has the typical Mexican hat shape responsible for spontaneous chiral symmetry breaking. We adopt the linear sigma model ($\mathrm{L}\sigma\mathrm{M} $) realization of the chiral potential, as discussed and provided in  Ref.~\cite{Somasundaram:2021hna}: 
\begin{eqnarray}
   V(s)&\equiv& V_{\rm L \sigma M}(s) \nonumber \\
    &=&\frac{m^{2}_{s }}{2}\,s ^{2}\,+\,
\frac{m^{2}_{s}\,-\,m^{2}_{\pi }}{2\,f_\pi}\,s^3\,+\,
\frac{m^{2}_{s}\,-\,m^{2}_{\pi }}{8\,f_\pi^2}\,s^4 ~. \nonumber\\
 & \label{eq:V_LSM}
 \end{eqnarray}

The vector meson couplings (MCs) are considered by the potential $V_{\rm MCs}$, which accounts for the cross couplings among the vector ($\omega$) and iso-vector ($\mathbf{\rho}$) mesons and/or vector mesons' self-couplings~\cite{Horowitz2010} as,
\begin{eqnarray}
V_{\text{MCs}}(\omega,\rho) &=& V_{\omega \rho}+V_{\omega^4}\nonumber\\
&=&-\lambda_{\omega \rho}(g_{\rho}^2 \vec{\rho}_{\mu}\cdot\vec{\rho}^{\mu})
(g_{\omega}^2 \omega_{\mu}\omega^{\mu}) \nonumber \\
&\quad& - \frac{\zeta}{4!}\left(g_{\omega}^{2}\,\omega_{\mu}\omega^{\mu}\right)^{2}, \label{eq:V_NL}
\end{eqnarray}
which will be the topic of \Cref{sec:vector-isovector}. Though the $\delta$ meson field is often omitted in RMF models because of its large degeneracy with the $\rho$ meson, the isovector-scalar $\delta$ meson field is included here to account for the Dirac mass splitting between neutrons and protons in dense matter. In addition, this splitting is known to influence the isovector nuclear properties~\cite{KUBIS1997191,PenaArteaga2016}, the NS properties~\cite{Kumar2023,Miyatsu_2022,Scurto2025}, and the direct Urca cooling processes~\cite{Scurto2025}.

In the presence of the nuclear scalar field, the nucleon mass is modified as:
\begin{equation}
\label{eq:nucleon_mass}
M_N(s)=M_N+g_s s+\frac{1}{2}\kappa_\mathrm{NS}\left(s^2+\frac{s^3}{3f_\pi}\right),
\end{equation}
where, $M_N$ is the physical nucleon mass ($\simeq 938.9\ \rm MeV$). 
Here we do not take for the first order response, namely the scalar coupling constant, the value $g_s=M_N/f_\pi$ of the L$\sigma M$, but take it as a parameter possibly fixed by an underlying nucleon model. The nucleon polarisability $\kappa_\mathrm{NS}$, incorporates the effect of the nucleon response, i.e., the central ingredient of the quark-meson coupling model (QMC) introduced in the pioneering work of P. Guichon~\cite{Guichon1988}. As in Ref.~\cite{Chanfray2005} we consider an additional susceptibility term, with quadratic and quartic terms dependent on the scalar field $s$ as:
\begin{equation}
\label{eq:kappa_tilde}
\tilde\kappa_\text{NS}(\bar{s})=\frac{\partial^2M_N(s)}{\partial s^2}\vert_{s=\bar{s}}=
\kappa_\text{NS}\left(1+\frac{\bar{s}}{{f_\pi}}\right) \, ,
\end{equation} 
which vanishes at full chiral restoration, i.e., ${\bar s}=-f_\pi$, where $\bar s$ represents the value taken by the $s$ field in the ground state. 
The dimensionless parameter $C_{\rm NS}$ is defined  as~\cite{Chamseddine2023,Chamseddine_NJL}, 
\begin{equation}
C_\text{NS}\equiv \frac{\kappa_\text{NS} f_{\pi}^2}{2M_N} \, .
\end{equation}
This parameter is expected to be lower than 1~\cite{Chanfray2025,Chanfray2023}, but it is sometimes taken above 1 in phenomenological adjustments~\cite{Chanfray2025,Chamseddine_NJL,Chamseddine2023,Somasundaram:2021hna}.



The nucleon and meson masses are fixed constants, and they are given in \Cref{tab:parameters}. 
The scalar coupling constant $g_s$
The coupling constant $g_\rho$ is a free parameter fixed by the symmetry energy condition. The pion and $\rho$-tensor do not contribute at the Hartree approximation. In contrast to the earlier works ~\cite{Somasundaram:2021hna,Chamseddine2023,Chamseddine_NJL}, where $g_{\delta}$ is fixed to $g_\delta=1$, we adopt a uniform prior in the range $0 \le g_\delta \le 20$.

In the following, we also consider the well-known RMF $\sigma\omega$ model, which was first introduced by Walecka and collaborators~\cite{SerotWalecka1986,Serot1997}. The RMF model is based on meson exchange between nucleons. A non-linear $\sigma$ potential is often added to soften the EoS~\cite{BOGUTA1977} and the isovector meson $\rho$ has also been added to describe isospin asymmetric matter. Even if it resembles the L$\sigma$M chiral potential from the RMF-CC model, see Eq.~\eqref{eq:V_LSM}, the non-linear potential in the RMF approach is entirely adjusted to reproduce nuclear physics data and takes the following form:
\begin{equation}\label{eq:VRMF}
V_{\rm RMF}(s)\equiv\frac{m_s^{2}}{2}\, s^{2}+ \frac{c_2 M_N}{3} \,(g_s s)^3
+ \frac{c_3}{4}\, (g_s s)^4 \, .
\end{equation}
The parameters $c_2$ and $c_3$ are additional free parameters of the RMF model, while the equivalent coefficients in front of $s^3$ and $s^4$ are fixed in RMF-CC models. Another difference between the RMF and RMF-CC models is the Dirac mass. In the RMF approach, it is given by:
\begin{equation}\label{eq:MN_RMF}
M_{N,\rm RMF}(s) = M_N + g_s s \, .
\end{equation}
One can recover the RMF Dirac mass from the RMF-CC expression~\eqref{eq:nucleon_mass} by setting the dimensionless parameter $C_{\rm NS}=0$.
In RMF approach, the scalar field is associated with the exchange of a $\sigma$ meson, whose mass is taken in the range 450-550~MeV mimicking the exchange of two $\pi$. In RMF-CC, instead, $s$ is a field which is generated by the chiral symmetry breaking and its mass is usually larger than the nuclear physics $\sigma$ mass~\cite{Somasundaram:2021hna}. Despite these differences, it is possible to compare the predictions for RMF and RMF-CC models.

\begin{table}[t]
\begin{ruledtabular}
\caption{Nucleon and meson masses, which are taken to be constant in the present analysis.}
\label{tab:parameters}%
\begin{tabular}{c c c c c}
$M_N$ & $m_{\rho}$ & $m_{\delta}$ & $m_{\omega}$ & $m_\pi$ \\
MeV & MeV & MeV & MeV & MeV  \\
\hline
938.9 & 779.0 & 984.7 & 783.0 & 139.6\\
\end{tabular}
\end{ruledtabular}
\end{table}

\subsection{Neutron Star EoS}

NS matter is composed of baryons and leptons, whose fractions are ruled by conservation laws and the thermodynamical minimization principle. In the following, we distinguish the core, where matter is entirely liquid, from the crust, which contains a lattice of neutron-rich nuclei. The transition between the crust and the core is expected to occur at around half saturation density.

\subsubsection{Core EoS}

\label{sec:NS_EoS}

Throughout this work, the NS core is assumed to be a liquid of nucleons $N$ (neutrons and protons) and leptons $\ell$ (electrons $e^-$ and muons $\mu^-$) to describe the $\beta-$equilibrated and charge-neutral NS matter. Hence, the Lagrangian density for NS matter is constructed by the nuclear physics CCM along with the leptonic contribution as given by;
\begin{equation}
\mathcal{L}=\mathcal{L}_{N}+\mathcal{L}_m+\mathcal{L}_{\ell},
\end{equation}
where,
\begin{equation}
\mathcal{L}_{\ell}=\sum_{\ell=\{e^- , \ \mu^-\}} \bar{\psi}_l(i\gamma^{\mu}\partial_{\mu}-m_l){\psi}_l \, .
\end{equation}

In this work, we restrict to the mean field approximation and we follow the procedure described in Ref.~\cite{Chamseddine2023}; the mesons' equations of motion at the Hartree level (i.e., the mesons take their classical expectation value) are:
\begin{eqnarray}
\label{eq:classical_EOM}
&& V'(\bar s)=-g^*_s (n_{S,p} + n_{S,n})\, , \nonumber\\
&& m^{2}_{\omega} \, \bar\omega+2\lambda_{\omega \rho}\, g_{\omega}^2 g_{\rho}^2 \, \bar{\rho}^2 \bar{\omega} + \frac{\zeta}{6}g_\omega^4 \, \bar \omega^3=g_\omega (n_{p} + n_{n})\, , \nonumber\\
&& m^{2}_{\rho}\, \bar{\rho}+2\lambda_{\omega \rho}g_{\omega}^2 g_{\rho}^2\,\bar{\rho} \bar{\omega}^2=g_\rho\,(n_p-n_n)\, , \nonumber\\
&& m^{2}_{\delta} \, \bar\delta =-g_\delta (n_{S,p} - n_{S,n})	\, ,
\end{eqnarray}
where $g^*_s = \frac{\partial M_N(s)}{\partial s}\vert_{s=\bar{s}}$ and ($\bar s, \bar \omega, \bar \rho, \bar \delta)$ correspond to the expectation values of the mesons.
The densities $n_N$ and $n_{S,N}$ are the vector and scalar densities defined as:
\begin{align}
n_{S,N} &= \int\frac{2\,d^3k}{(2\pi)^3}\, \frac{M^*_N}{\sqrt{\mathrm{k}^2 + M^{*2}_N}} \,\theta(\mathrm{k}_{F_N} - \mathrm{k})\, , \ (N=p,n)\, , \nonumber \\
n_{N} &= \int\frac{2\,d^3k}{(2\pi)^3}\,\theta(k_{F_N} - k)\, .
\end{align}

The energy density $\varepsilon$ (in Hartree limit) is:
\begin{eqnarray}
\varepsilon &=& 2 \int \frac{d^3k}{(2\pi)^3} \sum_{N=p,n} \sqrt{k^2 + M_N^{*2}} \, \theta(k_{F_N} - k)  \nonumber \\
&&+ \frac{1}{2}
\left(
  m_{\omega}^{2}\,\bar{\omega}^{2}
+ m_{\delta}^{2}\,\bar{\delta}^{2}
+ m_{\rho}^{2}\,\bar{\rho}^{2}
\right) + V(s)
\nonumber \\
&&+ 3\lambda_{\omega\rho}\,\bar{\omega}^{2}\bar{\rho}^{2} + \frac{\zeta}{8}g_\omega^4\, \bar \omega^4 \, \nonumber \\
&& + 2 \int \frac{d^3k}{(2\pi)^3} \sum_\ell {\sqrt{k^2+{m_\ell}^2}} \theta(k_{F\ell} - k)  \, ~,
\end{eqnarray}
where $m_\ell$ and $k_{F\ell}$ are the mass and Fermi momenta of the lepton $\ell$ respectively. The starred quantities are the Dirac scalar nucleonic mass $M^*_N$ and the effective nucleonic energy $E^*_N$, defined as,
\begin{eqnarray}
M^*_N &\equiv& M_N + \Sigma_S  \ , \\
E^*_N &\equiv&\sqrt{k^2 + M^{*}_N } = \mu_N - \Sigma_0 \, , \nonumber
\end{eqnarray}
where $\Sigma_{S,N}$ and $\Sigma_0$ are the self-energies of scalar and time nature, defined as, 
\begin{align}
\Sigma_{S,N} &\equiv M_N(\bar s) - M_N + g_{\delta} \tau_3 \,\bar{\delta} \,, \label{eq:sigmaS_direct} \\ 
\Sigma_{0} &\equiv g_{\omega} \,\bar{\omega} + \tau_3 g_{\rho} \,\bar{\rho}\,,  \label{eq:sigma0_direct} 
\end{align}
where, $\tau_3\equiv +1 (-1) $ for proton (neutron). Given the nucleon density $n_b$, the meson field equations are solved ensuring the baryon number conservation, charge neutrality and the chemical equilibrium in NS matter. The pressure ($p$) can be obtained using the Gibbs-Duhem relation:
\begin{equation}
p=-\varepsilon + \sum_{i} \mu_i n_i \, .
\label{eqn:pres}
\end{equation} 
In Eq.~\eqref{eqn:pres}, the $\mu_i$ and $n_i$ are the chemical potentials and densities indexed by $i$, where $i$ runs over particle species ($i=\{ n,\ p,\ e^-,\ \mu^-\}$).

\subsubsection{Crust EoS and Crust-Core Matching}

The NS core EoS, see \Cref{sec:NS_EoS}, is matched to the SLy4 crust EoS from Ref.~\cite{SLy4_Gulminelli2015}. The crust-core matching plays a crucial role in the NS properties, especially for $M\sim 1-1.4M_{\odot}$~\cite{CUTER}. In this work, we adopt the same methodology as in Ref.~\cite{Margueron2017a} by considering a $\log{\varepsilon}-\log{p}$ cubic spline interpolation between the crust and the core EoS in the transition region. The spline is performed between the stopping point of crust EoS at $n_l=0.1n_{\rm sat}$ and the starting point of the core EoS at $n_h=n_{\rm sat}$. The choice for $n_l$ and $n_h$ and the choice of crust EoS for crust-core matching on the NS observables are similar to that of Ref.~\cite{Margueron2017a}, and they have a negligible effect compared to observational uncertainties, especially for NS with $M>1.4M_\odot$. The spline smoothly matches the crust and core EoS, avoiding sharp jumps and unphysical kinks, preventing numerical instabilities in the crust-core region. Before proceeding further, we ensure that the resulting EoS satisfies causality and hydrostatic stability. 


\subsection{Neutron Star Macroscopic Properties}

The mass ($M$) and radius ($R$) of non-rotating and non-magnetized NSs at hydrostatic equilibrium are solutions of the general relativity hydrostatic equilibrium equations, also called the Tolman-Oppenheimer-Volkoff (TOV) equations~\cite{Tolman1939,Oppenheimer1939}:
\begin{eqnarray}
    \dv{m(r)}{r} &=& 4\pi \varepsilon(r) r^{2}, \nonumber \\
    \dv{p(r)}{r} &=& -\frac{[p(r) + \varepsilon(r)][m(r) + 4\pi r^{3} p(r)]}{r [r - 2m(r)]}.
    \label{eq:MR_relation}
\end{eqnarray}

For a given set of EoS, $p = p(\varepsilon)$, and central pressure $p_c$, the TOV equations are integrated outward from the center ($r = 0$) until the pressure vanishes, $p(r) = 0$, defining the stellar radius $R$ and the total $M = m(R)$. Varying the central pressure $p_c$ generates a family of NS configurations for an EoS. This family predicts a maximum mass: $M_{\rm max}(\mathcal{E})$.

For BNS or black hole-neutron star (BHNS) systems, the NS undergoes a tidal deformation induced by the companion's gravitational field. The dominant contribution to the GW signal arises from the quadrupolar gravitoelectric tidal response, which is characterized by the dimensionless tidal deformability $\Lambda_2$. It depends on the NS EoS through the tidal Love number $k_2$ and the stellar compactness $C = M/R$ via the relation:
\begin{equation}
    \Lambda \equiv \Lambda_2 = \frac{2}{3} k_2 C^{-5}.
    \label{eq:diml_tidaldef}
\end{equation}
Hereafter, the subscript `$2$' from $\Lambda_2$ is omitted, and $\Lambda$ denotes the quadrupolar dimensionless tidal deformability. The  Love number $k_2$, which encodes the stiffness of the NS EoS, is obtained by simultaneously solving a supplementary set of stellar perturbation equations along with the TOV equations, as detailed in \cite{Hinderer,Hinderer2008,Hinderer2010}.

\section{Statistical Analysis: Bayesian Methodology and Constraints}

Within the Bayesian framework, the posterior probability assigned to the EoS parameters $\mathcal{E}$, given a set of independent datasets $D=\{D_i\}$, is obtained via Bayes’ theorem and can be written as,
\begin{equation}\label{eq:Bayes}
P(\mathcal{E}\mid D)\equiv
\frac{L(D\mid\mathcal{E})\,\pi(\mathcal{E})}{P(D)}
\propto {\prod_i L(D_i\mid\mathcal{E})\,\pi(\mathcal{E})} \, ,
\end{equation}
where $\pi(\mathcal{E})$ denotes the prior probability distribution of the EoS parameters, $L(D_i\mid\mathcal{E})$ is the likelihood function corresponding to the dataset $D_i$, and $P(D)$ is the Bayesian evidence and acts as a normalization constant. 

We consider different distributions for the priors $\pi(\mathcal{E})$: $\mathcal{U}[x_l,x_u]$, represents the uniform distribution for $x$ in the range $x_l$ and $x_u$, $\mathcal{N}(\mu_x,\sigma_x)$ the Gaussian (or normal) distribution for the variable $x$ with mean $\mu_x$ and standard deviation $\sigma_x$, and $\mathcal{N} (\mu_x, \sigma_x) T [x_l, x_u ]$ the truncated normal distribution. These distributions reflect our current knowledge on the prior parameters and are given in Tables~\ref{tab:Lattice_data} and \ref{tab:NEPs}.

The Bayesian approach enables consideration of a set of experimental, observational and theoretical constraints, which are described in detail in the following subsections.

\subsection{Lattice QCD constraints (LQCD) :} 

Several investigations~\cite{Leinweber2004,Thomas2004,ARMOUR2010} suggest that the nucleon mass can be expressed in terms of the pion mass with an analytic contribution depending upon the powers of $m_{\pi}^2$ and non-analytic pion self-energy ($\Sigma_{\pi}$) as:
\begin{equation}
M_N(m_{\pi}^2)=a_0+a_2m_\pi^2+a_4m_{\pi}^4+\cdots +\Sigma_{\pi} \, .
\label{eq:Mnofpi}
\end{equation}

In addition to the nucleon mass, the other key entity that is measurable from the LQCD is the sigma commutator ($\sigma_N$) defined as:
\begin{equation}
\sigma_N \equiv m_\pi^2\dv{M_N(m_\pi^2)}{m_\pi^2}=a_2m_\pi^2+a_4m_\pi^4+\cdots+m_\pi^2 \dv{\Sigma_\pi}{m_\pi^2} \, .
\label{eq:sigma}
\end{equation}

The pion self-energy $\Sigma_{\pi}$ and its derivative can be obtained in a model-independent way, and then, separating the contribution of $\Sigma_{\pi}$, the uncertainties in the  LQCD parameters ($a_2$, $a_4$) are obtained from LQCD measurements~\cite{Leinweber2004}. The uncertainties of the parameters $a_2$ and $a_4$ are tabulated in \Cref{tab:Lattice_data}.

\begin{table}[t]
\begin{ruledtabular}
\centering
\caption{LQCD priors expressed in terms of the parameters $a_2$ and $a_4$~\cite{Khan2002,Leinweber2004}, see Eqs.~\eqref{eq:Mnofpi} and \eqref{eq:sigma}.
}
\label{tab:Lattice_data}
\begin{tabular}{ c  c }
Parameter& prior distribution \\
\hline
$a_2$ (GeV$^{-1}$)&   $\mathcal{U}[1.397,1.669]$\\
$a_4$ (GeV$^{-3}$)&   $\mathcal{U}[-0.563,-0.455]$\\
\end{tabular}        
\end{ruledtabular}
\end{table}

The parameters $a_2$ and $a_4$ can be expressed in terms of the CCM model parameters, and the information about $a_2$ and $a_4$ can be used to constrain the model parameters. We will discuss the explicit dependency of $a_2$ and $a_4$ on the CCM model parameters later in this work. 

\subsection{The Nuclear Empirical Parameters (NEP)}

The energy per nucleon in asymmetric nuclear matter can be expressed using the standard parabolic approximation~\cite{Margueron2017a}:
\begin{equation}
e(n_b,\delta_{\rm asym}) = e_{\mathrm{is}}(n_b) + \delta_{\rm asym}^2 e_{\mathrm{iv}}(n_b),
\end{equation}
where $\delta_{\rm asym}= (n_n - n_p)/n_b$ is the asymmetry parameter, with $n_n$, $n_p$, and $n_b = n_n + n_p$ denoting the neutron, proton, and total nucleon
densities, respectively. The (iso-spin) symmetric nuclear matter (SNM) and pure neutron matter (PNM) can be explained with $\delta_{\rm asym}$  being 0 and 1, respectively. The quantities $e_{\mathrm{is}}(n_b)$ and $e_{\mathrm{iv}}(n_b)$ represent the isoscalar and isovector contributions to the energy per nucleon.

Near the saturation density $n_{\rm sat}$, the isoscalar $e_{\rm is}$ and isovector $e_{\rm iv}$ energies can be expressed in terms of the nuclear empirical parameters (NEP) defined from the polynomial expansion function of the density parameter $x = (n_b - n_{\mathrm{sat}})/(3n_{\mathrm{sat}})$:
\begin{eqnarray}
e_{\mathrm{is}}(n_b) &=& E_{\mathrm{sat}} 
+ \frac{1}{2} K_{\mathrm{sat}}\, x^2 
+ \frac{1}{6} Q_{\mathrm{sat}}\, x^3 
+ \frac{1}{24} Z_{\mathrm{sat}}\, x^4 + \cdots, \nonumber \\
\label{eq:e_is} \\[6pt]
e_{\mathrm{iv}}(n_b) &=& E_{\mathrm{sym}} + L_{\mathrm{sym}}\, x 
+ \frac{1}{2} K_{\mathrm{sym}}\, x^2 
+ \frac{1}{6} Q_{\mathrm{sym}}\, x^3 + \cdots\, , \nonumber\\
\label{eq:e_iv}
\end{eqnarray}
where, the isovector energy, also named the symmetry energy, is defined as $e_{\mathrm{iv}}(n_b) \equiv e_{\mathrm{PNM}}(n_b) - e_{\mathrm{SNM}}(n_b)$ or equivalently, $e_{\mathrm{iv}}(n_b) = e(n_b,\delta_{\rm asym}=1) - e(n_b,\delta_{\rm asym}=0)$.

In Eq.~\eqref{eq:e_is}, $E_{\mathrm{sat}}$ is the binding energy per nucleon, $K_{\mathrm{sat}}$ the incompressibility, $Q_{\mathrm{sat}}$ the skewness, and $Z_{\mathrm{sat}}$ the kurtosis of symmetric matter at $n_{\rm sat}$, while in Eq.~\eqref{eq:e_iv} $E_{\mathrm{sym}}$, $L_{\mathrm{sym}}$, $K_{\mathrm{sym}}$, and $Q_{\mathrm{sym}}$ are the corresponding quantities for the density dependence of the symmetry energy: its energy, slope, curvature, and skewness parameters.

In the Hartree approximation, nuclear matter is described within the nuclear physics CCM by the Lagrangian density $\mathcal{L}=\mathcal{L}_N+\mathcal{L}_m$. For each EoS parameter set $\mathcal{E}$, we first verify the existence of a physically acceptable saturation point, defined as the nucleon density ($n_{\rm sat}$) at which the pressure of SNM vanishes. If so, we extract $n_{\mathrm{sat}}$, otherwise the parameter set $\mathcal{E}$ is rejected and the probability is set to zero: ${P(\mathcal{E}|\rm{NEP})=0}$. 

\begin{table}[t]
\begin{ruledtabular}
\centering
\caption{Priors considered for the NEP.}
\label{tab:NEPs}
\begin{tabular}{ c  c }
Parameter & prior distribution \\
\hline
$n_{\rm sat}\  (\rm fm^{-3})$&  $\mathcal{N} (0.16,0.01) T[0.14,0.18]$ \\
$E_{\rm sat}\  (\rm MeV)$&$ \mathcal{N} (-16,0.40) T[-17,-15]$ \\
$K_{\rm sat} \  (\rm MeV) $&$ \mathcal{U}(170,350)$\\
$E_{\rm sym} \  (\rm MeV) $&$\mathcal{N} (32,3) T[26,38]$\\
$L_{\rm sym} \  (\rm MeV) $&$ \mathcal{U}(1,150)$ \\
\end{tabular}        
\end{ruledtabular}
\end{table}

In this work, prior constraints are applied only to the primary empirical parameters $n_{\mathrm{sat}}$, $E_{\mathrm{sat}}$, $K_{\mathrm{sat}}$, $E_{\mathrm{sym}}$, and $L_{\mathrm{sym}}$, as listed in Table~\ref{tab:NEPs}. We consider broad prior ranges for $K_{\mathrm{sat}}$ and $L_{\mathrm{sym}}$ to account for the significant uncertainties in these quantities and to give the model sufficient freedom to explore the effects of various observational constraints. No explicit priors are imposed on the higher-order parameters ($Q_{\mathrm{sat}}$ or $K_{\mathrm{sym}}$) because they are experimentally unknown~\cite{Margueron2026}, though their predicted values under different constraint scenarios are analyzed later.

\subsection{Astrophysical Constraints}  

In the following, we review the set of astrophysical constraints considered in our analysis:

\begin{itemize}
\item \textbf{Massive $2M_\odot$ PSRs measurements from radio-astronomy (\textbf{Astro}\_$\, \bm{2M_{\odot}}$):}  
Precise radio timing of PSRs in binary systems has established the existence of massive NSs with $M\geq 2M_\odot$. Accordingly, a necessary condition to explain this astrophysical observation is to require the maximum mass $M_{\rm max}(\mathcal{E}) \geq 2.0M_\odot$ for a given EoS $\mathcal{E}$. While one could assign additional statistical weight using the precisely measured mass of PSR J0740+6620 ($2.08 \pm 0.07 M_\odot$) through a Gaussian likelihood, see for instance Ref.~\cite{nucleardatapy2026}, we adopt the more conservative approach of treating all EoSs reaching $2M_\odot$ as equally probable with respect to the current maximum-mass constraint.
\item \textbf{$MR$ measurements from NICER observatory:}  
We incorporate the joint $MR$ constraints inferred from the NICER observations for PSR J0030+0451~\cite{Miller_2019,NICER0030_Riley} and PSR J0740+6620~\cite{Miller_2021,NICER0740_Riley}. The  likelihood is then computed as follows:
\begin{align}
L(D_{MR}|\mathcal{E}) &\propto \int_{M_{\rm min}}^{M_{\rm max}(\mathcal{E})}  dM \int  dR \, 
      P(D_{MR}|M,R) \nonumber \\
    & \times \delta \bigl(R - R(M,\mathcal{E})\bigr) P(M|\mathcal{E})~,
\label{eq:MR_likelihood}
\end{align}
where $P(D_{MR}|M,R)$ is the observational posterior and $P(M|\mathcal{E})$ is the prior distribution of stellar masses for a given parameter set $\mathcal{E}$. We consider $P(M|\mathcal{E})$ as uniform mass distribution between $M_{\rm min} = 1 M_{\odot}$ and the maximum mass $M_{\rm max}(\mathcal{E})$.

In this work, we do not include the source PSR J0437-4715, which has also been considered in several publications by the NICER collaboration, owing to the discrepancy between the independent radius estimates in Refs.~\cite{Choudhury_2024} and \cite{Miller_2026}. For the same reason, the source PSR J0614-3329~\cite{Mauviard_2025} is also not included in our analysis.
\item \textbf{GWs measurements from LIGO-Virgo collaboration (GW):}
We consider the measurement from GW170817 with the joint posterior samples for the masses $M_1$, $M_2$ and tidal deformabilities $\Lambda_1$, $\Lambda_2$ obtained within the \texttt{IMRPhenomPv2\_NRTidal} waveform model under the low-spin prior~\cite{Abbott2019}. The corresponding likelihood contribution is evaluated as:
\begin{align}
L(D_{GW}|\mathcal{E}) & \propto \ \int\  dM_1 \int^{M_1}  dM_2 \int  d\Lambda_1 \, d\Lambda_2 \nonumber \\
&\times P(D_{GW}|M_1,M_2,\Lambda_1,\Lambda_2)  P(M_1,M_2|\mathcal{E})\nonumber \\
&\times \delta \bigl(\Lambda_1 - \Lambda_1(M_1,\mathcal{E})\bigr) 
 \delta \bigl(\Lambda_2 - \Lambda_2(M_2,\mathcal{E})\bigr)
\label{eq:GW_likelihood}
\end{align}
where $P(D_{GW}|M_1,M_2,\Lambda_1,\Lambda_2)$ represents the GW posterior for GW170817 measurement.
\end{itemize}

Considerable effort has been devoted to constraining the NS EoS by joint multimessenger astrophysical observations, including the BNS GW170817, NICER $MR$ measurements, and precise radio timing of massive pulsars. In this work, our most comprehensive astrophysical dataset incorporates all three inputs: the GW170817 constraints, some NICER $MR$ posteriors, and the $2M_\odot$ requirement. We collectively denote this combined set of astrophysical constraints as \textbf{Astro\_\,Full}, i.e., \textbf{Astro\_\,Full}=\{$2M_{\odot}$, NICER, GW\}. Assuming statistical independence, the corresponding joint likelihood is constructed as the product of the individual likelihoods,
$L({\rm Astro\_\,Full}\mid\mathcal{E})=\prod_{i} L(D_i\mid\mathcal{E})$, for $D_i$ = $\{ 2M_{\odot}$, NICER, GW\}.
  
In cases where only a subset of constraints is employed (e.g., $D=\{\mathrm{LQCD}\}$ or $D=\{\mathrm{LQCD},\mathrm{NEP}\}$), the specific combination adopted is stated explicitly at the relevant position in the text. Furthermore, in the later sections, when we consider only the $2M_{\odot}$ requirement, we explicitly denote this as `` \textbf{Astro}\_$\, \bm{2M_{\odot}}$” in the corresponding figures, tables, and discussion. Similarly, we denote `` \textbf{Astro}\_$\, \bm{2M_{\odot}\mathrm{GW}}$” the cases where only the $2M_{\odot}$ and GW constraints are considered instead of the full ones, i.e., excluding the constraint from NICER measurements.

\subsection{Bayes factor for model comparison}

Given two competing EoS models, denoted as $M_i$ and $M_j$, and a set of observational data $D$, the odds ratio $\mathcal{O}_j^i$ in favor of model $M_i$ over $M_j$ is defined as,
\begin{equation}
    \mathcal{O}_j^i \equiv \frac{P(M_i \mid D)}{P(M_j \mid D)} \, .
\end{equation}
Using Bayes’ theorem, the odds ratio can be written as
\begin{equation}
    \mathcal{O}_j^i
    = \frac{P(D \mid M_i)}{P(D \mid M_j)}
      \frac{\pi(M_i)}{\pi(M_j)} \, ,
\end{equation}
where $\pi(M_i)$ denotes the prior probability assigned to model $M_i$ before any data are considered. In this work, we assume that all models are \emph{a priori} equally probable, such that $\pi(M_i)/\pi(M_j)=1$. Under this assumption, the odds ratio reduces to the Bayes factor $\mathcal{B}_j^i$, defined as
\begin{equation}
    \mathcal{B}_j^i \equiv \frac{P(D \mid M_i)}{P(D \mid M_j)} \, .
\end{equation}


When multiple independent datasets $D = \{D_m\}$ are considered, the joint evidence is obtained by,
\begin{equation}  
\mathcal{Z}_i \equiv P(D \mid M_i) = \int \pi_i(\mathcal{E}) \prod_m L(D_m|\mathcal{E}) \, d\mathcal{E} \, .
\end{equation}

The log-Bayes factor is then obtained using the joint evidences as,
\begin{equation}\label{eq:joint_Bayes}
\log\mathcal{B}_j^i = \log\mathcal{Z}_i - \log\mathcal{Z}_j \, .
\end{equation}

Following Ref.~\cite{Kass1995}, we interpret the Bayes factors $\log_{10}(\mathcal{B}^{i}_{j})$ as follows:
\begin{itemize}
\item $\log_{10}(\mathcal{B}^{i}_{j}) \leq -2$: decisive evidence for $M_j$ over $M_i$;
\item $-2 < \log_{10}(\mathcal{B}^{i}_{j}) \leq -1$: strong evidence for $M_j$ over $M_i$;
\item $-1 < \log_{10}(\mathcal{B}^{i}_{j}) < -0.5$: substantial evidence for $M_j$ over $M_i$;
\item $\log_{10}(\mathcal{B}^{i}_{j}) \geq -0.5$: inconclusive evidence.
\end{itemize}

Parameter estimation is performed using the nested sampling algorithm \textsc{MultiNest}~\cite{Multinest2009,Multinest2011}, as implemented in the publicly available Python package \textsc{PyMultiNest}~\cite{Pymultinest2014}.\textsc{MultiNest} specifically computes the Bayesian evidence $\mathcal{Z}$ with a prescribed numerical precision, as a convergence criterion~\cite{Multinest2009}. The algorithm iteratively evolves a set of live points to sample the prior volume, terminating when the remaining unexplored evidence falls below a user-defined tolerance. This allows \textsc{MultiNest} to reliably estimate $\mathcal{Z}$ for each model, including the joint evidences $\mathcal{Z}_i$ defined above. Consequently, the log-Bayes factors $\log\mathcal{B}_j^i$, obtained as the difference in log-evidence, benefit from robust and accurate estimates of the evidence, thereby improving the reliability of model comparison across multiple independent datasets.

\begin{table*}
\centering
\caption{Median values with the upper and lower bounds for the 90\% CI of the recovered posterior of EoS model parameters for different constraint sets. }
\label{tab:posteriors_RMFCC_VLSM}
\setlength{\tabcolsep}{0.18em}
\renewcommand{\arraystretch}{1.25}
\begin{tabular}{l|c|c|c|c|c|c|c|c}
\hline\hline
\diagbox{Constraints}{Parameters} 
 & $m_s$  [MeV] & $g_s$  & $C_{\rm NS}$  & $g_\omega$  & $g_\rho$  & $g_\delta$  & $\lambda_{\omega\rho}$  & $\zeta$ \\
\hline\hline
Prior & $\mathcal{N}(860,400)$ & $\mathcal{N}(12,6)$ & $\mathcal{N}(1.4,0.7)$ & $\mathcal{N}(8,5)$ & $\mathcal{N}(6,4)$ & $\mathcal{U}(0,20)$ & $\mathcal{U}(0,1)$ & $\mathcal{U}(0,0.05)$ \\
\hline\hline

& \multicolumn{8}{c}{\textbf{RMF-CC}} \\
\hline

NEP+LQCD & \MsLatNEP & \gsigmaLatNEP & \CnsLatNEP & \gomegaLatNEP & \grhoLatNEP & \gdeltaLatNEP & -- & -- \\
+Astro$\_\,2M_\odot$ & \MsLatNEPtwoMsun & \gsigmaLatNEPtwoMsun & \CnsLatNEPtwoMsun & \gomegaLatNEPtwoMsun & \grhoLatNEPtwoMsun & \gdeltaLatNEPtwoMsun & -- & -- \\
+Astro$\_\,2M_\odot\rm GW$ & \MsLatNEPGWtwoMsun & \gsigmaLatNEPGWtwoMsun & \CnsLatNEPGWtwoMsun & \gomegaLatNEPGWtwoMsun & \grhoLatNEPGWtwoMsun & \gdeltaLatNEPGWtwoMsun & -- & -- \\
+Astro$\_\,$Full & \MsLatNEPAstro & \gsigmaLatNEPAstro & \CnsLatNEPAstro & \gomegaLatNEPAstro & \grhoLatNEPAstro & \gdeltaLatNEPAstro & -- & -- \\
NEP & \MsNEP & \gsigmaNEP & \CnsNEP & \gomegaNEP & \grhoNEP & \gdeltaNEP & -- & -- \\
 +Astro$\_\,2M_\odot\rm GW$ & \MsRMFCCNEPGWtwoMsun & \gsigmaRMFCCNEPGWtwoMsun & \CnsRMFCCNEPGWtwoMsun & \gomegaRMFCCNEPGWtwoMsun & \grhoRMFCCNEPGWtwoMsun & \gdeltaRMFCCNEPGWtwoMsun & -- & -- \\
+Astro$\_\,$Full & \MsNEPAstro & \gsigmaNEPAstro & \CnsNEPAstro & \gomegaNEPAstro & \grhoNEPAstro & \gdeltaNEPAstro & -- & -- \\
\hline\hline

& \multicolumn{8}{c}{\textbf{RMF-CC$+\rm V_{\omega\rho}$}} \\
\hline
NEP+LQCD & \MsRMFCCLwLatNEP & \gsigmaRMFCCLwLatNEP & \CnsRMFCCLwLatNEP & \gomegaRMFCCLwLatNEP & \grhoRMFCCLwLatNEP & \gdeltaRMFCCLwLatNEP & \LambdaomegaRMFCCLwLatNEP & -- \\

+Astro$\_\,2M_\odot\rm GW$ & \MsRMFCCLwLatNEPGWtwoMsun & \gsigmaRMFCCLwLatNEPGWtwoMsun & \CnsRMFCCLwLatNEPGWtwoMsun & \gomegaRMFCCLwLatNEPGWtwoMsun & \grhoRMFCCLwLatNEPGWtwoMsun & \gdeltaRMFCCLwLatNEPGWtwoMsun & \LambdaomegaRMFCCLwLatNEPGWtwoMsun & -- \\
+Astro$\_\,$Full & \MsRMFCCLwLatNEPAstro & \gsigmaRMFCCLwLatNEPAstro & \CnsRMFCCLwLatNEPAstro & \gomegaRMFCCLwLatNEPAstro & \grhoRMFCCLwLatNEPAstro & \gdeltaRMFCCLwLatNEPAstro & \LambdaomegaRMFCCLwLatNEPAstro & -- \\
NEP+Astro$\_\,2M_\odot\rm GW$& \MsRMFCCLwNEPGWtwoMsun & \gsigmaRMFCCLwNEPGWtwoMsun & \CnsRMFCCLwNEPGWtwoMsun & \gomegaRMFCCLwNEPGWtwoMsun & \grhoRMFCCLwNEPGWtwoMsun & \gdeltaRMFCCLwNEPGWtwoMsun & \LambdaomegaRMFCCLwNEPGWtwoMsun & -- \\
NEP+Astro$\_\,$Full & \MsRMFCCLwNEPAstro & \gsigmaRMFCCLwNEPAstro & \CnsRMFCCLwNEPAstro & \gomegaRMFCCLwNEPAstro & \grhoRMFCCLwNEPAstro & \gdeltaRMFCCLwNEPAstro & \LambdaomegaRMFCCLwNEPAstro & -- \\
\hline\hline

& \multicolumn{8}{c}{\textbf{RMF-CC$+\rm V_{\omega\rho}+V_{\omega^4}$}} \\
\hline
NEP+LQCD & \MsRMFCCLwZetaLatNEP & \gsigmaRMFCCLwZetaLatNEP & \CnsRMFCCLwZetaLatNEP & \gomegaRMFCCLwZetaLatNEP & \grhoRMFCCLwZetaLatNEP & \gdeltaRMFCCLwZetaLatNEP & \LambdaomegaRMFCCLwZetaLatNEP & \zetaRMFCCLwZetaLatNEP \\

 +Astro$\_\,2M_\odot\rm GW$& \MsRMFCCLwZetaLatNEPGWtwoMsun & \gsigmaRMFCCLwZetaLatNEPGWtwoMsun & \CnsRMFCCLwZetaLatNEPGWtwoMsun & \gomegaRMFCCLwZetaLatNEPGWtwoMsun & \grhoRMFCCLwZetaLatNEPGWtwoMsun & \gdeltaRMFCCLwZetaLatNEPGWtwoMsun & \LambdaomegaRMFCCLwZetaLatNEPGWtwoMsun & \zetaRMFCCLwZetaLatNEPGWtwoMsun \\

+Astro$\_\,$Full & \MsRMFCCLwZetaLatNEPAstro & \gsigmaRMFCCLwZetaLatNEPAstro & \CnsRMFCCLwZetaLatNEPAstro & \gomegaRMFCCLwZetaLatNEPAstro & \grhoRMFCCLwZetaLatNEPAstro & \gdeltaRMFCCLwZetaLatNEPAstro & \LambdaomegaRMFCCLwZetaLatNEPAstro & \zetaRMFCCLwZetaLatNEPAstro \\
NEP+Astro$\_\,$Full & \MsRMFCCLwZetaNEPAstro & \gsigmaRMFCCLwZetaNEPAstro & \CnsRMFCCLwZetaNEPAstro & \gomegaRMFCCLwZetaNEPAstro & \grhoRMFCCLwZetaNEPAstro & \gdeltaRMFCCLwZetaNEPAstro & \LambdaomegaRMFCCLwZetaNEPAstro & \zetaRMFCCLwZetaNEPAstro \\
\hline\hline
\end{tabular}
\end{table*}

\section{Results}\label{sec:results}

Although we primarily focus on the impacts of the astrophysical constraints on our model, for the sake of completeness and comparison, we also provide the statistical analysis of the CCM model under LQCD and NEP constraints in this section. We then consider an extension of the model including $\omega\rho$ coupling and non-linear $\omega$ couplings. 


\subsection{Chiral Confining Model within mean field approximation (RMF-CC)}
\label{sec:RMFCC_VLSM}


We start our analysis considering the CCM model and the Hartree approach (RMF-CC), in the absence of $V_{\rm MCs}$, as in Ref.~\cite{Somasundaram:2021hna}, for instance. That is, we fix $\lambda_{\omega \rho} = \zeta = 0$ in~\Cref{eq:L_m}. We refer to this minimal model with the choice for the scalar potential $V(s)\equiv V_{\rm L \sigma M}$ and $V_{\rm MCs}=0$ as ``RMF-CC". It is consistent with Ref.~\cite{Somasundaram:2021hna}.


This particular choice of the potentials leads us to the following model parameters $\mathcal{E}_{\rm RMF\textrm{-}CC }=\{m_s, g_s, C_{\rm NS}, g_{\omega}, g_{\rho}, g_{\delta}\}$. In addition, we fix the pion decay constant $f_{\pi} = 94$~MeV throughout the analysis to account for the $V_{\rm L \sigma M}$ given in \Cref{eq:V_LSM}. 

The LQCD parameters $a_2$ and $a_4$ can be expressed in the closed-form connecting the three scalar parameters ($m_s, g_s, C_{\rm NS}$) of the RMF-CC model within the following two relations:
\begin{align}
    a_2&=\frac{f_{\pi}g_s}{m_s^2} \label{eq:a2_Vlsm}\\
    a_4&=-\frac{f_\pi g_s}{2m_s^4}(3-2C_{\rm NS})  \label{eq:a4_Vlsm}
\end{align}

\begin{table*}[htbp]
\centering
\caption{Median values and upper and lower bounds for 90\% CI of the reconstructed LQCD and iso-scalar NEPs.}
\label{tab:isoscalarNEP_posteriors_RMFCC_VLSM}
\setlength{\tabcolsep}{0.2em}
\renewcommand{\arraystretch}{1.25}
\begin{tabular}{l|c|c|c|c|c|c|c}
\hline\hline
\diagbox{Constraints}{Parameters} 
 & $a_2\rm \ [GeV^{-1}]$ 
 & $a_4\rm \ [GeV^{-3}]$ 
 & $n_{\rm sat}$ [MeV]
 & $E_{\rm sat}$ [MeV]
 & $K_{\rm sat}$[MeV] 
 & $Q_{\rm sat}$ [MeV]
 & $M^*_{\rm SNM}(n_{\rm sat})\  [M_N]$ \\
\hline\hline

& \multicolumn{7}{c}{\textbf{RMF-CC}} \\
\hline

NEP+LQCD & \atwoLatNEP & \afourLatNEP & \nsatLatNEP & \EsatLatNEP & \KsatLatNEP & \QsatLatNEP & \MeffecLatNEP \\
+Astro$\_\,2M_{\odot}$ & \atwoLatNEPtwoMsun & \afourLatNEPtwoMsun & \nsatLatNEPtwoMsun & \EsatLatNEPtwoMsun & \KsatLatNEPtwoMsun & \QsatLatNEPtwoMsun & \MeffecLatNEPtwoMsun \\
+Astro$\_\,2M_{\odot}\rm GW$ & \atwoLatNEPGWtwoMsun & \afourLatNEPGWtwoMsun & \nsatLatNEPGWtwoMsun & \EsatLatNEPGWtwoMsun & \KsatLatNEPGWtwoMsun & \QsatLatNEPGWtwoMsun & \MeffecLatNEPGWtwoMsun \\
+Astro$\_\,$Full & \atwoLatNEPAstro & \afourLatNEPAstro & \nsatLatNEPAstro & \EsatLatNEPAstro & \KsatLatNEPAstro & \QsatLatNEPAstro & \MeffecLatNEPAstro \\

NEP & \atwoNEP & \afourNEP & \nsatNEP & \EsatNEP & \KsatNEP & \QsatNEP & \MeffecNEP \\
+Astro$\_\,2M_{\odot}\rm GW$ & \atwoRMFCCNEPGWtwoMsun & \afourRMFCCNEPGWtwoMsun & \nsatRMFCCNEPGWtwoMsun & \EsatRMFCCNEPGWtwoMsun & \KsatRMFCCNEPGWtwoMsun & \QsatRMFCCNEPGWtwoMsun & \MeffecRMFCCNEPGWtwoMsun \\
+Astro$\_\,$Full & \atwoNEPAstro & \afourNEPAstro & \nsatNEPAstro & \EsatNEPAstro & \KsatNEPAstro & \QsatNEPAstro & \MeffecNEPAstro \\

\hline\hline

& \multicolumn{7}{c}{\textbf{RMF-CC$+\rm V_{\omega\rho}$}} \\
\hline
NEP+LQCD & \atwoRMFCCLwLatNEP & \afourRMFCCLwLatNEP & \nsatRMFCCLwLatNEP & \EsatRMFCCLwLatNEP & \KsatRMFCCLwLatNEP & \QsatRMFCCLwLatNEP & \MeffecRMFCCLwLatNEP \\

+Astro$\_\,2M_{\odot}\rm GW$ & \atwoRMFCCLwLatNEPGWtwoMsun & \afourRMFCCLwLatNEPGWtwoMsun & \nsatRMFCCLwLatNEPGWtwoMsun & \EsatRMFCCLwLatNEPGWtwoMsun & \KsatRMFCCLwLatNEPGWtwoMsun & \QsatRMFCCLwLatNEPGWtwoMsun & \MeffecRMFCCLwLatNEPGWtwoMsun \\

+Astro$\_\,$Full & \atwoRMFCCLwLatNEPAstro & \afourRMFCCLwLatNEPAstro & \nsatRMFCCLwLatNEPAstro & \EsatRMFCCLwLatNEPAstro & \KsatRMFCCLwLatNEPAstro & \QsatRMFCCLwLatNEPAstro & \MeffecRMFCCLwLatNEPAstro \\
NEP+Astro$\_\,2M_\odot\rm GW$ & \atwoRMFCCLwNEPGWtwoMsun & \afourRMFCCLwNEPGWtwoMsun & \nsatRMFCCLwNEPGWtwoMsun & \EsatRMFCCLwNEPGWtwoMsun & \KsatRMFCCLwNEPGWtwoMsun & \QsatRMFCCLwNEPGWtwoMsun & \MeffecRMFCCLwNEPGWtwoMsun \\
NEP+Astro$\_\,$Full & \atwoRMFCCLwNEPAstro & \afourRMFCCLwNEPAstro & \nsatRMFCCLwNEPAstro & \EsatRMFCCLwNEPAstro & \KsatRMFCCLwNEPAstro & \QsatRMFCCLwNEPAstro & \MeffecRMFCCLwNEPAstro \\
\hline\hline

& \multicolumn{7}{c}{\textbf{RMF-CC$+\rm V_{\omega\rho}+V_{\omega^4}$}} \\
\hline
NEP+LQCD & \atwoRMFCCLwZetaLatNEP & \afourRMFCCLwZetaLatNEP & \nsatRMFCCLwZetaLatNEP & \EsatRMFCCLwZetaLatNEP & \KsatRMFCCLwZetaLatNEP & \QsatRMFCCLwZetaLatNEP & \MeffecRMFCCLwZetaLatNEP \\

+Astro$\_\,2M_\odot\rm GW$ & \atwoRMFCCLwZetaLatNEPGWtwoMsun & \afourRMFCCLwZetaLatNEPGWtwoMsun & \nsatRMFCCLwZetaLatNEPGWtwoMsun & \EsatRMFCCLwZetaLatNEPGWtwoMsun & \KsatRMFCCLwZetaLatNEPGWtwoMsun & \QsatRMFCCLwZetaLatNEPGWtwoMsun & \MeffecRMFCCLwZetaLatNEPGWtwoMsun \\
+Astro$\_\,$Full & \atwoRMFCCLwZetaLatNEPAstro & \afourRMFCCLwZetaLatNEPAstro & \nsatRMFCCLwZetaLatNEPAstro & \EsatRMFCCLwZetaLatNEPAstro & \KsatRMFCCLwZetaLatNEPAstro & \QsatRMFCCLwZetaLatNEPAstro & \MeffecRMFCCLwZetaLatNEPAstro \\
NEP+Astro$\_\,$Full & \atwoRMFCCLwZetaNEPAstro & \afourRMFCCLwZetaNEPAstro & \nsatRMFCCLwZetaNEPAstro & \EsatRMFCCLwZetaNEPAstro & \KsatRMFCCLwZetaNEPAstro & \QsatRMFCCLwZetaNEPAstro & \MeffecRMFCCLwZetaNEPAstro \\
\hline\hline

& \multicolumn{7}{c}{\textbf{RMF}} \\
\hline
NEP+Astro$\_\,$Full & - & - & \nsatRMFNEPAstro & \EsatRMFNEPAstro & \KsatRMFNEPAstro & \QsatRMFNEPAstro & \MeffecRMFNEPAstro \\
\hline\hline
\end{tabular}
\end{table*}

\begin{table}[t]
\centering
\caption{Median values with the upper and lower bounds for 90\% SCI of the reconstructed iso-vector NEPs for different models resulting from various constraints.}
\label{tab:isovectorNEP_posteriors_RMFCC_VLSM}
\setlength{\tabcolsep}{0.5em}
\renewcommand{\arraystretch}{1.25}
\begin{tabular}{l|c|c|c}
\hline\hline
\diagbox{Constraints}{NEP}
 & $E_{\rm sym}$ & $L_{\rm sym}$ & $K_{\rm sym}$ \\
 & [MeV]  & [MeV]  & [MeV]\\
\hline\hline
& \multicolumn{3}{c}{\textbf{RMF-CC}} \\
\hline
NEP+LQCD & \JsymLatNEP & \LsymLatNEP & \KsymLatNEP \\
+Astro$\_\,2M_{\odot}$ & \JsymLatNEPtwoMsun & \LsymLatNEPtwoMsun & \KsymLatNEPtwoMsun \\
+Astro$\_\,2M_{\odot}\rm GW$ & \JsymLatNEPGWtwoMsun & \LsymLatNEPGWtwoMsun & \KsymLatNEPGWtwoMsun \\
+Astro$\_\,$Full & \JsymLatNEPAstro & \LsymLatNEPAstro & \KsymLatNEPAstro \\
NEP & \JsymNEP & \LsymNEP & \KsymNEP \\
+Astro$\_\,2M_{\odot}\rm GW$ & \JsymRMFCCNEPGWtwoMsun & \LsymRMFCCNEPGWtwoMsun & \KsymRMFCCNEPGWtwoMsun \\
+Astro$\_\,$Full & \JsymNEPAstro & \LsymNEPAstro & \KsymNEPAstro \\
\hline\hline

& \multicolumn{3}{c}{\textbf{RMF-CC$+\rm V_{\omega\rho}$}} \\
\hline
NEP+LQCD & \JsymRMFCCLwLatNEP & \LsymRMFCCLwLatNEP & \KsymRMFCCLwLatNEP \\
+Astro$\_\,2M_{\odot}\rm GW$ & \JsymRMFCCLwLatNEPGWtwoMsun & \LsymRMFCCLwLatNEPGWtwoMsun & \KsymRMFCCLwLatNEPGWtwoMsun \\
+Astro$\_\,$Full & \JsymRMFCCLwLatNEPAstro & \LsymRMFCCLwLatNEPAstro & \KsymRMFCCLwLatNEPAstro \\

NEP+Astro$\_\,2M_{\odot}\rm GW$ & \JsymRMFCCLwNEPGWtwoMsun & \LsymRMFCCLwNEPGWtwoMsun & \KsymRMFCCLwNEPGWtwoMsun \\
NEP+Astro$\_\,$Full & \JsymRMFCCLwNEPAstro & \LsymRMFCCLwNEPAstro & \KsymRMFCCLwNEPAstro \\
\hline\hline

& \multicolumn{3}{c}{\textbf{RMF-CC$+\rm V_{\omega\rho}+V_{\omega^4}$}} \\
\hline
NEP+LQCD & \JsymRMFCCLwZetaLatNEP & \LsymRMFCCLwZetaLatNEP & \KsymRMFCCLwZetaLatNEP \\
+Astro$\_\,2M_{\odot}\rm GW$ &\JsymRMFCCLwZetaLatNEPGWtwoMsun & \LsymRMFCCLwZetaLatNEPGWtwoMsun & \KsymRMFCCLwZetaLatNEPGWtwoMsun \\
+Astro$\_\,$Full & \JsymRMFCCLwZetaLatNEPAstro & \LsymRMFCCLwZetaLatNEPAstro & \KsymRMFCCLwZetaLatNEPAstro \\
NEP+Astro$\_\,$Full & \JsymRMFCCLwZetaNEPAstro & \LsymRMFCCLwZetaNEPAstro & \KsymRMFCCLwZetaNEPAstro \\
\hline\hline

& \multicolumn{3}{c}{\textbf{RMF}} \\
\hline
NEP+Astro$\_\,$Full & \JsymRMFNEPAstro & \LsymRMFNEPAstro & \KsymRMFNEPAstro \\
\hline\hline
\end{tabular}
\end{table}

The posteriors of the model parameters for RMF-CC are given in \Cref{tab:posteriors_RMFCC_VLSM} for different constraints: NEP+LQCD only ; NEP+LQCD+different astrophysical data, labelled as Astro\_$2M_\odot$, Astro\_$2M_\odot$GW or Astro\_Full ; NEP only ; NEP+different astrophysical data. Note that the value for the scalar mass is systematically larger than the mass for the nuclear physics $\sigma$ meson (typically taken between 450 and 550~MeV). This is related to the fact that the scalar field $s$ is not a meson field, as discussed in Ref.~\cite{Somasundaram:2021hna}. The parameter $C_{\rm NS}$ is also larger than 1, which is due to the use of the L$\sigma$M chiral potential instead of the NJL potential, as discussed in Ref.~\cite{Chamseddine_NJL}.

In \Cref{tab:posteriors_RMFCC_VLSM}, the mean value and the uncertainty for the parameter $g_\rho$ are in close agreement with the results obtained in Ref.~\cite{Somasundaram:2021hna}, for RMF-CC calibrated to reproduce the symmetry energy NEP $E_{\rm sym}$. Unlike $g_\rho$, the coupling $g_\delta$ remains poorly known. It is, however, taken as a running parameter, while it was fixed to $g_\delta=1$ in Ref.~\cite{Somasundaram:2021hna}.
It is poorly constrained in our analysis because we have not imposed constraints on the isovector channel, other than $E_{\rm sym}$. NEP like $L_{\rm sym}$ or $K_{\rm sym}$ are free to vary within the uniform prior distribution for $L_{\rm sym}$. 
Note that when astrophysical constraints are included, the posterior for $g_{\delta}$ is rather unchanged (given the statistical uncertainties). There is, however, a systematic shift up of the parameters $g_s$ and $g_\omega$, rendering the EoS stiffer.

The posteriors given in \Cref{tab:posteriors_RMFCC_VLSM} are employed to reconstruct the nuclear and NS matter properties summarized in \Cref{tab:isoscalarNEP_posteriors_RMFCC_VLSM} for the LQCD parameters and iso-scalar NEPs and in \Cref{tab:isovectorNEP_posteriors_RMFCC_VLSM} for the iso-vector NEPs. The sign and the order of magnitude for the LQCD parameters are independent of the LQCD prior, although the uncertainties are larger for the cases where LQCD priors are not considered. Given the prior for the incompressibility modulus $K_\sat$, see Table~\ref{tab:NEPs}, our analysis prefers large values for $K_\sat$, i.e., above 250~MeV, as often obtained with non-linear RMF approaches~\cite{Lalazissis1996}. The reduction of the incompressibility modulus is achieved with density-dependent coupling parameters~\cite{TypelWolter:1999}.
In addition, we also provide the nucleon effective mass~\footnote{Note that there is a difference in the nomenclature for the effective nucleon mass in the present work and in Ref.~\cite{Somasundaram:2021hna}. In Ref.~\cite{Somasundaram:2021hna}, the nucleon effective mass is discussed in terms of the Dirac mass and denoted by $M_D^*$ in SNM. In the present work, however, the Dirac mass in PNM and NS in $\beta$-equilibrium matter receives additional contributions from the $\delta$-meson field. To avoid confusion, we explicitly label the Dirac (effective ) mass in SNM as $M_{\mathrm{SNM}}^*$ throughout this work.} $M^*_{\rm SNM}(s)\equiv M_N(s)$ in SNM at $n_{\rm sat}$, i.e, $M^*_{\rm SNM}(n_{\rm sat})$ in \Cref{tab:isoscalarNEP_posteriors_RMFCC_VLSM}. 
The values obtained for $M^*_{\rm SNM}(n_{\rm sat})$ are large compared to the values expected to reproduce the ground state of finite nuclei~\cite{Roca-Maza2018}, see also the discussion in Ref.~\cite{Chamseddine_2026}.

The values obtained for $L_\sym$, see \Cref{tab:isovectorNEP_posteriors_RMFCC_VLSM}, are large, i.e., $L_\sym\sim 90$~MeV, and for $K_\sym$ we obtain values around $K_\sym\sim 0$~MeV.

\begin{figure}[htbp]
     \centering
     \includegraphics[width=\linewidth]{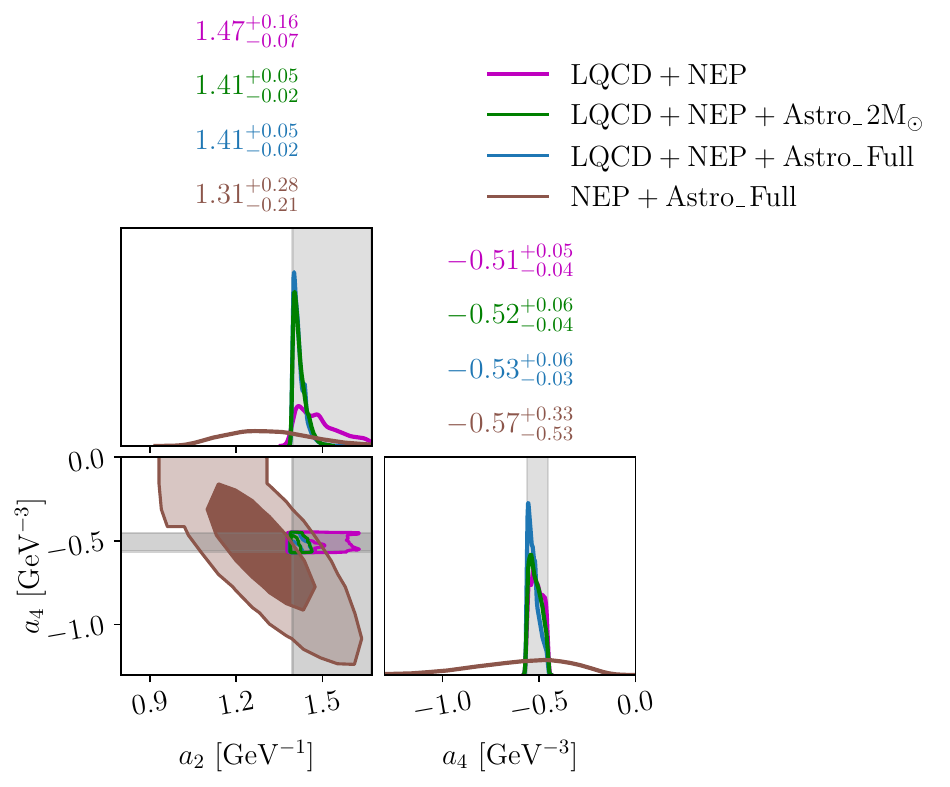}
     \caption{Corner plot showing the posteriors for the LQCD parameters $a_2$ and $a_4$. The gray bands show the allowed regions for the LQCD parameters.}
     \label{fig:LQCD_RMFCC}
\end{figure}

\begin{figure*}[t]
\centering
\includegraphics[width=0.95\linewidth]{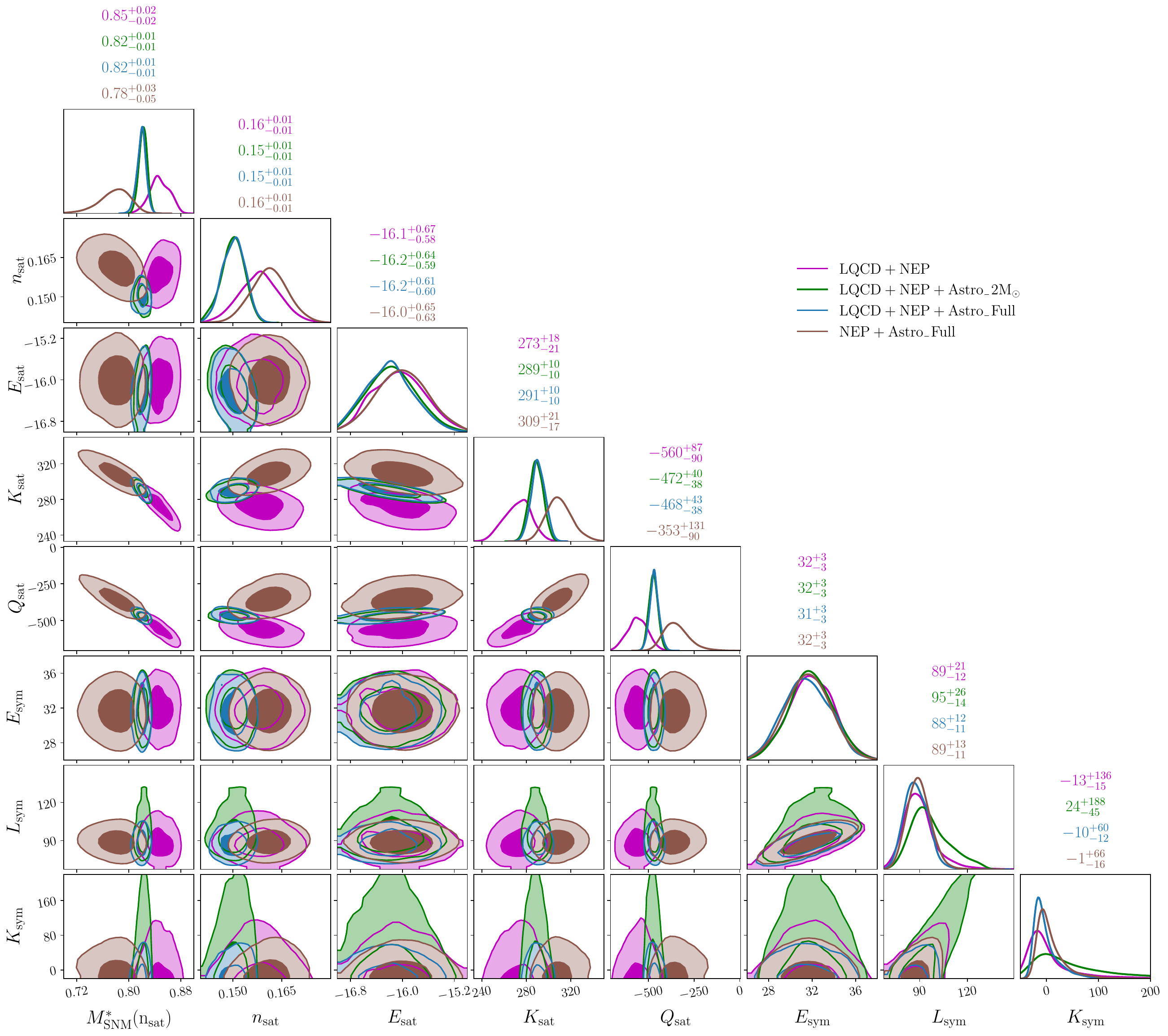}
\caption{Corner plot showing the joint (off-diagonal panels) and marginalized (diagonal panels) posterior distributions of the NEPs and $M^*_{\rm SNM}(n_{\rm sat})$ within the RMF-CC model under different sets of constraints as given in the legend. The saturation density $n_{\rm sat}$ is shown in the unit of $\rm fm ^{-3}$, $M^*_{\rm SNM}(n_{\rm sat})$ in the unit of $M_N$ and the empirical parameters $E_{\rm sat}, \ K_{\rm sat},\ Q_{\rm sat},\ E_{\rm sym}, \ L_{\rm sym},$ and $K_{\rm sym}$ are displayed in the units of MeV. To avoid clutter, the units are omitted in the figure.}
\label{fig:NEPs_RMFCC}
\end{figure*}

The posteriors given in \Cref{tab:posteriors_RMFCC_VLSM} are also employed to calculate the marginal and joint posterior distribution for the LQCD parameters and the NEPs, which are displayed in \Cref{fig:LQCD_RMFCC} and \Cref{fig:NEPs_RMFCC}, respectively, for the RMF-CC model.
As observed in the discussion of \Cref{tab:isoscalarNEP_posteriors_RMFCC_VLSM}, the sign and order of magnitude obtained for the LQCD parameters $a_2$ and $a_4$ are similar with and without the LQCD priors. However, imposing the LQCD prior largely reduces the values accepted for $a_2$ and $a_4$ as shown in  \Cref{fig:LQCD_RMFCC}, where the gray bands represent the allowed regions for the LQCD parameters. Imposing the $2M_\odot$ astrophysical constraint has a significant impact on the distribution for $a_2$, favoring the lower region compatible with Table~\ref{tab:Lattice_data}, i.e., $a_2\sim 1.4$~GeV$^{-1}$.

The distributions for the iso-scalar NEP shown in \Cref{fig:NEPs_RMFCC} are narrower considering the constraints from LQCD, NEPs, and Astro\_Full. There is almost no difference whether Astro\_Full or Astro\_$2M_\odot$ constraints are considered, showing that the astrophysical constraint is mostly due to the constraint imposed by the $2M_\odot$ observation. We also note that some quantities are in tension with the constraints from LQCD and from the astrophysical measurements, i.e., $M^*_{\rm SNM}(n_{\rm sat})$, $K_\sat$, and $Q_\sat$. The iso-vector NEPs, i.e., $E_\sym$, $L_\sym$, and $K_\sym$ are not very sensitive to the different constraints considered in \Cref{fig:NEPs_RMFCC}, provided the ones for the prior distribution of NEPs are considered.
Note the tight correlation between $K_\sat$ and $Q_\sat$, which is mostly due to the fact that these two NEPs are tightly correlated with $M^*_{\rm SNM}(n_{\rm sat})$. Note also the correlation between $E_\sym$ and $L_\sym$ induced by the constraints considered in our analysis.

\begin{figure*}[htbp]
\centering
\begin{subfigure}{.45\textwidth}
  \centering
  \includegraphics[width=\linewidth]{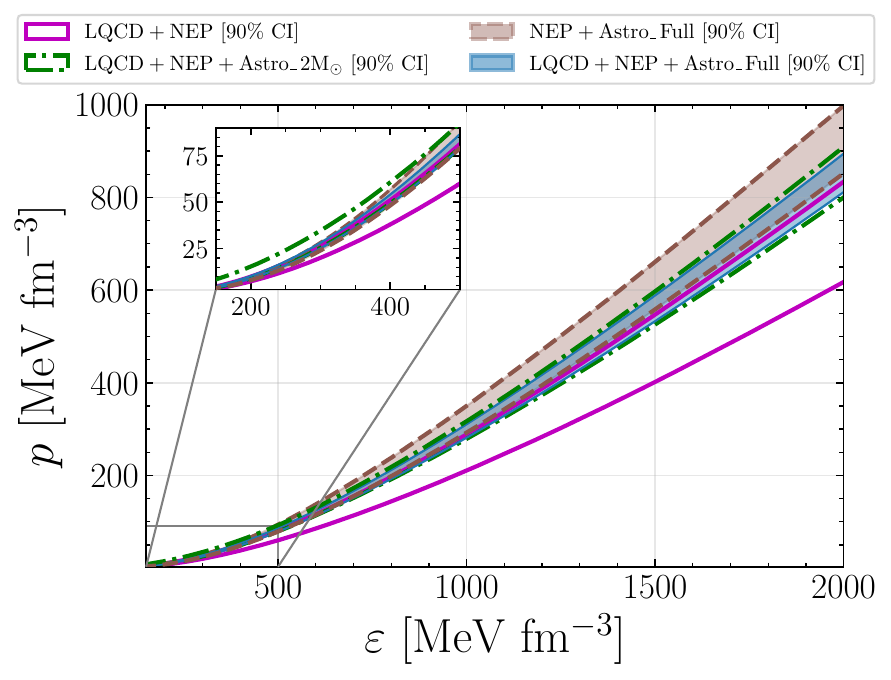}
  \caption{} 
  \label{fig:EoS_RMFCCVLSM}
\end{subfigure}%
\begin{subfigure}{.45\textwidth}
  \centering
  \includegraphics[width=0.97\linewidth]{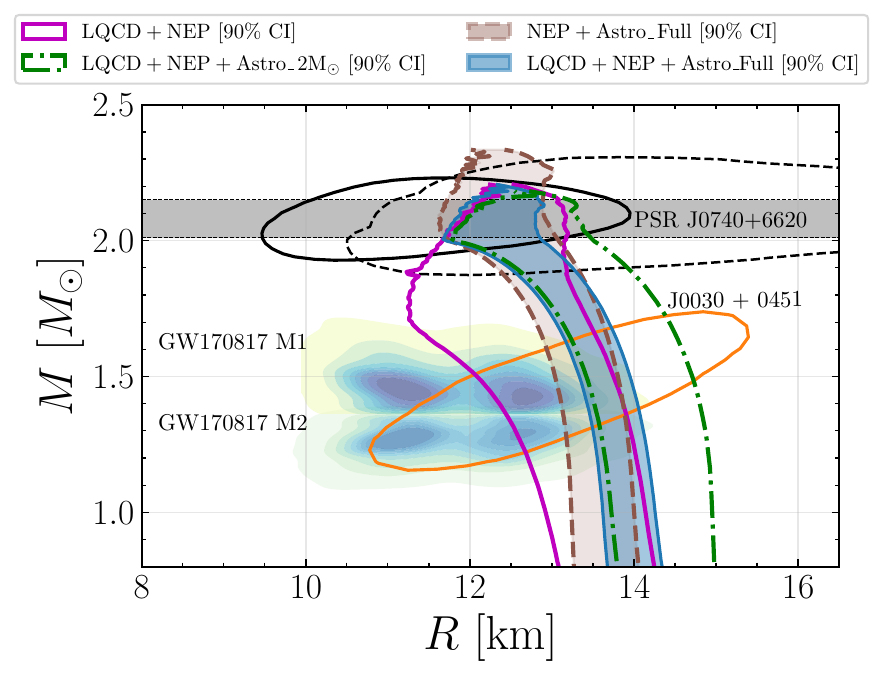}
  \caption{}
  \label{fig:MR_RMFCCVLSM}
\end{subfigure}
\caption{
(a) Uncertainty in the NS EoS: $p$ as a function of $\varepsilon$ at 90\% CI for RMF-CC model under different constraints. (b) Uncertainty in $MR$ relation: $R$ as a function of $M$ at 90\% CI  for RMF-CC model under different constraints. In (b), the grey horizontal band corresponds to $M=2.072^{+0.067}_{-0.066} M_{\odot}$ of PSR J0740$+$6620,  the 90\% contours for PSR J0740+6620 corresponding to Ref.~\cite{Miller_2021} are shown in a dashed black line, while the uncertainty corresponding to Ref.~\cite{NICER0740_Riley} is shown in a solid black line. The solid orange contour shows the $MR$ measurement for PSR J0030+0451. The $MR$ estimates of the two companion neutron stars of the merger event GW170817 are indicated by the shaded area labeled with GW170817 M1 (M2) in (b).}
\end{figure*}

\begin{figure}[htbp]
\centering
\includegraphics[width=\linewidth]{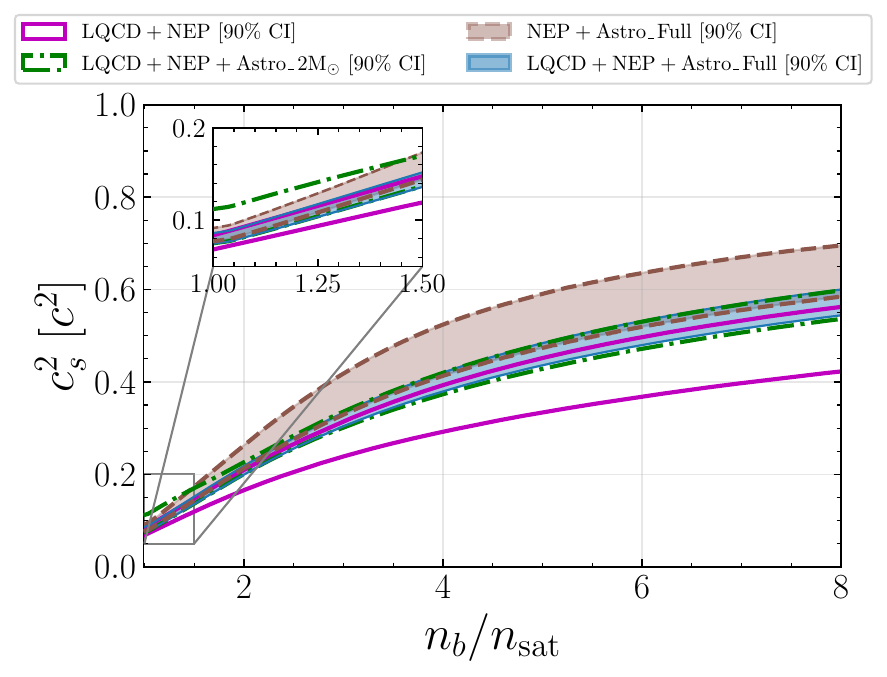}
\caption{$c^2_s$ (in units of $c^2$) of NS matter as a function of scaled nucleon density $n_b/n_{\rm sat}$ obtained with the posteriors of RMF-CC model parameters with different constraints given in the legend.} 
\label{fig:cs2_RMFCCVLSM}
\end{figure}

Let us now discuss the properties of single NSs.
The prediction for the NS EoS is shown in \Cref{fig:EoS_RMFCCVLSM}, while the prediction for the
$MR$ relations and the sound speed $c^2_s$ in NS matter are displayed in \Cref{fig:MR_RMFCCVLSM,fig:cs2_RMFCCVLSM}, respectively. The constraints associated with the contours are given in the legend of the figures. Note from \Cref{fig:EoS_RMFCCVLSM} and \Cref{fig:MR_RMFCCVLSM} that, LQCD and the astrophysical constraints (especially the $2M_{\odot}$ constraints), are in tension: the overlap between models satisfying LQCD and astrophysical constraints is small.
There is also a tension between the constraint from the GW170817 measurement and the other astrophysical constraints (especially the $2M_{\odot}$), see \Cref{fig:MR_RMFCCVLSM}.

Under the consideration of LQCD and NEP constraints, the resulting posteriors of the model parameters ($\mathcal{E}_{\rm RMF\textrm{-}CC}$) as well as for the NEPs such as $K_{\rm sat}$ and $M_{\rm SNM}^*(n_{\rm sat})$ are in great agreement to that of the results reported in Ref.~\cite{Somasundaram:2021hna}. The presence of LQCD constraints imposes strong bounds and a strong degeneracy among the scalar parameters $ m_s$, $g_s$ and $C_{\rm NS}$, thereby restricting the parameter space (or the posterior volume). 

\begin{figure}[htbp]
\centering
  \includegraphics[width=\linewidth]{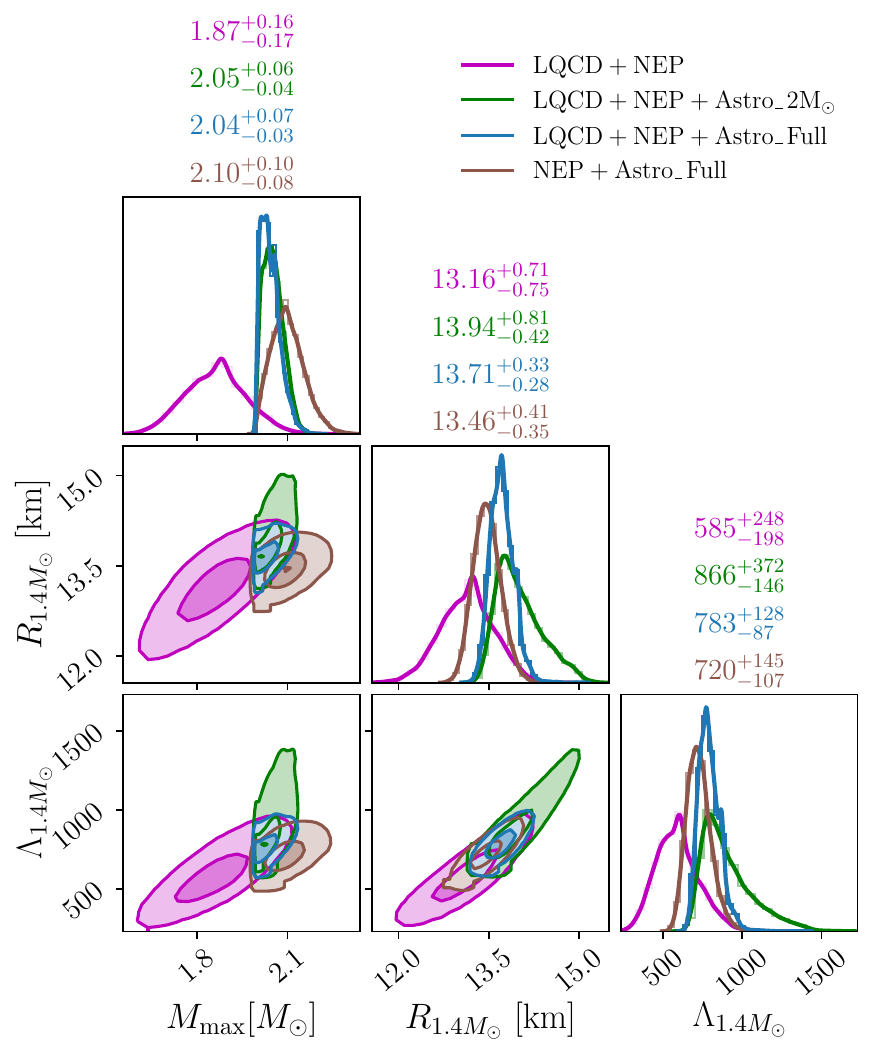}
\caption{Corner plot showing the joint (off-diagonal panels) and marginalized (diagonal panels) posterior distributions of the key NS properties reconstructed within the RMF-CC model under different sets of constraints. The median values and the corresponding 90\% CI lower and upper bounds are indicated for each quantity. For a given constraint set, a single color consistently represents the associated density contours, medians, and 90\% credible regions; e.g., results corresponding to $D_3=\{\text{LQCD, NEP, } \mathrm{Astro\_\,Full}\}$ are shown in blue.}
\label{fig:R14_L14_Mmax_RMFCCVLSM}
\end{figure}

We display in \Cref{fig:R14_L14_Mmax_RMFCCVLSM} the posterior distribution for $M_{\rm max}$, $R_{1.4M_{\odot}}$ (the radius of a 1.4$M_{\odot}$ NS) and $\Lambda_{1.4M_{\odot}}$ (the tidal deformability of a 1.4$M_{\odot}$).
Considering the LQCD and NEP constraints we obtain soft EoSs at high density, predicting a median of $M_{\rm max}\sim 1.87 M_{\odot}$ for the RMF-CC model, as shown in \Cref{fig:R14_L14_Mmax_RMFCCVLSM}. 

Though one of the motivations of the RMF-CC model is to better reproduce the LQCD constraints by including confinement or polarisation effects in the nucleon mass, the inclusion of the LQCD constraints strongly constrains the parameter space of the scalar parameters ($g_s, m_s, C_{\rm NS}$) that control the stiffness of the EoSs. Hence, we analyze the constraints on the model parameters and the RMF-CC model's predictions by relaxing the LQCD constraints. As shown earlier and illustrated in \cref{fig:EoS_RMFCCVLSM,fig:MR_RMFCCVLSM,fig:cs2_RMFCCVLSM,fig:R14_L14_Mmax_RMFCCVLSM}, excluding the LQCD data leads to better agreement between the RMF-CC model predictions and astrophysical constraints. The findings in this scenario can then also be compared to the other class of RMF models under similar constraints, as we discuss in ~\Cref{sec:compare_RMF}. 

\begin{figure}[htpb]
    \centering
    \includegraphics[width=\linewidth]{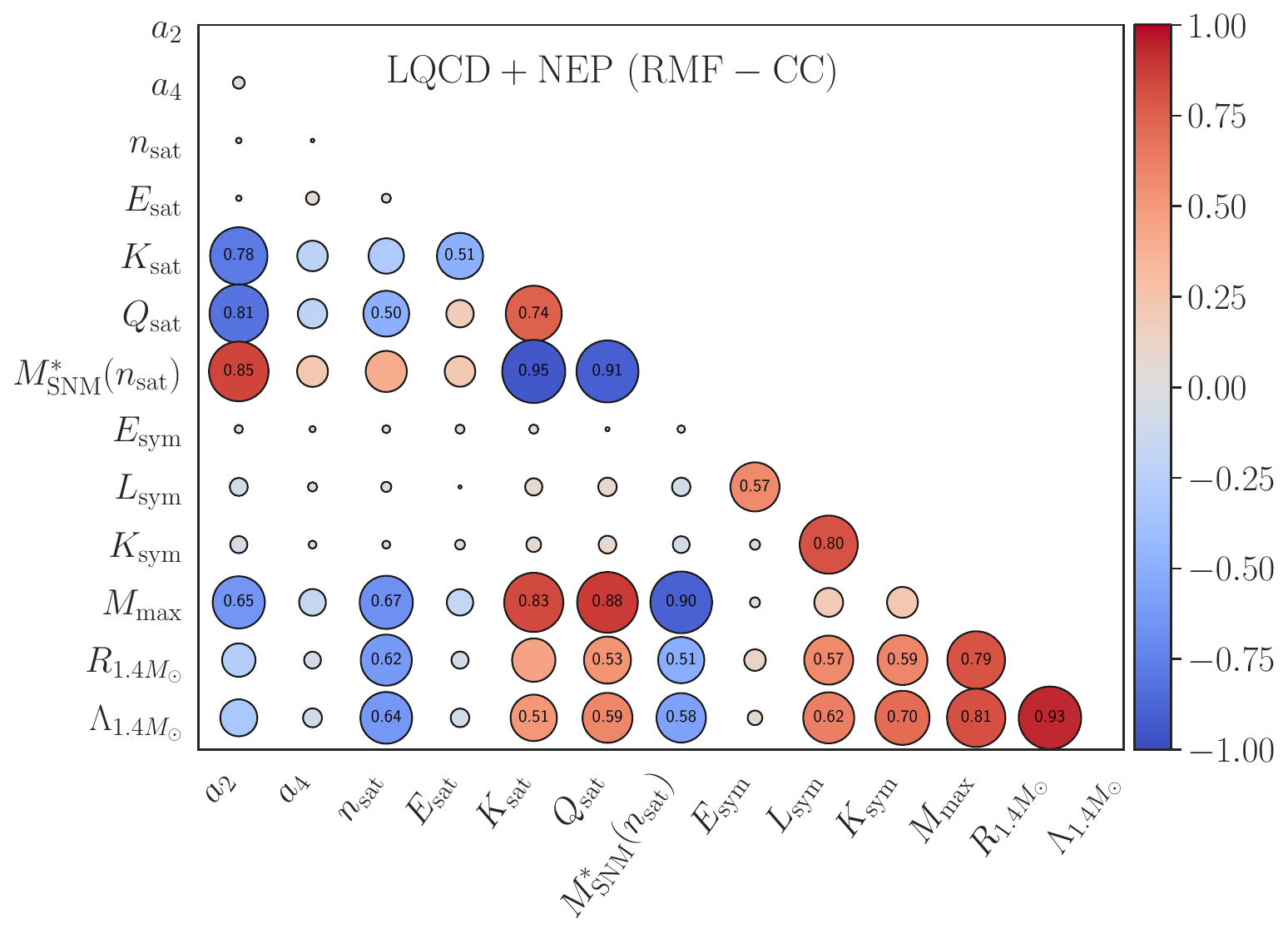}
    \caption{Pearson's correlation coefficient among the Lattice parameters,  NEPs, and NS properties with the constraints $D=\{\rm LQCD, NEP\}$. The size of the circle is relative to the absolute magnitude of the correlation coefficient. For the absolute correlation coefficient $\geq 0.5$, the values are marked inside the circle. The matrix is resulting from the analysis of baseline RMF-CC model~\cite{Somasundaram:2021hna}.}
    \label{fig:corr_lat_NEP}
\end{figure}

\begin{figure}[htbp]
    \centering
    \includegraphics[width=\linewidth]{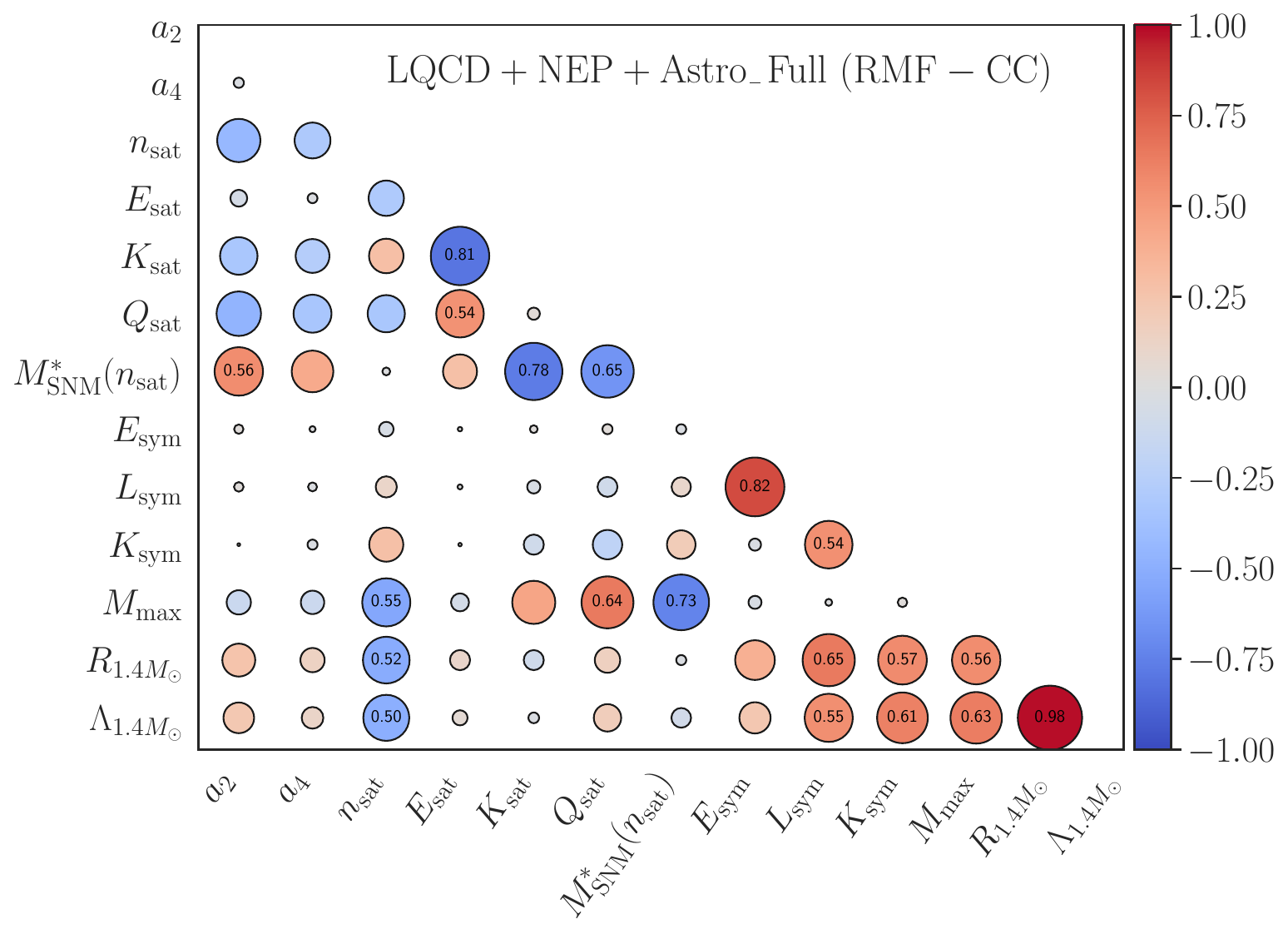}
    \caption{ Same as ~\Cref{fig:corr_lat_NEP}, but for the joint consideration of astrophysical constraints along with the LQCD and NEPs: i.e., $D_3=\{\rm LQCD,\ NEP,\ Astro\_\,Full\}$.
 }
    \label{fig:corr_lat_NEP_Astro}
\end{figure}

For a better understanding of the results, we display the Pearson correlation coefficient among the key NEPs and NS properties obtained for the dataset $D=\{{\rm LQCD},{\rm NEP}\}$, i.e., in absence of astrophysical constraints, and for the dataset $D=\{{\rm LQCD},\rm {NEP},{\rm Astro\_Full}\}$, in  \Cref{fig:corr_lat_NEP,fig:corr_lat_NEP_Astro}, respectively. In \Cref{fig:corr_lat_NEP,fig:corr_lat_NEP_Astro}, positive and negative correlations are shown in red and blue, respectively, while the color intensity indicates the strength of the correlation. For clarity, only correlations with an absolute value $\geq 0.5$ are explicitly labeled in the figures.

In the presence of LQCD's constraint, the model prefers a higher value for $M^*_{\rm SNM}$ and the strong inverse correlation among the $M^*_{{\rm SNM}}(n_{\rm sat})$ and $M_{\rm max}$, see ~\Cref{fig:corr_lat_NEP_Astro}, explains the non-existence of enough parameter space to to reach $2M_{\odot}$. However, unlike the conventional RMF models in the absence of nucleon polarization, where most of the qualitative conclusions regarding the stiffness of the NS EoSs are explained with $M^*_{\rm SNM}(n_{\rm sat})$, in the RMF-CC model there is additional influence coming from the confinement effect at high-density, which will be discussed latter in this work.  To reproduce the stiff EoSs for $2M_{\odot}$ NSs, the model prefers $K_{\rm sat}\sim 300$ MeV, whether or not LQCD constraints are considered. As shown in \Cref{fig:LQCD_RMFCC}, for the simultaneous reproduction of LQCD and NEP along with the $2M_{\odot}$ NS, the RMF-CC model prefers lower values of the lattice parameter $a_2$. 

\begin{figure*}[htbp]
\centering
\begin{subfigure}{.45\textwidth}
  \centering
  \includegraphics[width=\linewidth]{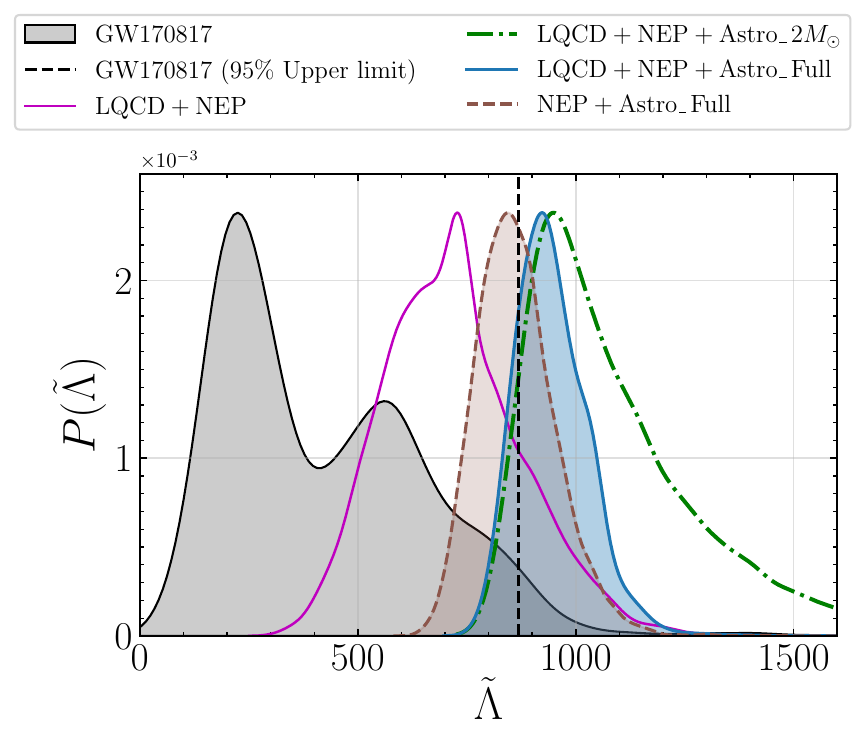}
  \caption{} 
  \label{fig:ltilde_pdf}
\end{subfigure}%
\begin{subfigure}{.45\textwidth}
  \centering
  \includegraphics[width=\linewidth]{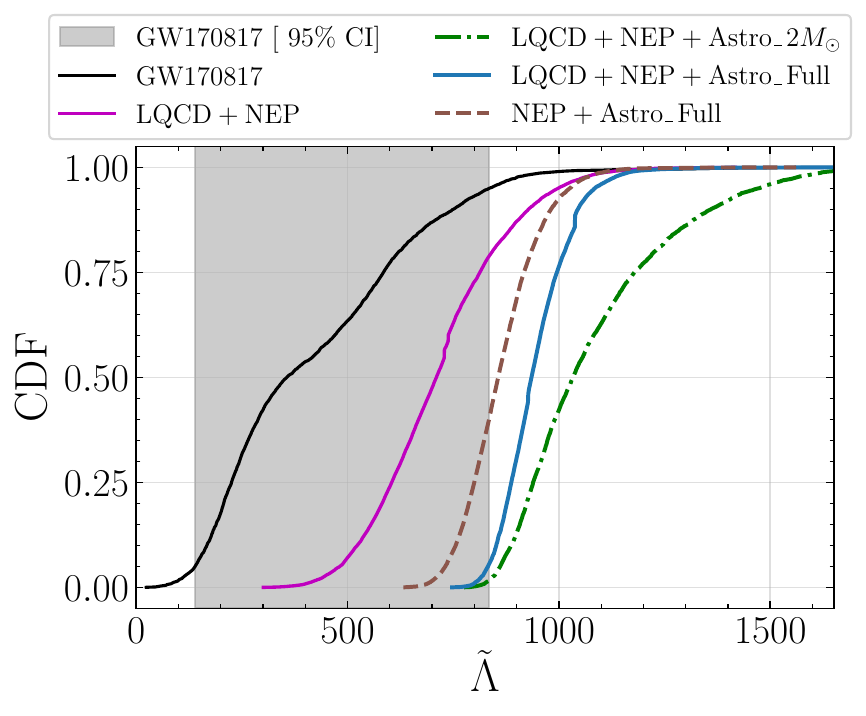}
  \caption{}
  \label{fig:ltilde_cdf}
\end{subfigure}
\caption{(a) The probability distribution function (PDF) of $\tilde{\Lambda}$. The grey band displays the observational $P(\tilde{\Lambda})$ for the BNS event GW170817. The 95\% upper bound is also shown with a dashed line. The distribution of $\tilde{\Lambda}$, for the RMF-CC model under different constraints, has also been displayed. The $P(\tilde{\Lambda})$s for the RMF-CC model are scaled to fit in a single figure for better comparison. (b) The CDFs corresponding to the PDFs shown in (a).}
\end{figure*}

Let us now present quantities related to binary systems.
To assess the compatibility of the model with GW170817, we compute the effective tidal deformability $\tilde{\Lambda}$ defined as:
\begin{equation}
\tilde{\Lambda} = \frac{16}{13}
\frac{(M_1 + 12 M_2) M_1^4 \Lambda_1 + (M_2 + 12 M_1) M_2^4 \Lambda_2}{(M_1 + M_2)^5}~,
\end{equation}
where $M_i$ and $\Lambda_i$ $(i=1,2)$ are the masses and dimensionless quadrupolar tidal deformabilities of the two binary components in their rest frame. We evaluate $\tilde{\Lambda}(\mathcal{E})$ assuming a unique EoS $\mathcal{E}$ for all compact stars and using posterior samples of the model RMF-CC with the assumption of equal mass binary $M_1 = M_2 = 1.36\,M_\odot$. Note that a different choice of the mass ratio, or equivalently of $M_1$ and $M_2$, may modify the quantitative results discussed in this section; however, the qualitative conclusions remain unaltered. The choice of masses corresponds to the measured chirp mass $\mathcal{M}_c = 1.186\,M_\odot$ for GW170817 in the source frame. 
Note that the likelihood probability accounts for the observational uncertainties in the individual component masses and tidal deformabilities. 
The posterior distribution for $\tilde{\Lambda}(\mathcal{E})$ is compared to the observed LIGO-VIRGO posterior $\tilde{\Lambda}_{\rm GW170817}$, see \Cref{fig:ltilde_pdf} for the probability density functions (PDFs) and \Cref{fig:ltilde_cdf} for the cumulative 
distribution functions (CDFs). 

The statistical comparisons are summarized as follows:
\begin{itemize}
\item \textbf{For $D_1 = \{{\rm LQCD, NEP}\} $:}
The overlap of $\tilde{\Lambda}(\mathcal{E})$ posterior with the 95\% CI of GW170817 is $\sim$ 0.8. This is also visible in ~\Cref{fig:ltilde_cdf}, where the model’s CDF (magenta line) intersects the upper bound of observed 95\% CI at a CDF $\sim 0.8 $, i.e $\rm CDF(\tilde{\Lambda}_{\rm RMF\textrm{-}CC}^{D_1} \leq \tilde{\Lambda}^{95\%  \ \rm Upper \ bound}_{\rm GW170817})\sim 0.8$. This result suggests that the RMF-CC model under $D_1$'s constraint is consistent with GW170817: a significant portion of its $\tilde{\Lambda}$ posterior overlaps with the observational constraints.
\item \textbf{For $D_2 = \{{\rm LQCD, NEP, \mathrm{Astro}\_\,2M_{\odot}}\} $:}
It is clear that, from \Cref{fig:ltilde_cdf,fig:ltilde_pdf}, with the further restrictions from the $2M_{\odot}$ constraints poses a loose agreement to the GW170817 data with a $\rm{CDF}(\tilde{\Lambda}_{\rm RMF\textrm{-}CC}^{D_2} \leq \tilde{\Lambda}^{95\%  \ \rm Upper \ bound}_{\rm GW170817})\sim 0.03$. 
\item \textbf{For $D_3 = \{{\rm LQCD, NEP, \mathrm{Astro\_\,Full}}\} $:}
Even with the inclusion of the tidal deformability constraints from  GW170817, the model under $ D_3$'s constraints is still inconsistent with GW170817. We found that $\rm{CDF}(\tilde{\Lambda}_{\rm RMF\textrm{-}CC}^{D_3} \leq \tilde{\Lambda}^{95\%  \ \rm Upper \ bound}_{\rm GW170817})\sim 0.07$. 
\item \textbf{For $D_4 = \{{\rm NEP, \mathrm{Astro\_\,Full}}\} $:}
The exclusion of the LQCD data or consideration of the `NEP' and `Astro' only results in a higher median for $M_{\rm max}$ (see \Cref{fig:R14_L14_Mmax_RMFCCVLSM}) and also improves the model's compatibility with the tidal deformability inferred from GW170817 with $\rm CDF(\tilde{\Lambda}_{\rm RMF\textrm{-}CC}^{D_4} \leq \tilde{\Lambda}^{95\%  \ \rm Upper \ bound}_{\rm GW170817})\sim 0.5$ (see \Cref{fig:ltilde_cdf,fig:ltilde_pdf}).
\end{itemize}

\begin{table*}[t]
\centering
\caption{Median values with the upper and lower bounds for 90\% CI for NS observable and  EoS properties for different models under $D_4=\{ \rm NEP, Astro\_\,Full\} $ and $D_3=\{\text{LQCD, NEP, } \mathrm{Astro\_\,Full}\} $ constraints. In the table, $n_b(M),\ p_c(M)$ and $\ c_s^2(M)$ represent the  nucleon density, pressure, and the $c^2_s$ at the center of a NS with mass $M$, respectively.}
\label{tab:NS_properties}
\renewcommand{\arraystretch}{1.25}
\begin{tabular}{l|c|c|c|c|c}
\hline
 & \multicolumn{3}{c|}{\textbf{NEP+Astro$\_\,$Full}} 
 & \multicolumn{2}{c}{\textbf{LQCD+NEP+Astro$\_\,$Full}} \\
\cline{2-6}
Quantity
 & RMF & RMF-CC &  RMF-CC$+\rm V_{\omega\rho}$
 & RMF-CC &  RMF-CC$+\rm V_{\omega\rho}$ \\
\hline
$M_{\rm max}\,[M_\odot]$ & \MmaxRMFNEPAstro & \MmaxRMFCCNEPAstro & \MmaxRMFCClwNEPAstro & \MmaxRMFCCLatNEPAstro & \MmaxRMFCClwLatNEPAstro \\
$R (M_{\rm max})\,[\rm km]$ & \RmaxRMFNEPAstro & \RmaxRMFCCNEPAstro & \RmaxRMFCClwNEPAstro & \RmaxRMFCCLatNEPAstro & \RmaxRMFCClwLatNEPAstro \\
$n_b(M_{\rm max})\,[\mathrm{fm^{-3}}]$ & \nbMmaxRMFNEPAstro & \nbMmaxRMFCCNEPAstro & \nbMmaxRMFCClwNEPAstro & \nbMmaxRMFCCLatNEPAstro & \nbMmaxRMFCClwLatNEPAstro \\
$p_c(M_{\rm max})\,[\mathrm{MeV\,fm^{-3}}]$ & \pcmaxRMFNEPAstro & \pcmaxRMFCCNEPAstro & \pcmaxRMFCClwNEPAstro & \pcmaxRMFCCLatNEPAstro & \pcmaxRMFCClwLatNEPAstro \\
$c^2_s(M_{\rm max}) \ [c^2]$ & \CsMmaxRMFNEPAstro & \CsMmaxRMFCCNEPAstro & \CsMmaxRMFCClwNEPAstro & \CsMmaxRMFCCLatNEPAstro & \CsMmaxRMFCClwLatNEPAstro \\
$R_{1.4}\,[\mathrm{km}]$ & \RadiusonepointfourRMFNEPAstro & \RadiusonepointfourRMFCCNEPAstro & \RadiusonepointfourRMFCClwNEPAstro & \RadiusonepointfourRMFCCLatNEPAstro & \RadiusonepointfourRMFCClwLatNEPAstro \\
$\Lambda_{1.4}$ & \TidalonepointfourRMFNEPAstro & \TidalonepointfourRMFCCNEPAstro & \TidalonepointfourRMFCClwNEPAstro & \TidalonepointfourRMFCCLatNEPAstro & \TidalonepointfourRMFCClwLatNEPAstro \\
$n_b(1.4\,M_\odot)\,[\mathrm{fm^{-3}}]$ & \nbOnepointFourMsunRMFNEPAstro & \nbOnepointFourMsunRMFCCNEPAstro & \nbOnepointFourMsunRMFCClwNEPAstro & \nbOnepointFourMsunRMFCCLatNEPAstro & \nbOnepointFourMsunRMFCClwLatNEPAstro \\
$p_c(1.4\,M_\odot)\,[\mathrm{MeV\,fm^{-3}}]$ & \pcOnepointFourmsunRMFNEPAstro & \pcOnepointFourmsunRMFCCNEPAstro & \pcOnepointFourmsunRMFCClwNEPAstro & \pcOnepointFourmsunRMFCCLatNEPAstro & \pcOnepointFourmsunRMFCClwLatNEPAstro \\
$c^2_s(1.4\,M_\odot) \ [c^2]$ & \csOnepointFourMsunRMFNEPAstro & \csOnepointFourMsunRMFCCNEPAstro & \csOnepointFourMsunRMFCClwNEPAstro & \csOnepointFourMsunRMFCCLatNEPAstro & \csOnepointFourMsunRMFCClwLatNEPAstro \\
$p(n_b=0.3\rm\,fm^{-3})\,[\mathrm{MeV\,fm^{-3}}]$ & \pThreeZeroRMFNEPAstro & \pThreeZeroRMFCCNEPAstro & \pThreeZeroRMFCClwNEPAstro & \pThreeZeroRMFCCLatNEPAstro & \pThreeZeroRMFCClwLatNEPAstro \\
$c^2_s(n_b=0.3\rm\,fm^{-3}) \ [c^2]$ & \csThreeZeroRMFNEPAstro & \csThreeZeroRMFCCNEPAstro & \csThreeZeroRMFCClwNEPAstro & \csThreeZeroRMFCCLatNEPAstro & \csThreeZeroRMFCClwLatNEPAstro \\
\hline
\end{tabular}
\end{table*}

In~\Cref{tab:NS_properties} we tabulated the uncertainties in the relevant NS observables and NS EoS properties after incorporating the astrophysical constraints. Following the current interest and future experimental prospects, we also include the EoS properties, specifically the pressure $p$ and the squared speed of sound $c_s^2$ of NS matter at $n_b = 0.3~\rm fm^{-3}$ in~\Cref{tab:NS_properties}.

\begin{figure}[t]
\centering
\includegraphics[width=\linewidth]{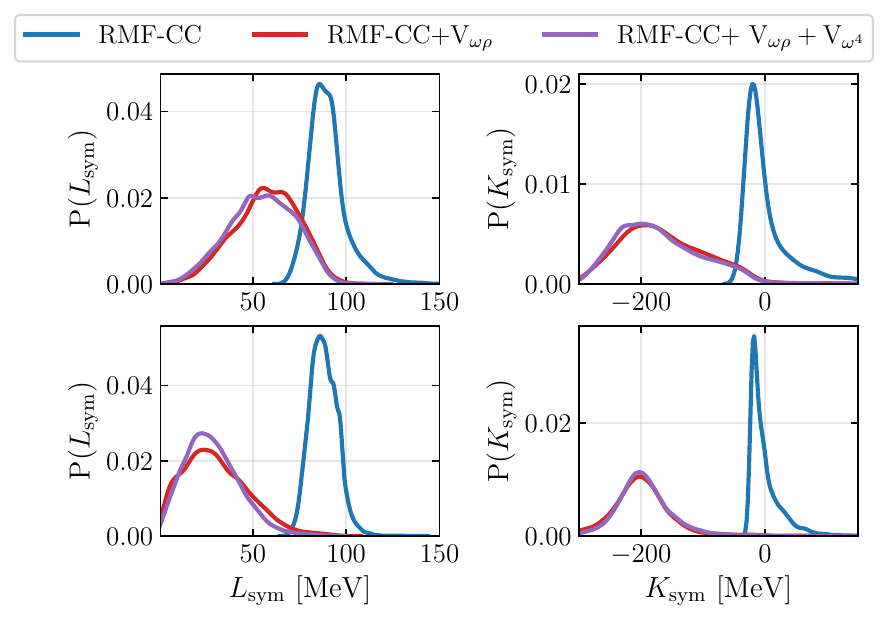}
\caption{Posterior probability distributions of $L_{\rm sym}$ (left) and $K_{\rm sym}$ (right). The upper panels show the posteriors obtained using joint constraints from LQCD and NEPs. The lower panels display the posteriors after imposing astrophysical constraints along with LQCD and NEPs.} 
\label{fig:lsym_ksym_lamomega_rho}
\end{figure}

\begin{figure*}[t]
\centering
\includegraphics[width=0.9\linewidth]{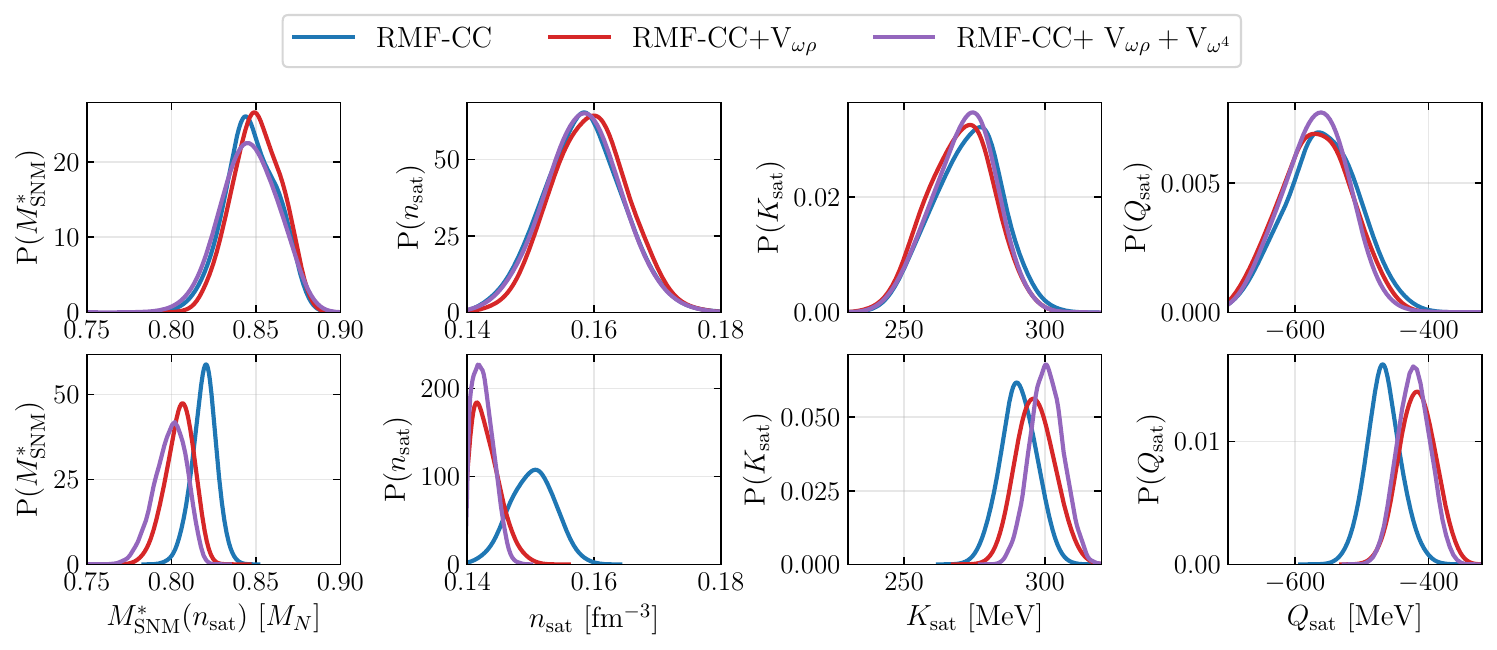}
\caption{Same as ~\Cref{fig:lsym_ksym_lamomega_rho}, but for posterior probability distributions of $M^*_{\rm SNM}$, $n_{\rm sat}$ ,  $K_{\rm sat}$ , and  $Q_{\rm sat}$ (from left to right).} 
\label{fig:nsat_ksat_qsat_lamomega_rho}
\end{figure*}

In summary, although the RMF-CC model successfully reproduces LQCD parameters and NEPs when only these are considered, imposing astrophysical observations sharply reduces the posterior volume and makes the model incompatible with GW170817. The dominant source of this incompatibility is the requirement $M_{\rm max} \geq 2\, M_\odot$. This constraint enforces a stringent constraint on the stiffer EoS, predominantly resulting in the  $\tilde{\Lambda}$ distribution to values well above the  95\% CI  upper bound of   $\tilde{\Lambda}_{\rm GW170817}$, as clearly visible from \Cref{fig:ltilde_pdf,fig:ltilde_cdf}). The resulting stiffening due to the $2M_{\odot}$ constraint, also results in higher values for ($R_{1.4}$ and $\Lambda_{1.4}$), as shown in \Cref{fig:R14_L14_Mmax_RMFCCVLSM}. These larger values can be explained from the model's systematic preference for a high $L_{\rm sym} \sim 85$~MeV, (which persists even without astrophysical constraints) and moderately strong positive correlation among $L_{\rm sym}$ and both $R_{1.4}$ and $\Lambda_{1.4}$ under astrophysical constraints, as displayed in ~\Cref{fig:corr_lat_NEP_Astro}. Furthermore, we find that excluding the NICER $MR$ measurements does not significantly affect the inferred model parameters or NEPs (see ~\Cref{tab:posteriors_RMFCC_VLSM,tab:isoscalarNEP_posteriors_RMFCC_VLSM,tab:isovectorNEP_posteriors_RMFCC_VLSM}). The dominant constraints arise from the $2\, M_\odot$ pulsar and GW measurements. 

In the following, we show that the extensions of the RMF-CC model by including additional non-linear couplings contribute to reducing the tension described in this section.

\subsection{RMF-CC with vector-isovector \texorpdfstring{$\omega\rho$}{omega-rho} coupling}
\label{sec:vector-isovector}

The RMF-CC model in its original formulation~\cite{Chamseddine2023,Chanfray2001,Massot2008,Massot2009}, \emph{i.e.}, in the absence of $V_{\rm MCs}$, predicts a relatively large symmetry energy slope  $L_{\rm sym}$ and, consequently, high tidal deformabilities $\tilde{\Lambda}_{\rm GW170817}$, as discussed in \Cref{sec:RMFCC_VLSM}. In contrast, subsequent developments in RMF models have demonstrated that cross-meson couplings are essential for improving agreement with neutron skin thickness measurements and astrophysical observations of NSs~\cite{Horowitz2001,Todd2005,Chen2014}. In particular, the vector-isovector $\omega\rho$ coupling, parameterized by $\lambda_{\omega \rho}$ plays a central role in reducing the value of $L_{\rm sym}$~\cite{Horowitz2001} and, as a result, in lowering the predicted values of $R_{1.4M_{\odot}}$ and $\Lambda_{1.4M_{\odot}}$, which makes it more compatible with GW170817 measurement. Note that in this work, we have adopted a large range of values for $\lambda_{\omega \rho}$ going from $0$ up to $1$. In this section, we disregard other higher-order interaction terms, such as $\omega^4$ and $\rho^4$, discussed in Refs.~\cite{Horowitz2001,Horowitz2010}, since the $\omega\rho$ coupling is expected to dominate to these additional corrections.

While the scalar sector of RMF-CC model is constrained by chiral symmetry breaking (the $s$ meson is the physical indicator of chiral symmetry breaking), as discussed in \Cref{sec:Formalism}, the situation is different for the vector channels: they are indeed allowed by chiral symmetry, but they do not play any dynamical role in its breaking, so their coefficients are not fixed or constrained by it. It is thus natural to treat theses terms in a phenomenological approach. We therefore implement the coupling to vector mesons following the parameterization approach proposed in Ref.~\cite{Horowitz2010}. The modified Lagrangian density is then given by,
\begin{equation}
\mathcal{L}_{\rm RMF\textrm{-}CC+V_{\omega\rho}}=
\mathcal{L}_{\rm RMF\textrm{-}CC} + \lambda_{\omega \rho}
(g_{\rho}^2 \vec{\rho}_{\mu}\cdot\vec{\rho}^{\mu})
(g_{\omega}^2 \omega_{\mu}\omega^{\mu}) \, .
\end{equation}
We refer to this extended model as RMF-CC$+\rm V_{\omega\rho}$, where we set $\zeta = 0$ in the potential term $V_{\rm MCs}$. The model is defined by the parameter set $\mathcal{E}_{\rm RMF\textrm{-}CC+V_{\omega\rho}} = \{ m_s, g_s, C_{\rm NS}, g_{\omega}, g_{\rho}, g_{\delta}, \lambda_{\omega \rho} \}$. The posterior medians and corresponding $90\%$ CIs for these parameters, obtained from the Bayesian analysis under different scenarios, are presented in \Cref{tab:posteriors_RMFCC_VLSM}. While there is no significant impact of this new coupling on $m_s$, $g_s$, $C_{\rm NS}$, $g_\omega$, and $g_\rho$, the value for $g_\delta$ is reduced by almost a factor of 2. As detailed before, we deduce from \Cref{tab:posteriors_RMFCC_VLSM}, the medians and uncertainties of the isoscalar and isovector NEPs given in ~\Cref{tab:isoscalarNEP_posteriors_RMFCC_VLSM,tab:isovectorNEP_posteriors_RMFCC_VLSM}. Again, most of the NEPs are insensitive to the new coupling, except $L_\sym\sim 30$-$60$~MeV and $K_\sym\sim -220$-$-150$~MeV. The uncertainties of a few relevant NS and NS EoS properties after incorporating the astrophysical constraints are tabulated in \Cref{tab:NS_properties}. Note the impact of the $\omega\rho$ coupling on the reduction of the effective tidal deformability, $\Lambda_{1.4}\sim 470$ and the pressure at 0.3~fm$^{-3}$, which is $\sim 19.80$~MeV~fm$^{-3}$.

The effect of the $\omega\rho$ coupling and of the constraints on the iso-vector NEPs $L_\sym$ and $K_\sym$ is shown in \Cref{fig:lsym_ksym_lamomega_rho}.
As discussed before, the inclusion of $\omega\rho$ coupling substantially reduces the predicted value for $L_{\rm sym}$, which goes from $\sim 85$ MeV in the RMF-CC model down to $\sim 60$~MeV (considering only LQCD and NEP constraints) or $\sim 30$~MeV (with additional astrophysical constraints) in RMF-CC$+\rm V_{\omega\rho}$, see left panels shown in \Cref{fig:lsym_ksym_lamomega_rho}. The $\omega\rho$ interaction also leads to a lower predicted value of the curvature parameter $K_{\rm sym} \sim -170$ MeV compared to $\sim -10 $ MeV within the RMF-CC model (considering only LQCD and NEP constraints, see ~\Cref{tab:isovectorNEP_posteriors_RMFCC_VLSM}). It is clear from \Cref{fig:lsym_ksym_lamomega_rho} and \Cref{tab:isovectorNEP_posteriors_RMFCC_VLSM} that, for the RMF-CC$+\mathrm{V}{\omega\rho}$ model, the inclusion of astrophysical constraints significantly reduces the uncertainty in $K{\rm sym}$, while favoring values $\sim -200$ MeV.

In a similar way, the effect on the isoscalar NEPs is shown in \Cref{fig:nsat_ksat_qsat_lamomega_rho}.
As discussed before, the inclusion of the $\omega\rho$ interaction has only a marginal effect on the isoscalar NEPs before the application of astrophysical observational constraints, as illustrated in the upper panels of \Cref{fig:nsat_ksat_qsat_lamomega_rho}: the PDFs are essentially unchanged compared to the RMF-CC model.
The application of astrophysical observational constraints has an impact on the isoscalar NEPs: $M^*_{\rm SNM}(n_\sat)$ is slightly shifted down and reaches $\sim 0.81$, $n_\sat$ is shifted down while $K_\sat$ and $Q_\sat$ are sligthly larger. The shifts are, however, small, since they are not larger than $1\sigma$ away from the values obtained without the $\omega\rho$ coupling.


\begin{figure*}[htbp]
\centering
\includegraphics[width=0.9\linewidth]{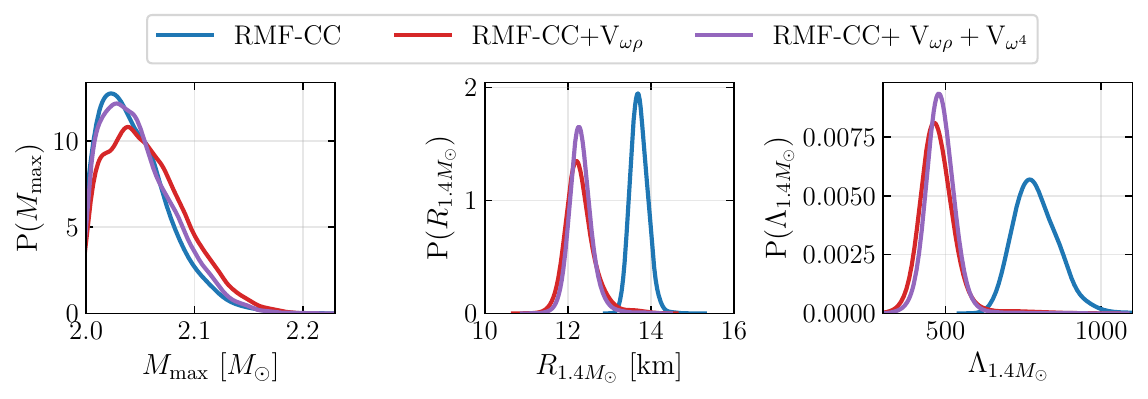}
\caption{The posterior probability distribution for the different models explored in our analysis. From left to right, PDF are shown for $M_{\rm max}$, $R_{1.4M_{\odot}}$ and $\Lambda_{1.4M_{\odot}}$ and they include constraints from LQCD, NEPs, and astrophysics.} 
\label{fig:Mmax_R14_L14_lamomega_rho}
\end{figure*}

We assess the model's overall performance against the combined set of constraints (LQCD, NEPs and astrophysics) and results are shown in \Cref{fig:Mmax_R14_L14_lamomega_rho}. The RMF-CC$+\rm V_{\omega\rho}$ model predicts lower values for $R_{1.4M_{\odot}}$ and $\Lambda_{1.4M_{\odot}}$ compared to the RMF-CC model. This leads to improved agreement with tidal deformability measurements from the BNS event GW170817. In particular, the cumulative distribution function yields,
\begin{equation}
\mathrm{CDF}\left(
\tilde{\Lambda}_{\rm RMF\textrm{-}CC+V_{\omega\rho}}^{D_3}
\leq \tilde{\Lambda}_{\rm GW170817}^{95\%\,\rm upper\ bound}
\right) \sim 0.9 \, .
\end{equation}


\begin{figure*}[htbp]
\centering
\begin{subfigure}{.45\textwidth}
\centering
\includegraphics[width=\linewidth]{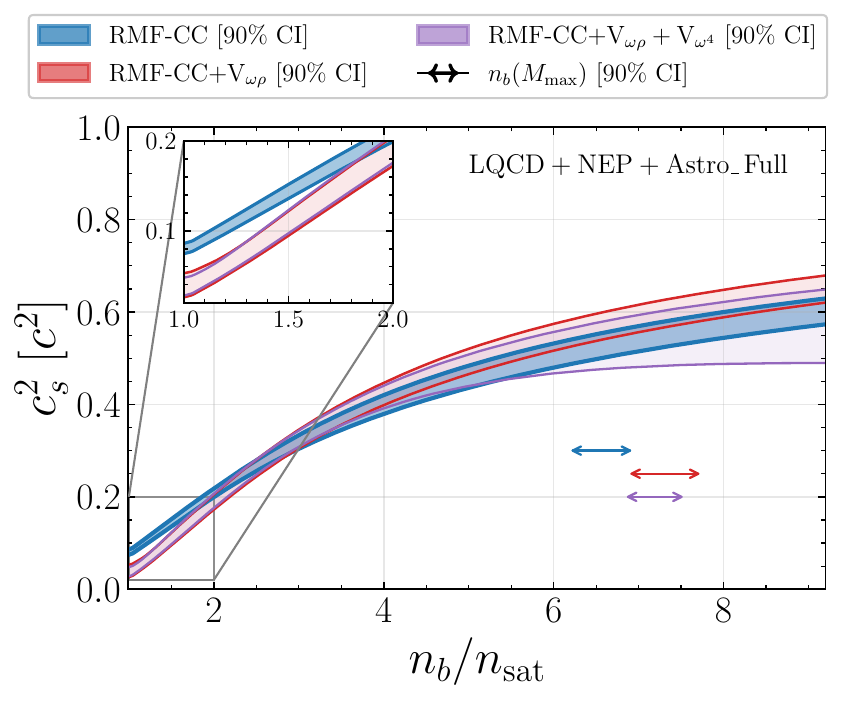}
\caption{} 
\label{fig:cs2_nb_lamomegarho}
\end{subfigure}%
\begin{subfigure}{.45\textwidth}
\centering
\includegraphics[width=0.95\linewidth]{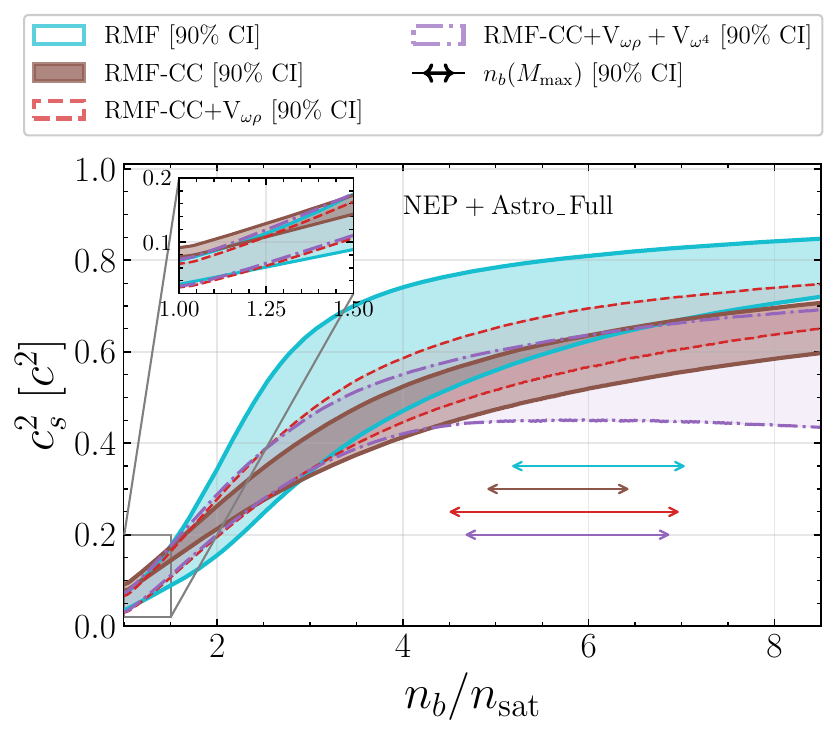}
\caption{}
\label{fig:cs2_nb_NEPAstro}
\end{subfigure}
\caption{ Same as ~\Cref{fig:cs2_RMFCCVLSM}, but for a different class of RMF-CC models as labeled in the figure. (a) obtained using joint constraints from LQCD, NEPs, and astrophysical observations. (b)  obtained using joint constraints from NEPs, and astrophysical observations. The arrows in each panel indicate the uncertainty in the scaled central density $n_b(M_{\rm max})$ corresponding to the maximum NS masses.).
}
\label{fig:Cs2}
\end{figure*}

\begin{figure*}[t]
\centering
\begin{subfigure}{.48\textwidth}
  \centering
  \includegraphics[width=\linewidth]{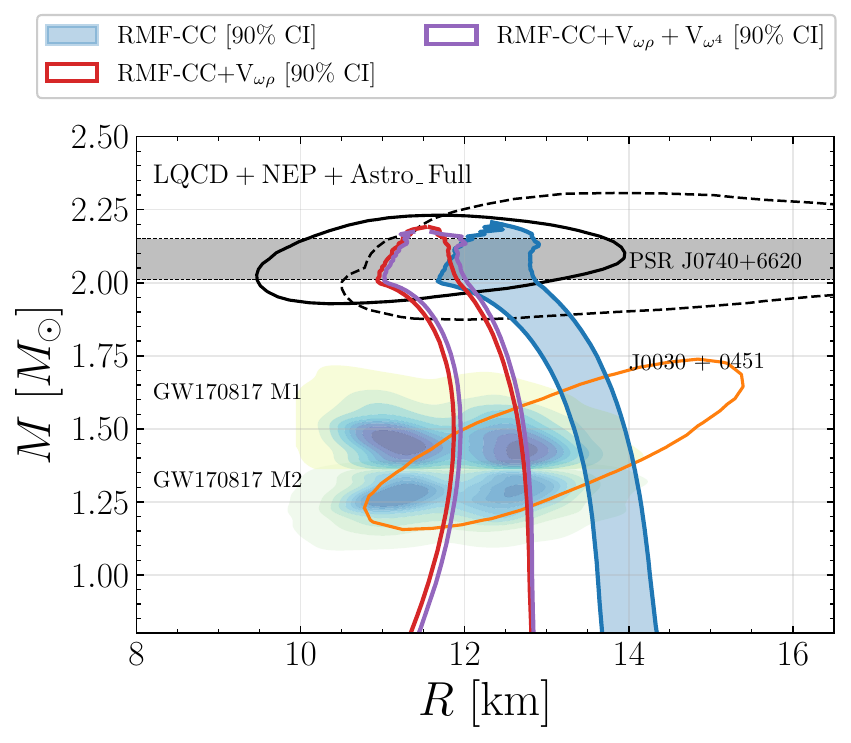}
  \caption{} 
  \label{fig:MR_lamomegarho}
\end{subfigure}%
\begin{subfigure}{.48\textwidth}
  \centering
  \includegraphics[width=\linewidth]{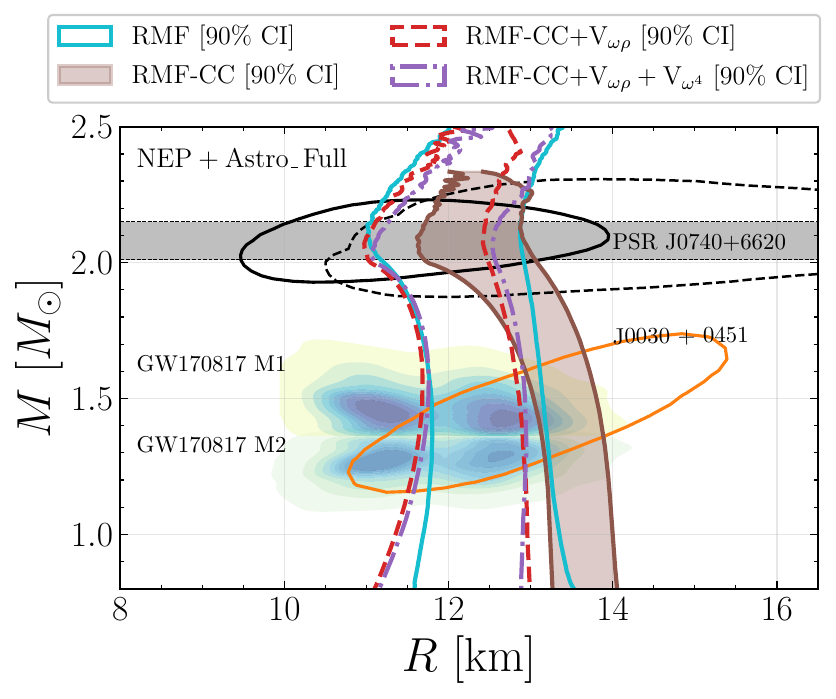}
  \caption{}
  \label{fig:MR_NEPAstro}
\end{subfigure}
\caption{ Uncertainty in $MR$ relation: $R$ as a function of $M$ at 90\% CI for a different  models as labeled in the figure. (a) obtained using joint constraints from LQCD, NEPs, and astrophysical observations. (b)  obtained using joint constraints from NEPs, and astrophysical observations.
}
\label{fig:MR}
\end{figure*}

We show the 90\% credible intervals of the sound speed squared ($c_s^2$) as a function of density under the dataset $D_3=\{ \mathrm{LQCD, \, NEP, \, Astro\_Full}\}$  and  $D_4=\{ \mathrm{ NEP, \, Astro\_Full}\}$ : i.e with and without LQCD constraints in  \Cref{fig:cs2_nb_lamomegarho,fig:cs2_nb_NEPAstro}. The corresponding $MR$ relations for the RMF-CC and RMF-CC+$\mathrm{V}_{\omega\rho} $ models under the dataset $D_3=\{ \mathrm{LQCD, \, NEP, \, Astro\_Full}\}$ and $D_4=\{ \rm{NEP, \, Astro\_Full}\}$ in  \Cref{fig:MR_lamomegarho,fig:MR_NEPAstro}, respectively. In ~\Cref{fig:Cs2,fig:MR} , we have also included the uncertainty resulting from  RMF-CC under same constraints for comparison.

 The inclusion of the $\omega\rho$ coupling softens the EoS in the 1-2$\,n{\rm sat}$ density range (see the inset in \Cref{fig:cs2_nb_lamomegarho} and \Cref{fig:cs2_nb_NEPAstro}), which is relevant for compatibility with GW170817 constraints (see \Cref{fig:MR}). To simultaneously satisfy the $2\, M_\odot$ mass constraint, only those models with sufficiently stiff high-density  EoS are preferred. As a result,  the posterior distribution of $c_s^2$ shifts toward higher values at higher densities compared to the RMF-CC model. Consequently, the posteriors for $K_{\rm sat}$ and $Q_{\rm sat}$ shift toward larger values relative to the RMF-CC model, while the effective nucleon mass $M^*_{\rm SNM}(n{\rm sat})$ shifts to smaller values, favoring high-density stiffness. 

It is expected that the $\omega\rho$ coupling also impacts the properties of finite nuclei, i.e. charge radii and isovector dependence of the binding energy. So, incorporating nuclear structure constraints within the RMF-CC framework as explored in complementary work~\cite{Chamseddine_2026}, can substantially tighten the bounds on $\lambda_{\omega\rho}$ and $g_{\delta}$, and consequently on the iso-vector NEPs $L_{\rm sym}$ and $K_{\rm sym}$. In a future extension of this analysis, we will investigate the combine constraints from finite size nuclei and astrophysics properties on our predictions.

\subsection{RMF-CC with $\omega\rho$ Coupling and Vector Self-Interaction}

In many RMF extension models, particularly those in the FSU family~\cite{Todd2005,Chen2014,Tolos2017}, vector self-interactions are also considered to modify the high-density behavior of the EoS. These interactions can arise from higher-order terms in the bosonization of, an underlying CCM, where next-to-leading-order contributions generate nonlinear vector-meson couplings. In the present work, we adopt a phenomenological approach, consistent with the discussion in the previous section, and explore the impact of a higher-order vector self-interaction term parameterized by the coupling $\zeta$. The resulting Lagrangian density is given by,
\begin{equation}
\mathcal{L}_{\rm RMF\textrm{-}CC+V_{\omega\rho}+V_{\omega^4}}
=
\mathcal{L}_{\rm RMF\textrm{-}CC+V_{\omega\rho}}
+
\frac{\zeta}{4!}\left(g_{\omega}^{2}\,\omega_{\mu}\omega^{\mu}\right)^{2}.
\end{equation}

It is expected that the vector self-interaction  significantly softens the EoS at high densities. The softening occurs because the repulsive $\omega$ field is reduced at high density: instead of increasing linearly with nucleon density ($n_b$), the field $\bar{\omega}$ acquires an approximate cube-root dependence when $\zeta \neq 0$ (see Refs. \cite{Horowitz2001PRC,Pradhan2022NPA}). 
At the same time, the parameter $\zeta$ has a negligible impact around saturation density in SNM and PNM.

Since a nonzero $\zeta$ substantially softens the EoS and lowers the maximum stable NS mass, the coupling constant $\zeta$ is limited to ensure compatibility with the existence of $M_{\rm max}\geq 2\, M_{\odot}$ NSs. Systematic studies, including Bayesian analyses that incorporate astrophysical observations, generally indicate that $\zeta$ is small, prefering either $\zeta\simeq 0$, or at most a small finite value ($\simeq 0.03$~\cite{Pradhan2022NPA}). Consequently, the prior for $\zeta$ is limited to values from $0$ to $0.05$, see \Cref{tab:posteriors_RMFCC_VLSM}.

The values for $\zeta$ in the model RMF-CC$+\rm V_{\omega\rho}+V_{\omega^4}$ are given in \Cref{tab:posteriors_RMFCC_VLSM} for various constraints. The other parameters are not very sensitive to the choice for the parameter $zeta$.
Consistent with previous analyses, we find that when astrophysical observations are applied, the posterior strongly favors $\zeta \simeq 0$ with an maximum upper bound  $\zeta_{\rm max}\sim 0.025$.

The impact of the model RMF-CC$+\rm V_{\omega\rho}+V_{\omega^4}$ on the isoscalar and isovector NEPs are shown in \Cref{tab:isoscalarNEP_posteriors_RMFCC_VLSM,tab:isovectorNEP_posteriors_RMFCC_VLSM}. There is no significant difference between the NEPs predicted by the model RMF-CC$+\rm V_{\omega\rho}$ and the model RMF-CC$+\rm V_{\omega\rho}+V_{\omega^4}$. It is confirmed for the PDF for the NEPs shown in \Cref{fig:lsym_ksym_lamomega_rho,fig:nsat_ksat_qsat_lamomega_rho} and for the global NS properties shown in \Cref{fig:Mmax_R14_L14_lamomega_rho}. The posterior distributions of the leading isoscalar, isovector NEPs, and NS properties with joint consideration of LQCD,NEP and astrophysical costraints are displayed in 
\Cref{fig:nsat_ksat_qsat_lamomega_rho,fig:lsym_ksym_lamomega_rho,fig:Mmax_R14_L14_lamomega_rho}, respectively. The posteriors of the NEPs and NS properties obtained with RMF-CC$+\rm V_{\omega\rho}+V_{\omega^4}$ are comparable with those from the RMF-CC$+\rm V_{\omega\rho}$  model, since the dominant effect is already captured by the $\omega\rho$ interaction and the Bayesian inference favors $\zeta\simeq 0$. 

The sound speed ($c^2_s$) as a function of density resulting from the data-set $D_3$ and $D_4$ are shown in \Cref{fig:cs2_nb_lamomegarho} and \Cref{fig:cs2_nb_NEPAstro}, respectively.  The uncertainties on the $MR$ relations with the consideration of $D_3$ and $D_4$ are displayed in ~\Cref{fig:MR_lamomegarho} and ~\Cref{fig:MR_NEPAstro}, respectively. The resulting uncertainties on the $c^2$ as well as in $MR$ relation obtained with RMF-CC$+\rm V_{\omega\rho}+V_{\omega^4}$ model  are comparable to those resulting from  the RMF-CC$+\rm V_{\omega\rho}$  model.

In summary, we found that the RMF-CC model present tension incorporating the constraints from LQCD, NEPs and astrophysics.
Introducing $\omega\rho$ coupling is very important to reduce this tension, while the vector self-interaction does not change the predictions obtained for RMF-CC$+\rm V_{\omega\rho}$. 

In the following, we develop statistical tools to systematically compare the different models. Among  RMF-CC, RMF-CC$+\rm V_{\omega\rho}$ and RMF-CC$+\rm V_{\omega\rho}\!+\!V_{\omega^4}$,  the  RMF-CC$+\rm V_{\omega\rho}$ model exhibits the largest $\log(\mathcal{Z})$ when astrophysical constraints irrespective of consideration of LQCD constraints. Hence, the RMF-CC and  RMF-CC$+\rm V_{\omega\rho}\!+\!V_{\omega^4}$ models are compared against  RMF-CC$+\rm V_{\omega\rho}$. The Bayes factors
$\mathcal{B}^i_{\rm RMF\textrm{-}CC+V_{\omega\rho}}, \quad i = \{ \rm RMF\text{-}CC,\  RMF\textrm{-}CC\!+\!\rm V_{\omega\rho}\!+\!V_{\omega^4} \}$, are tabulated in  \Cref{tab:Bayesfactor_LQCDNEPASTRO}.

\begin{table}[t]
\centering
\caption{Bayes factors, $\log_{10}\mathcal{B}^i_j$, evaluated using the two sets of constraints mentioned  in the table: NEP+Astro\_$\,{\rm Full}$ and LQCD+NEP+Astro\_$\,{\rm Full}$. See text for more details.}
\label{tab:Bayesfactor_LQCDNEPASTRO}
\setlength{\tabcolsep}{0.5em}
\renewcommand{\arraystretch}{1.25}
\begin{tabular}{l|c|c}
\hline\hline
\diagbox{$\mathcal{B}^i_j$}{Constraints}
 & LQCD+NEP& NEP  \\
 & +Astro$\_\,$Full  & +Astro$\_\,$Full\\
 \hline\hline
$\log_{10}\left[\mathcal{B}^{\rm RMF-CC}_{ \rm RMF-CC+V_{\omega\rho}} \right]$ & -1.13 & -1.24 \\
$\log_{10}\left[\mathcal{B}^{\rm RMF-CC+V_{\omega\rho}+V_{\omega^4}}_{ \rm RMF-CC+V_{\omega\rho}} \right]$ & -0.26 & -0.19 \\

$\log_{10}\left[\mathcal{B}^{\rm RMF-CC}_{ \rm RMF} \right]$ & -- & -1.91 \\
$\log_{10}\left[\mathcal{B}^{\rm RMF-CC+V_{\omega\rho}}_{ \rm RMF} \right]$ & -- & -0.67 \\
$\log_{10}\left[\mathcal{B}^{\rm RMF-CC+V_{\omega\rho}+V_{\omega^4}}_{ \rm RMF} \right]$ & -- & -0.86 \\

\hline\hline
\end{tabular}
\end{table}

From the Bayes factors tabulated  in \Cref{tab:Bayesfactor_LQCDNEPASTRO}, one can conclude that, in the vicinity of the astrophysical constraints, there is strong evidence in favor of the RMF-CC$+\mathrm{V}_{\omega\rho}$ model over the baseline RMF-CC model without the $\omega\rho$ interaction. In contrast,  RMF-CC$+\rm V_{\omega\rho}$ and RMF-CC$+\rm V_{\omega\rho}\!+\!V_{\omega^4}$ exhibit similar preference, as the Bayes factor comparison yields insubstantial evidence.

\subsection{Comparison of RMF-CC models with RMF model}\label{sec:compare_RMF}


\begin{figure*}[t]
    \centering
    \includegraphics[width=0.99\linewidth]{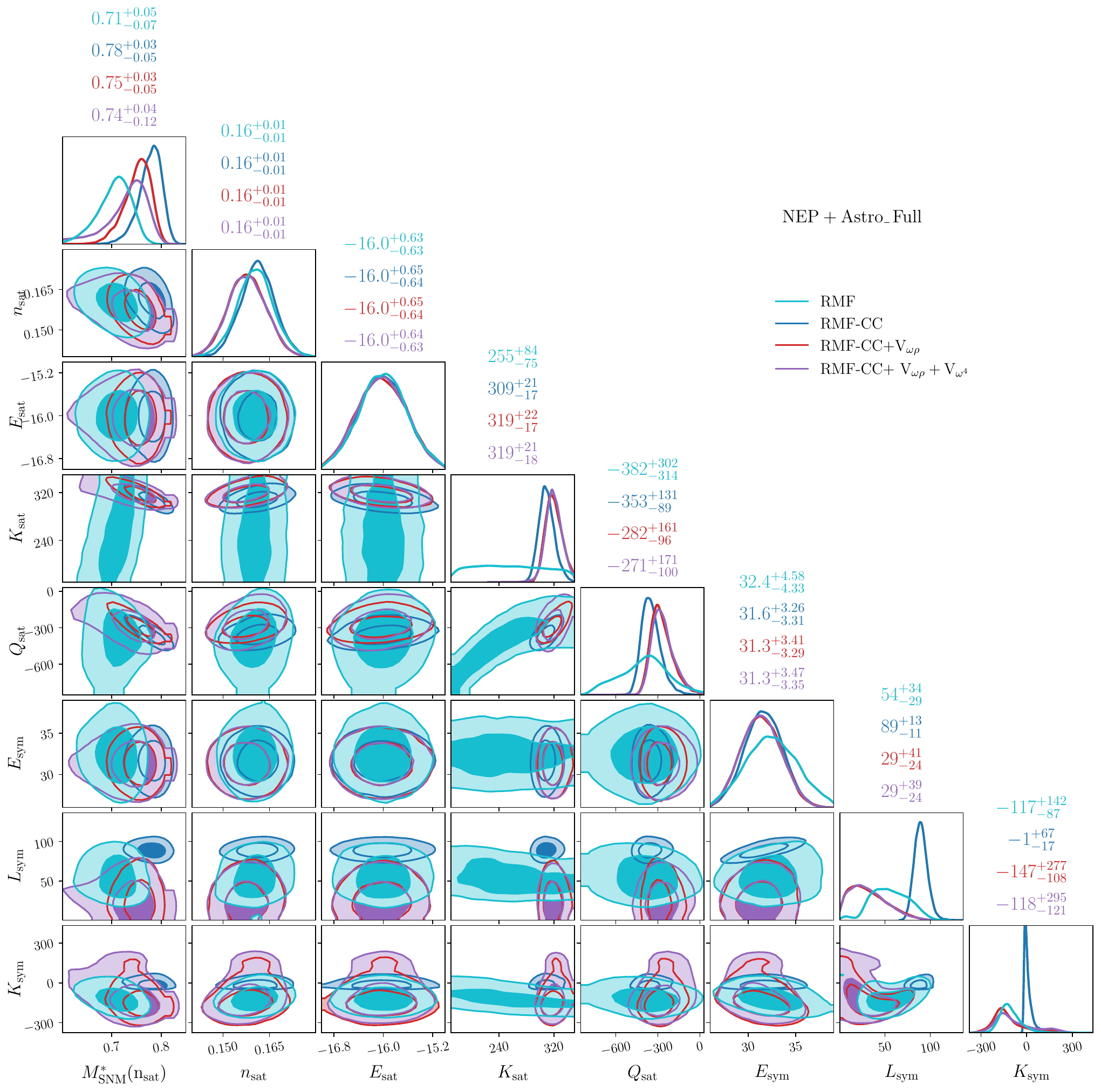}
    \caption{Same as ~\Cref{fig:NEPs_RMFCC}, but for the resulting posteriors from different RMF-CC models, labeled by color. The posteriors are obtained with $D_4=\{ \rm NEP, Astro\_\,Full\}$. For comparison, the predictions for the conventional RMF model with the same constraints are also shown. }
    \label{fig:NEPs_NEPAstro}
\end{figure*}

The comparison between the RMF and the aforementioned RMF-CC models is performed without imposing LQCD constraints on any of them, since the RMF model is not calibrated to LQCD. 

There have been extensive investigations in the literature aimed at improving the RMF framework in light of astrophysical observations, particularly by examining the role of non-linear vector meson interactions like $\omega^4$. Numerous studies have demonstrated that such non-linear vector couplings play a crucial role in enhancing the consistency of RMF models with current multi-messenger astrophysical data. In the present work, however, we do not aim to provide a detailed Bayesian inference or a systematic development of the RMF model itself. Instead, we adopt a representative RMF model incorporating non-linear vector interactions ($V_{\rm MCs}$ defined in Eq.~\eqref{eq:V_NL}), together with specific choices of the scalar potential $V_{\rm RMF}(s)$ defined in Eq.~\eqref{eq:VRMF} and the effective nucleon mass $M_{N,\rm RMF}(s)$ as defined in Eq.~\eqref{eq:MN_RMF}. This framework is used to facilitate a direct comparison between the RMF and RMF-CC models, focusing on their respective predictions and based on the same constraints from NEPs and astrophysical observations.

\begin{figure}[tb]
    \centering
    \includegraphics[width=\linewidth]{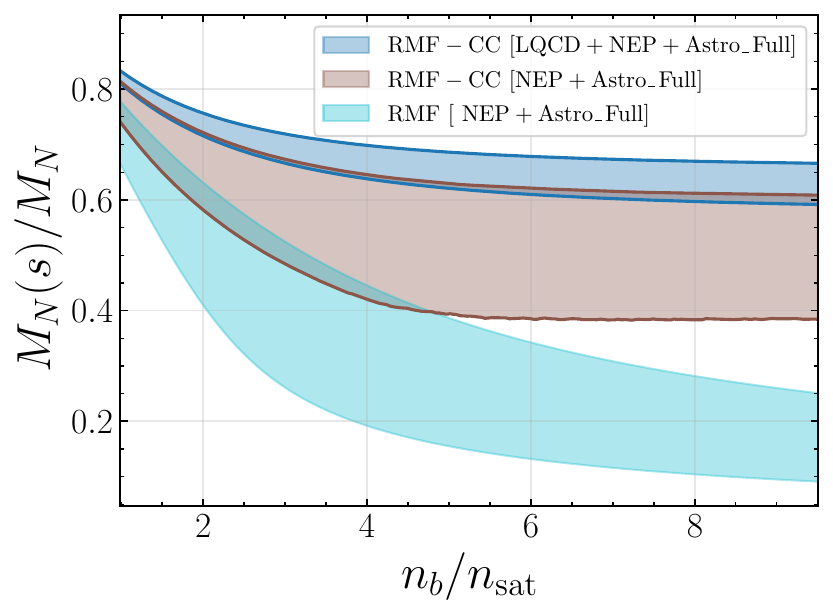}
    \caption{Variation of $M_N(s)$  as a function of scaled density at a 90\% CI for RMF-CC and RMF models in NS matter.}
    \label{fig:Mstar_compare_NEPAstro}
\end{figure}

\begin{figure}[t]
    \centering
    \includegraphics[width=0.95\linewidth]{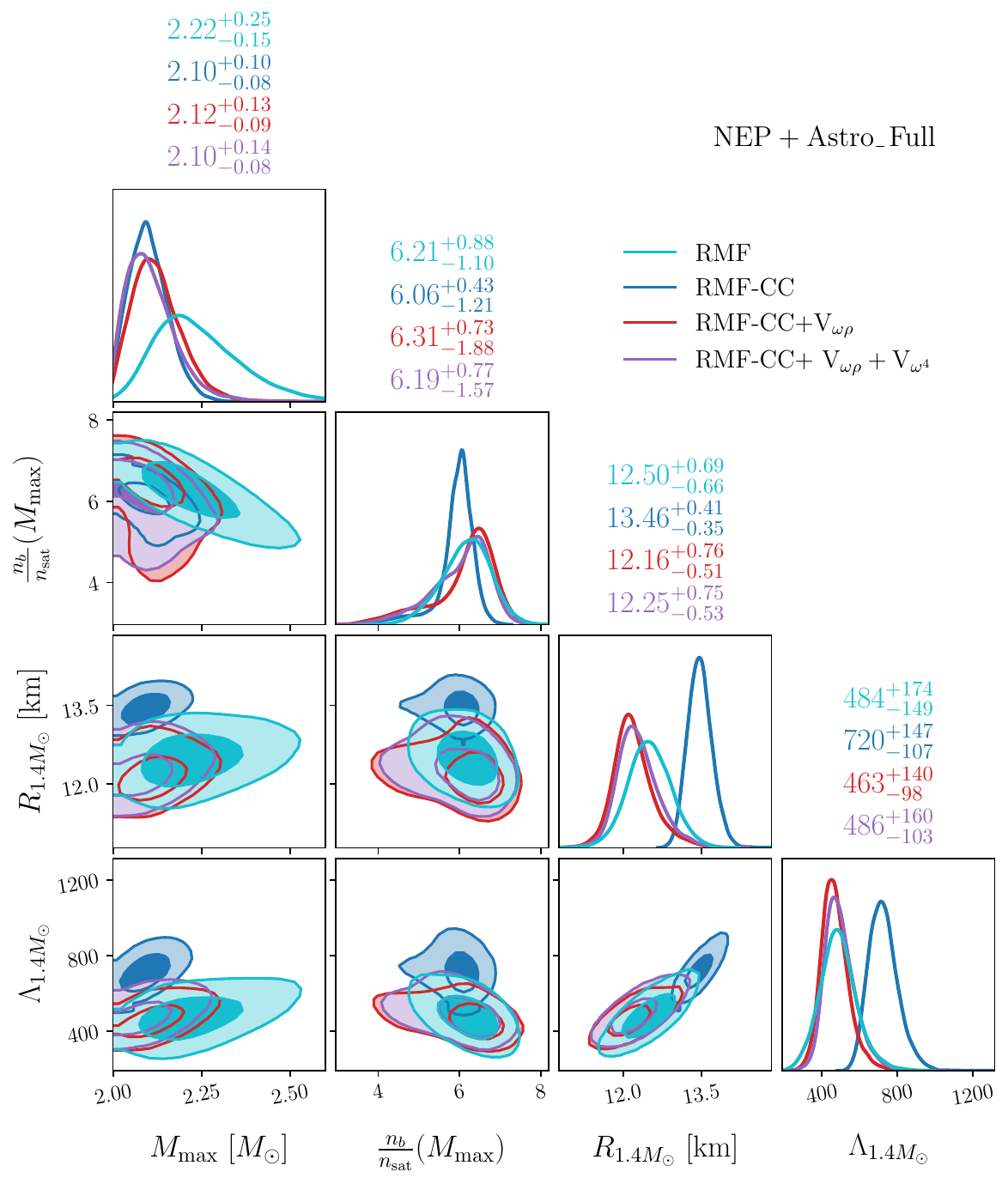}
    \caption{Same as ~\Cref{fig:R14_L14_Mmax_RMFCCVLSM}, but for the  posteriors resulting  with $D_4=\{ \rm NEP, Astro\_\,Full\}$ and different RMF-CC models, labeled by color. For comparison, the predictions for the conventional RMF model with the same constraints are also shown.}
    \label{fig:NS_Props_NEPAstro}
\end{figure}

Within the RMF model, the median values and $90\%$ CIs of the NEPs obtained from the posterior distributions are tabulated in \Cref{tab:isoscalarNEP_posteriors_RMFCC_VLSM} and \Cref{tab:isovectorNEP_posteriors_RMFCC_VLSM} for the isoscalar and isovector sectors, respectively. For comparison, we present the joint and marginalized posterior distributions of the NEPs and NS properties for the RMF, RMF-CC, RMF-CC$+\mathrm{V}_{\omega\rho}$, and RMF-CC$+\rm V_{\omega\rho}+V_{\omega^4}$ models in \Cref{fig:NEPs_NEPAstro} and \Cref{fig:NS_Props_NEPAstro}, respectively. The resulting predictions for NS properties and NS EoS quantities, with consideration of  NEP and astrophysical constraints for RMF model  are tabulated in \Cref{tab:NS_properties}.  We display the uncertainty of the $c^2_s$ as a function of density and $MR$ relation for the  RMF model along with the RMF-CC, RMF-CC$+\mathrm{V}_{\omega\rho}$, and RMF-CC$+\rm V_{\omega\rho}+V_{\omega^4}$ models models in ~\Cref{fig:cs2_nb_NEPAstro} and ~\Cref{fig:MR_NEPAstro}, respectively. 

The RMF-CC model produces stiffer EoSs than the RMF model in the density range $1-2\,n_{\rm sat}$. At higher densities, however, the RMF model becomes stiffer, driven by a more rapid increase of the sound speed with density, whereas the RMF-CC model exhibits a smoother rise. The softer behavior of the RMF EoS and the gentler slope of $c_s^2(n_b)$ in the intermediate density range $1-2\,n_{\rm sat}$ play an essential role in reproducing the GW170817 constraints. 

The inclusion of the $\omega\rho$ interaction in RMF-CC, \emph{i.e.}  RMF-CC$+\rm V_{\omega\rho}$, reduces the sound speed $c_s^2(n_b)$ in the $1\textrm{-}1.5\,n_{\rm sat}$ region, bringing it into closer agreement with both the RMF model and astrophysical data. Nevertheless, the high-density behavior of the RMF-CC models remains constrained by the scalar potential and nucleon polarization effects. This difference is reflected in the predicted maximum masses: the RMF model supports more massive NSs and exhibits a broader posterior distribution for $M_{\rm max}$ at higher values than the RMF-CC models, see \Cref{fig:NS_Props_NEPAstro}.

We find that within the RMF-CC framework, the reproduction of massive $2 M_{\odot}$ NSs results in a relatively narrow distribution for $K_{\rm sat}$ around a higher value $\sim300$ MeV, see \Cref{fig:NS_Props_NEPAstro}. In contrast, the RMF model prefers a broader distribution for $K_{\rm sat}$, including, for instance, $K_{\rm sat} \sim 250$~MeV and even lower. Although $Q_{\rm sat}$ remains weakly constrained in all cases, with large uncertainties, RMF-CC models systematically favor larger values compared to the RMF model.

While  RMF-CC$+\rm V_{\omega\rho}$ improves agreement with astrophysical data, it predicts lower values of $L_{\rm sym} \sim 30$ MeV compared to $L_{\rm sym} \sim 55$ MeV in the RMF model, despite a substantial overlap of their respective posterior distributions, see ~\Cref{fig:NEPs_NEPAstro}. 



In RMF-CC models, $M_N^*(n_{\rm sat})$ is constrained to relatively large values to reproduce the NEPs as discussed in Ref.~\cite{Somasundaram:2021hna}, see also \Cref{fig:NEPs_NEPAstro}. Even with the inclusion of astrophysical constraints, $M_{\rm SNM}^*(n_{\rm sat})$ is predicted with higher values in RMF-CC models compared to the RMF model, as it is required to reproduce NEPs simultaneously.

To further illustrate the role of nucleon polarization, we compare $M_N^*(s)$ in NS matter for RMF and RMF-CC  in \Cref{fig:Mstar_compare_NEPAstro}. The confining mechanism in RMF-CC prevents the nucleon mass from decreasing at high densities $n_b>4n_\sat$, predicting a constant value for $M_N$ for high densities. As a consequence, the repulsive kinetic contribution to the EoS is limited in RMF-CC models, leading to softer EoSs. Hence, RMF-CC models predict the $M_{\rm max}$ distribution towards smaller values compared to the RMF model, see \Cref{fig:NS_Props_NEPAstro}. 

With the inclusion of the LQCD constraints the nucleon mass $M_N(s)$ at high densities is strongly restricted to larger values, see \Cref{fig:Mstar_compare_NEPAstro} and also \Cref{fig:nsat_ksat_qsat_lamomega_rho} showing the Dirac mass at $n_\sat$, which in turn restricts the maximum mass $M_{\rm max}$ to lower values within the RMF-CC model, see \Cref{fig:Mmax_R14_L14_lamomega_rho}.

With the data set $D_4=\{ \rm NEP, Astro\_\,Full\}$, we found maximum $\log_{10}{\mathcal{Z}}$ for the RMF model compared to the RMF-CC,  RMF-CC$+\rm V_{\omega\rho}$ and  RMF-CC$+\rm V_{\omega\rho}\!+\!V_{\omega^4}$ models. Hence, we provide the Bayes factor comparison against the RMF model in \Cref{tab:Bayesfactor_LQCDNEPASTRO}. Following Ref.~\cite{Kass1995} and the results given in \Cref{tab:Bayesfactor_LQCDNEPASTRO}, we find strong evidence in favor of the RMF model over the RMF-CC discussed in ~\Cref{sec:RMFCC_VLSM}. Although, we have noticed significant improvements within RMF-CC$+\mathrm{V}_{\omega\rho}$ compared to the RMF-CC model of Ref.~\cite{Somasundaram:2021hna}, the statistical analysis still predicts a substantial evidence in favor of RMF model compared to RMF-CC$+\rm V_{\omega\rho}$ and  RMF-CC$+\rm V_{\omega\rho}\!+\!V_{\omega^4}$  models in the presence of astrophysical constraints. We remind the reader that this result ignores the possible impact of phase transitions.

\section{Conclusion}\label{sec:conclusion} 

In this work, we performed a comprehensive investigation of the CCM within the Hartree framework, constrained by nuclear physics inputs, multi-messenger astrophysical observations, and/or LQCD predictions of the nucleon mass.

We find that the RMF-CC model discussed in Ref.~\cite{Somasundaram:2021hna} provides a consistent and accurate description of low-density nuclear matter, successfully reproducing lattice QCD results and NEPs within uncertainties. The RMF-CC framework provides a microphysically constrained framework, where chiral symmetry and confinement-induced nucleon response provides constraints to the scalar sector, which  gives the model interpretive power but reduces predictability at higher densities, i.e., stiffness of the EoS, effective-mass evolution, or density dependence of the isovector channel. Astrophysical data, therefore, expose a structural tension: supporting two-solar-mass stars requires a large incompressibility and a stiff high-density EoS, while GW170817 favors softer radii/tidal deformabilities. The $\omega\rho$ coupling provides the minimal isovector lever required to soften radii and tidal deformabilities, whereas the remaining maximum-mass tension reflects the high-density scalar/confinement sector. Thus, the paper identifies not only parameter constraints but also the missing degrees of freedom needed for the next generation of RMF-CC EoSs.

We further notice that the additional vector self-interaction does not change the predictions of RMF-CC$+\rm V_{\omega\rho}$. In the case of the RMF model, which predicts NS masses $> 2.5 M_{\odot}$, the vector self-interaction moderates the stiffness at high-density, which reduces the maximum NS mass and makes it more compatible with astrophysical predictions. However, within the RMF-CC framework, vector self-interaction is less significant, since the model inherently produces softening at high-density and predict maximum masses $\sim 2$-$2.2M_{\odot}$.

The additional LQCD constraint impacts not only the low density region of the EoS, but also its high density region. In particular, it restricts the scalar sector of the RMF-CC, more specifically the scalar response, which thereby indirectly constrains the high-density EoS. As shown in \Cref{fig:EoS_RMFCCVLSM,fig:cs2_RMFCCVLSM} and \Cref{tab:NS_properties}, the RMF-CC models including the LQCD constraint predict a pressure (respectively a sound speed $c_s^2$) at $n_b=0.3$~fm$^{-3}$ with a standard deviation of $\sim 10\%$ (9\%) compared to $\sim$20\% ($\sim$ 17\%) error in the absence of LQCD constraint.


In future extensions, this work can be further refined by simultaneously constraining the dense-matter EoS and model parameters using terrestrial experimental inputs, such as the neutron-skin measurements from CREX~\cite{Adhikari_CREX2022} and PREX~\cite{Adhikari_PREX2021}, theoretical predictions for PNM~(see Ref.~\cite{nucleardatapy2026} and references therein), and experimental constraints in intermediate density resulting from KaoS~\cite{Hartnack_KaoS2006}, FOPI~\cite{LEFEVRE_FOPI2016}, and ASY-EoS~\cite{Russotto_ASYEoS2016}. The present framework may also be extended to solve EoS beyond the Hartree approximation, or by incorporating additional degrees of freedom, such as including strange baryons (hyperons), or by allowing for a phase transition to deconfined quark matter at high densities~\cite{Pfaff2022}. Furthermore, the chiral confinement model can be compared with other microscopic EoS models such as  RMF models with density-dependent couplings~\cite{Frohaug2025,Passarella2025},  chiral RMF models in the absence of nucleon polarization~\cite{Fraga2019,Schmitt2020}, parity doublet model (PDM)~\cite{Dexheimer2008,Gao2024}, and the quark–meson coupling (QMC) model~\cite{GUICHON2018262,Motta2019,Antic2020}. The next step for this analysis will be the simultaneous reproduction of finite nuclei and dense matter within the same framework.

\section*{Acknowledgment}
The authors acknowledge the support the ANR project RELANSE ANR-23-CE31-0027-01 of the French National Research Agency (ANR), the CNRS-IN2P3 MAC2 masterproject, the European Union’s Horizon 2020 research and innovation program under grant agreement STRONG–2020-No824093.

\bibliographystyle{apsrev4-2}
\bibliography{Paper}

\end{document}